\documentclass[11pt]{article}
\usepackage[letterpaper,margin=0.8in]{geometry}
\usepackage{titling}
\usepackage{mdframed}
\usepackage[linesnumbered,lined,boxed,commentsnumbered]{algorithm2e}
\usepackage{listings}
\usepackage{tikz}
\usepackage{xcolor}
\usetikzlibrary{shapes.geometric, arrows}
\usepackage{amsmath}
\usepackage{color}
 \usepackage[noend]{algpseudocode}
\usepackage{amssymb}
\usepackage{faktor}
\usepackage{xfrac}
\usepackage[utf8]{inputenc}

\usetikzlibrary{matrix,shapes,arrows,positioning,chains}
\usepackage{scalerel}
\usetikzlibrary{fit,arrows,calc,positioning}
\usepackage{enumitem}
\usepackage{authblk}

\usepackage{hyperref}
\hypersetup{
    colorlinks,
    citecolor=black,
    filecolor=black,
    linkcolor=blue,
    urlcolor=black
}

\newlist{longenum}{enumerate}{5}
\setlist[longenum,1]{label=\roman*)}
\setlist[longenum,2]{label=\alph*)}
\setlist[longenum,3]{label=\arabic*)}
\setlist[longenum,4]{label=(\roman*)}
\setlist[longenum,5]{label=(\alph*)}

\tikzstyle{startstop} = [rectangle, rounded corners, minimum width=3cm, minimum height=1cm,text centered, text width=3cm, draw=black, fill=green!30]

\tikzstyle{io} = [trapezium, trapezium left angle=70, trapezium right angle=110, minimum width=3cm, minimum height=1cm, text centered,
text width=3.7cm, draw=black, fill=blue!30]

\tikzstyle{process} = [rectangle, minimum width=3cm, minimum height=1cm, text centered, text width=3cm, draw=black, fill=orange!30]
\tikzstyle{decision} = [diamond, minimum width=2.5cm, minimum height=1cm, text centered, text width=2.7cm, draw=black, fill=green!30]
\tikzstyle{arrow} = [thick,->,>=stealth]

\newcommand\reallywidehat[1]{\arraycolsep=0pt\relax%
\begin{array}{c}
\stretchto{
  \scaleto{
    \scalerel*[\widthof{\ensuremath{#1}}]{\kern-.5pt\bigwedge\kern-.5pt}
    {\rule[-\textheight/2]{1ex}{\textheight}} %WIDTH-LIMITED BIG WEDGE
  }{\textheight} %
}{0.5ex}\\           % THIS SQUEEZES THE WEDGE TO 0.5ex HEIGHT
#1\\                 % THIS STACKS THE WEDGE ATOP THE ARGUMENT
\rule{-1ex}{0ex}
\end{array}
}

\newcommand{\eqdef}{\stackrel{def}{=}}
\newcommand{\N}{\mathbb{N}}
\newcommand{\F}{\mathbb{F}}
\newcommand{\Q}{\mathbb{Q}}

\newcommand{\C}{\mathbb{C}}
\newcommand{\ML}{\mathcal{L}}
\newcommand{\MB}{\mathcal{B}}
\newcommand{\MX}{\mathcal{X}}

\newcommand{\ME}{\mathcal{E}}
\newcommand{\MS}{\mathcal{S}}
\newcommand{\MC}{\mathcal{C}}

\newcommand{\R}{\mathbb{R}}
\newcommand{\Z}{\mathbb{Z}}
\newcommand{\MP}{\mathbb{P}}

\newcommand{\B}[1]{\bar{#1}}
\newcommand{\ti}[1]{\tilde{#1}}

\newtheorem{theorem}{Theorem}[section]

\newtheorem{claim}[theorem]{Claim}

\newtheorem{lemma}[theorem]{Lemma}
\newtheorem{corollary}[theorem]{Corollary}

\newtheorem{definition}[theorem]{Definition}

\newtheorem{example}[theorem]{Example}

\begin{document}

\title{Reconstruction of depth-3, top fan-in two circuits over characteristic zero fields}
\author{Gaurav Sinha
\footnote{Department of Mathematics, California Institute of Technology, Pasadena CA 91106, USA. email : gsinha@caltech.edu}}

%\email{gsinha@caltech.edu}
\date{}
\maketitle

%%%%%%%%%%%%%%%%%%%%%%%%%%%%%%%%%%%%%
% BEGIN BODY of Document

\begin{abstract}
Reconstruction of arithmetic circuits has been heavily studied in the past few years and has connections to proving lower bounds and deterministic identity testing.
In this paper we present a polynomial time randomized algorithm for reconstructing $\Sigma\Pi\Sigma(2)$ circuits over $\F$ ($char(\F)=0$), i.e. depth$-3$ circuits 
with fan-in $2$ at the top addition gate and having coefficients from a field of characteristic $0$.

The algorithm needs only a blackbox query access to the polynomial $f \in \F[x_1,\ldots, x_n]$ of degree $d$, computable by a $\Sigma\Pi\Sigma(2)$ circuit $C$. 
In addition, we assume that the \emph{"simple rank"} of this polynomial (essential number of variables after removing the gcd of the two multiplication gates) is 
bigger than a fixed constant. Our algorithm runs in time $poly(n, d)$ and returns an equivalent $\Sigma\Pi\Sigma(2)$ circuit(with high probability).

The problem of reconstructing $\Sigma\Pi\Sigma(2)$ circuits over finite fields was first proposed by Shpilka \cite{Shpilka07}. 
The generalization to $\Sigma\Pi\Sigma(k)$ circuits, $k = O(1)$ (over finite fields) was addressed by Karnin and Shpilka in \cite{KarShp09}. 
The techniques in these previous involve iterating over all objects of certain kinds over the ambient field and thus the running time depends on the size 
of the field $\F$. Their reconstruction algorithm uses lower bounds on the lengths of Linear Locally Decodable Codes with $2$ queries. In our settings, such 
ideas immediately pose a problem and we need new ideas to handle the case of the
characteristic $0$ field $\F$.

Our main techniques are based on the use of Quantitative Syslvester Gallai Theorems from the work of Barak et.al. \cite{BDWY11} to find a small collection of 
\emph{"nice"} subspaces to project onto. The heart of our paper lies in subtle applications of the Quantitative Sylvester Gallai theorems to prove why projections 
w.r.t. the \emph{"nice"} subspaces can be ”glued”. We also use Brill's Equations from \cite{GKZ94} to construct a small set of candidate linear forms 
(containing linear forms from both gates). Another important technique which comes very handy is the polynomial time randomized algorithm for factoring multivariate 
polynomials given by Kaltofen \cite{KalTr90}.

\end{abstract}

\tableofcontents

\section{Introduction} \label{introduction}

The last few years have seen significant progress towards interesting problems dealing with arithmetic circuits.
Some of these problems include Deterministic Polynomial Identity Testing, Reconstruction of Circuits and
recently Lower Bounds for Arithmetic Circuits. There has also been work connecting these three different
aspects. In this paper we will primarily be concerned with the reconstruction problem.
Even though it's connections to Identity Testing and Lower Bounds are very exciting, the problem
in itself has drawn a lot of attention because of elegant techniques and connections to learning.
The strongest version of the problem requires that for any $f\in
\F[x_1,\ldots,x_n]$ with blackbox access given one wants to construct (roughly)
most succinct representation i.e. the smallest possible arithmetic circuit computing the polynomial. This general
problem appears to be very hard. Most of the work done has dealt with some special type of polynomials i.e. the ones
which exhibit constant depth circuits with alternating addition and
multiplication gates. Our result adds to this by looking at polynomials computed by circuits of this type (alternating addition/multiplication
gates but of depth $3$). Our circuits will have variables at the leaves, operations $(+,\times)$ at the gates and scalars at the edges.
We also assume that the top gate has only two children and the \emph{"simple rank"} of this polynomial (essential number of variables
after removing the gcd of the two multiplication gates) is bigger than a constant. The bottom most layer has addition gates and so computes linear forms, the middle layer then multiplies these
linear forms together and the top layer adds two such products. Later in Remark \ref{homogen} we discuss
that we may assume the linear forms computed at bottom level to be homogeneous and the in-degree of all gates at
middle level to be the same $(=$ degree of $f)$. Therefore these circuits compute polynomials with the following form :
\[
 f(x_1,\ldots,x_n) = G(x_1,\ldots,x_n)(T_0(x_1,\ldots,x_n) + T_1 (x_1,\ldots,x_n))
\]

where $T_i(x_1,\ldots,x_n) = \prod\limits_{j=1}^M l_{ij}$ and $G(x_1,\ldots,x_n) = \prod\limits_{j=1}^{d-M}G_j$ with the $l_{ij}$'s and $G_j$'s being linear forms
for $i\in \{0,1\}$. Also assume $gcd(T_0,T_1)=1$. Our condition about the essential number of variables (after removing gcd from
the multiplication gates) is called \emph{"simple rank"} of the polynomial and is defined as
dimension of the space
\[
 sp\{l_{ij} : i\in \{0,1\}, j\in \{1,\ldots,M\}\}
\]

\par{}
When the underlying field $\F$ is of characteristic $0$ ($\Q,\R$ or $\C$ for simplicity), we give an efficient randomized algorithm for reconstructing the circuit
representation of such polynomials. Formally our main theorem reads :

\begin{theorem}\label{maintheorem}[$\Sigma\Pi\Sigma_\F(2)$ Reconstruction Theorem]
 Let $f = G(T_0+T_1) \in \F[x_1,\ldots,x_n]$ be any degree $d$, $n-$ variate
polynomial (to which we have blackbox access) which can be computed by a depth
$3$ circuit with top fan-in
$2$ (i.e. a $\Sigma\Pi\Sigma(2)$ circuit) i.e. $G,T_i$ being products of affine forms. Assume
$gcd(T_0,T_1)=1$ and $span\{l : l\mid T_0T_1\}$ is bigger than  $s+1$ (a fixed constant defined below).
We give a randomized algorithm which runs
in time $poly(n,d)$ and computes the circuit for $f$ with high probability.
\end{theorem}

\begin{definition}\label{rvalue}
\begin{mdframed}
We fix $s$ to be any constant $> \max (C_{2k-1}+k, c_\F(4) )$ where :
\begin{enumerate}
\item $c_{\F}(l)=3l^2$ is the rank lower bound (see Theorem \ref{rankbound}) that guarantees nonzero-ness of any simple, minimal, $\Sigma\Pi\Sigma(l)$ circuit
with rank $> c_\F (l)$.
\item $k = c_{\F}(3) +2$.
\item $\delta$ is some fixed number in $(0,\frac{7-\sqrt{37}}{6})$.
 \item $C_k = \frac{C^k}{\delta}$ the constant that appears in Theorem \ref{bdwy}.

\end{enumerate}
\end{mdframed}
\end{definition}

From our discussion before the theorem about Remark \ref{homogen}, we can assume in the above theorem that the polynomial
and all linear forms involved are homogeneous.

As per our knowledge this is the first algorithm that efficiently
reconstructs such circuits (over the char $0$ fields). Over finite fields, the same problem has been considered by \cite{Shpilka07} and
our method takes inspiration from their work. They also generalized this finite field version to circuits with arbitrary (but constant)
top fan-in in \cite{KarShp09}. However we need many new tools and techniques as their methods don't generalize at
a lot of crucial steps. For eg:
\begin{itemize}
\item They iterate through linear forms in a finite field which
we unfortunately cannot do.
\item They use lower bounds for Locally Decodable Codes given in \cite{DS07} which again
does not work in our setup.
\end{itemize}
We resolve these issues by
\begin{itemize}
 \item Constructing candidate linear forms by solving simultaneous polynomial equations obtained from Brill's Equations (Chapter 4, \cite{GKZ94}).
 \item Using quantitative versions of the Sylvester
Gallai Theorems given in \cite{BDWY11} and \cite{DSW12}. This new method enables us to construct $nice$
subspaces, take projections onto them and glue the projections back to recover the circuit representation.
\end{itemize}

\subsection{Previous Work and Connections}

Efficient Reconstruction algorithms are known for some concrete class of circuits. We
list some here:
\begin{itemize}
 \item Depth-2 $\Sigma\Pi$ circuits (sparse polynomials) in \cite{KS01}
 \item Read-once arithmetic formulas in \cite{SV09}
 \item Non-commutative ABP's \cite{ArMS08}
 \item $\Sigma\Pi\Sigma(2)$ circuits over finite fields in \cite{Shpilka07}, extended to
 $\Sigma\Pi\Sigma(k)$ circuits (over finite fields) with $k=O(1)$ in \cite{KarShp09}.
\item Random Multi-linear Formulas in \cite{GuptaKL11}
\item Depth $4$ ($\Sigma\Pi\Sigma\Pi$) multi-linear circuits with top fan-in $2$ in \cite{GuptaKL12}
\item Random Arithmetic Formulas in \cite{GKY14}
\end{itemize}
All of the above work introduced new ideas and techniques and have been greatly
appreciated.  \\

It's straightforward to observe that a polynomial time
deterministic reconstruction algorithm for a circuit class $C$ also implies a
polynomial time Deterministic Identity Testing algorithm for the same class. From the works
\cite{Agr05} and \cite{HS80} it has been established that blackbox Identity Testing for certain circuit
classes imply super-polynomial circuit lower bounds for an
explicit polynomial. Hence the general problem of deterministic reconstruction
cannot be easier than proving
super-polynomial lower bounds. So one might first try and relax the requirements and demand a
randomized algorithm. Another motivation to consider the probabilistic version comes from Learning Theory.
A fundamental question called the \emph{exact learning problem using membership queries} asks
the following : {\bf Given oracle access to a Boolean
function, compute a small description for it.}
This problem has attracted a lot
of attention in the last few decades. For e.g. in \cite{Khar92}\cite{OGM86} and \cite{KV94} a negative
result stating that a class of boolean
circuits containing the trapdoor functions or pseudo-random functions has no
efficient learning algorithms. Among positive works \cite{SchSe96}, \cite{BBB00}, \cite{KS06}
show that when $f$ has a small circuit (inside some restricted class) exact
learning from
membership queries is possible.   Our problem
is a close cousin as we are looking for exact learning algorithms for algebraic
functions. Because of this connection with learning theory it makes sense to
also allow randomized algorithms for reconstruction.\\

\subsection{Depth 3 Arithmetic Circuits}

We will use the definitions from \cite{KayalSa09}.
 Let $C$ be an arithmetic circuit with coefficients
in the field $\F$. We say $C$ is a
$\Sigma\Pi\Sigma(k)$ circuit if it computes an expression of the form.
\[
 C(\B{x}) = \sum\limits_{i\in [k]} \prod\limits_{j\in [d]} l_{i,j}(\B{x})
\]

$l_{i,j}(\B{x}$) are linear forms of the type $l_{i,j}(\B{x}) = \sum
\limits_{s\in[n]}a_sx_s$ where $(a_1,\ldots,a_n)\in \F ^n$ and $(x_1,\ldots,x_n)$ is
an $n-$
tuple of variables.
For convenience we denote the multiplication gates in $C$ as
\[
 T_i = \prod\limits_{j\in [d]} l_{i,j}(\B{x})
\]

$k$ is the top fan-in of our circuit $C$ and $d$ is the
fan-in
of each multiplication gate $T_i$. With these definitions we will say that our
circuit
is of type $\Sigma\Pi\Sigma_\F(k,d,n)$. When most parameters are understood we will just call it a $\Sigma\Pi\Sigma(k)$ circuit.

\paragraph{Remark} \label{homogen} Note that we are considering homogeneous circuits. There are
two
basic assumptions:
\begin{enumerate}
 \item $l_{i,j}$'s have no constant term i.e. they are linear forms.
 \item Fan-in of each $T_i$ is equal to $d$.
\end{enumerate}
If these are not satisfied we can homogenize our circuit by considering
$Z^d(C(\frac{X_1}{Z},\ldots,\frac{X_n}{Z}))$. Now both the conditions will
be taken care of by reconstructing this new homogenized circuit. We need a rank condition on
our polynomial which remains essentially unchanged even after this substitution.

\begin{definition}[Minimal Circuit]
 We say that the circuit $C$ is minimal if no strict non empty subsets of the
$\Pi\Sigma$
 polynomials $\{T_1,\ldots , T_k\}$ sums to zero.
\end{definition}

\begin{definition}[Simple Circuit and Simplification]
 A circuit $C$ is called Simple if the gcd of the $\Pi\Sigma$ polynomials
 $gcd(T_1,\ldots,T_k)$ is equal to $1$ (i.e. is a unit). The simplification of a
$\Sigma\Pi\Sigma(k)$ circuit
 $C$ denoted as $Sim(C)$ is the $\Sigma\Pi\Sigma(k)$ circuit obtained by dividing
each
 term by the gcd of all terms i.e.
 \[
  Sim(C) \eqdef
\sum\limits_{i\in[k]}\frac{T_i}{gcd(T_1,\ldots,T_k)}
 \]

\end{definition}

\begin{definition}[Rank of a Circuit]
 Identifying each linear form $l(\B{x}) = \sum\limits_{s\in [n]}a_sx_s$ with the
 vector $(a_1, \ldots, a_n) \in \F^n$, we define the rank of $C$ to be the
dimension
 of the vector space spanned by the set $\{l_{i,j} | i \in [k], j \in [d]\}$.
\end{definition}

\begin{definition}[Simple Rank of a Circuit]
 For a $\Sigma\Pi\Sigma(k)$ circuit $C$ we define the \emph{Simple Rank} of $C$ as the
 rank of the circuit $Sim(C)$.
\end{definition}

Before we go further into the paper and explain our algorithm we state some results about
uniqueness of these circuits. In a nutshell for a $\Sigma\Pi\Sigma_\F(2,d,n)$ circuit $C$,  if one assumes
that the \emph{Simple rank} of $C$ is bigger than a constant ($c_\F(4) :$ defined later) then the circuit is essentially unique.

\subsection{Uniqueness of Representation}

Shpilka et. al. showed the uniqueness of circuit representation in
\cite{Shpilka07} using rank bounds for Polynomial Identity Testing. The bound they used were from
the work of Dvir et. al. in \cite{DS07}. It essentially states that
the rank of a simple, minimal $\Sigma\Pi\Sigma(k)$ circuit ($d\geq 2, k\geq 3$) which computes the identically zero polynomial
is $\leq 2^{O(k^2)}\log^{k-2}d$. For circuits over char $0$ fields improved rank bounds were given by Kayal et.al. in \cite{KayalSa09}.

In a series of following work the rank bounds for identically zero $\Sigma\Pi\Sigma(k)$ circuits got further improved. The best known bounds over
char $0$ fields were given by Saxena et. al. in \cite{SS10}. We rewrite Theorem 1.5 in \cite{SS10} here for completion.

\begin{theorem}[Theorem 1.5 in \cite{SS10}] \label{rankbound}
Let $C$ be a $\Sigma\Pi\Sigma(k,d,n)$ circuit over field $\F$ that is simple, minimal and zero. Then, $rk(C) < 3k^2$.
\end{theorem}
Let $c_{\F}(k) = 3k^2$. This gives us the following version of Corollary 7, Section 2.1 in
\cite{Shpilka07}.

\begin{theorem}[\cite{Shpilka07}]\label{uniqueness}
 Let $f(\B{x})\in \F[x]$ be a polynomial which exhibits a $\Sigma\Pi\Sigma(2)$
circuit
 \[
 C = G(A + B)
 \]
 $A = \prod\limits_{j\in [M]} A_j, B=\prod\limits_{j\in [M]} B_j, G =
\prod\limits_{i\in [d-M]} G_i$, where $A_i,B_j,G_k \in Lin_{\F}[\B{x}]$.
 $gcd(A, B)=1$, and $Sim(C) = A+B$ has rank $\geq c_{\F}(4) +1$ then the
 representation is unique. That is if:
 \[
 f=G(A+B) = \ti{G}(\ti{A} + \ti{B})
 \]
 where $A,B,\ti{A},\ti{B}$ are $\Pi\Sigma$ polynomials over $\F$ and
$gcd(\ti{A},\ti{B})=1$ then we have $G = \ti{G}$ and
 $(A,B)=(\ti{A},\ti{B})$ or $(\ti{B},\ti{A})$ (up to scalar multiplication).
\end{theorem}

\emph{Proof.}
Let $g= gcd(G, \ti{G})$ and let $G=gG_1, \ti{G} = g\ti{G_1}$. Then
$gcd(G_1,\ti{G_1})=1$ and we get
\[
G_1 A + G_1 B - \ti{G_1}\ti{A} - \ti{G_1} \ti{B} = 0
\]
This is a simple $\Sigma\Pi\Sigma(4)$ circuit with $rank$ bigger than
$c_{\F}(4)+1$ and is identically $0$ so
it must be not minimal. Considering the various cases one can easily prove the required
equality.

\subsection{Notation}\label{notation}

$[n]$ denotes the set $\{1,2,\ldots,n\}$. Throughout the paper we will work over
the field $\F$.
Let $V$ be a finite dimensional $\F$ vector space and $S\subset V$, $sp(S)$ will
denote the linear span of elements of $S$. $dim(S)$ is the dimension of
the subspace $sp(S)$. If $S=\{s_1,\ldots ,s_k\}\subset V$ is a set of linearly independent vectors
then $fl(S)$ denotes the affine subspace generated by points in $S$ (also called
a $(k-1)-flat$ or just $flat$ when dimension is understood). In particular:
\[
fl(S) = \{\sum\limits_{i=1}^k \lambda_i s_i : \lambda_i\in \F,
\sum\limits_{i=1}^k \lambda_i=1\}
\]
 Let $W\subset V$ be a subspace, then we can extend basis and get another
subspace
$W^\prime$ (called the complement of $W$) such that $W\oplus W^\prime = V$. Note
that the complement need not be unique.
Corresponding to each such decomposition of $V$ we may define orthogonal
projections $\pi_W,\pi_{W^\prime}$ onto $W,W^\prime$ respectively. Let
$v=w+w^\prime \in V, w\in W,w^\prime \in W^\prime$:
\[
\pi_{W}(v) = w, \pi_{W^\prime}(v)=w^\prime
\]

$(\B{x})$ will be used for the tuple $(x_1,\ldots,x_n)$.
\[
Lin_\F[\B{x}] = \{a_1x_1+\ldots + a_nx_n : a_i\in \F\}
\subset \F[\B{x}]
 \]
 is the vector space of all linear forms over the variables $(x_1,\ldots,x_n)$. For a linear form $l\in Lin_\F[\B{x}]$ and a polynomial $f\in \F[x]$
we write $l\mid f$ if $l$ divides $f$ and $l\nmid f$ if it does not. We say $l^d \mid\mid f$ if $l^d\mid f$ but $l^{d+1} \nmid f$.
 \[
 \Pi\Sigma^d_\F[\B{x}] = \{l_1(\B{x})\ldots l_d(\B{x}): l_i\in Lin_\F[\B{x}]\}\subset \F[\B{x}]
 \]
is the set of degree $d$ homogeneous polynomials which can be written as product of linear forms. This collection for all possible $d$ is called
the set
\[
 \Pi\Sigma_\F[\B{x}] = \bigcup\limits_{d\in \N}\Pi\Sigma^d_\F[\B{x}]
\]
also called $\Pi\Sigma$ polynomials for convenience.
Let $f(\B{x})\in\F[x]$ then
$Lin(f)\in \Pi\Sigma_\F[\B{x}]$ denotes the product of all linear factors of
$f(\B{x})$. Let $\ML(f)$ denote the set of all linear factors of $f$. For any set of polynomials $S\subset \C[\B{x}]$, we denote by $\mathbb{V}(S)$,
the set of all complex simultaneous solutions of polynomials in $S$ (this set is called the variety of $S$), i.e.
\[
 \mathbb{V}(S) =\{a\in \C : \text{ for all } f\in S, f(a)=0 \}
\]

Let $\MB=\{b_1,\ldots,b_n\}$ be an ordered basis for $V = Lin_\F[\B{x}]$.
We define maps $\phi_\MB : V\setminus\{0\} \rightarrow V$ as
\[
\phi_{\MB}(a_1b_1+\ldots+a_nb_n) = \frac{1}{a_k}(a_1b_1+\ldots+a_nb_n)
 \]
 where $k$ is such that $a_i=0$ for all $i<k$ and $a_k\neq 0$.\\

A non-zero linear form $l$ is called \underline{normal} with respect to $\MB$
if $l\in \Phi_\MB(V)$ i.e. the first non-zero coefficient is $1$. A polynomial
$P\in \Pi\Sigma_\F[\B{x}]$ is normal w.r.t. $\MB$ if it is a product of
normal linear forms. For two polynomials $P_1,P_2\in \Pi\Sigma_\F[\B{x}]$ we
define :
\[
gcd_{\MB}(P_1,P_2) =  P\in \Pi\Sigma_\F[\B{x}], P \text{ normal w.r.t. } \MB
\text{ such that } P\mid P_1, P\mid P_2
\]

When a basis is not mentioned we assume that the above definitions are with respect
to the standard basis.

We can represent any linear form in $Lin_\F[\B{x}]$ as a point in the vector
space $\F^n$ and vice versa. To be precise
we define the canonical map $\Gamma : Lin_{\F}[\B{x}] \rightarrow \F^n$ as
\[
\Gamma(a_1x_1+\ldots +a_nx_n) = (a_1,\ldots,a_n)
\]
$\Gamma$ is a linear isomorphism of vector spaces $Lin_\F[\B{x}]$ and $\F^n$.
Because of this isomorphism
we will interchange between points and linear forms whenever we can. We choose
to represent the linear form
$a(\B{x}) = a_1x_1+\ldots+a_nx_n$ as the point $a = (a_1,\ldots,a_n)$.\\

{\bf LI} will be the abbreviation for Linearly Independent and {\bf LD} will be
the abbreviation for Linearly Dependent.\\

\begin{definition}[Standard Linear Form]
A non zero vector $v$ is called $standard$ with respect to basis $\MB = \{b_1,\ldots,b_n\}$ if the coefficient of $b_1$ in
$v$ is $1$. When a basis is not mentioned we assume we're talking about the standard basis.
(Equivalently for linear forms the coefficient of $x_1$ is $1$). A $\Pi\Sigma$ polynomial
will be called $standard$ if it is a product of standard linear forms.
\end{definition}

We close this section with a lemma telling us when can
we replace the span of some vectors with the affine span or flat. We've used this several times in the paper.

\begin{lemma}\label{spantoflat}
 Let $l,l_1,\ldots,l_t \in Lin_\F[\B{x}]$ be \emph{standard} linear forms w.r.t. some basis $\MB =\{b_1,\ldots,b_n\}$ such that $l\in sp(\{l_1,\ldots,l_t\})$
 then
 \[
  l\in fl(\{l_1,\ldots,l_t\})
 \]

\end{lemma}
\emph{Proof.}
 Since $l\in sp(\{l_1,\ldots,l_t\})$, we know that $l=\sum\limits_{i\in [t]}\alpha_il_i$ for some scalars
 $\alpha_i\in \F$. All linear forms are $standard$ w.r.t. $\MB \Rightarrow$ comparing the coefficients of $b_1$ we get that
 $\sum\limits_{i\in [t]}\alpha_i=1$ and therefore $l\in fl(\{l_1,\ldots,l_t\})$.

Let $T\subset \F^n$, By a scaling of $T$ we mean a set where all vectors get scaled (possibly by different scalars).

\subsection{Summary of Technical Ideas}
This Subsection includes the very broad technical ideas we used. First we explain a technique to
reconstruct points from their projections. Then we give an overview of the {\bf  Project-Reconstruct-Lift} algorithm
and how we plan to execute it. After that we illustrate the algorithm in quite generality. In this illustration we
keep a lot of technicalities aside and try to motivate and visualize the algorithm through geometric intuition .

\subsubsection{A Simple Reconstruction Technique}\label{project}
We describe a method to recover points from their projections. A more rigorous treatment is in Appendix \ref{Identifier}. It
also contains details and proofs of the Algorithm that is used in this paper.
Suppose we have two disjoint sets of points $A = \{a_1,a_2\},B = \{b_1,\ldots,b_d\}$ in the projective space $\MP^{n+1}$ such that:
\begin{itemize}
 \item We know the set $A$.
 \item We know the projections of points in $B$ w.r.t. $a_1$ and $a_2$ i.e we know lines
 joining $L_{i,j}=\overrightarrow{a_i,b_j}$ for $i\in [2]$ and $j\in [d]$.
 \end{itemize}
 We want to use lines $L_{i,j} = \overrightarrow{a_ib_j}$ to find the set $\{b_1,\ldots,b_t\}$ in $O(poly(d))$ time. Note that
there are $\leq d$ lines through $a_1$ and $\leq d$ lines through $a_2$. The $b_j$'s lie at the intersection of these
lines and so we have $\leq d^2$ intersections. These intersections form a set of candidate points for $B$ but it is very hard to cut down
this set to $B$ in $poly(d)$ time. There is a trivial $O({d^2 \choose d})$ algorithm  - Go through all $d$ points in these
intersection points, make the lines and check if you get the same set of lines. This will give all sets of size $d$ which
could generate this configuration. Here is how the entire point configuration looks like. The green points $c_j$'s are intersections
of our lines which do not belong to $B$.

\includegraphics[width=6in,height=4in]{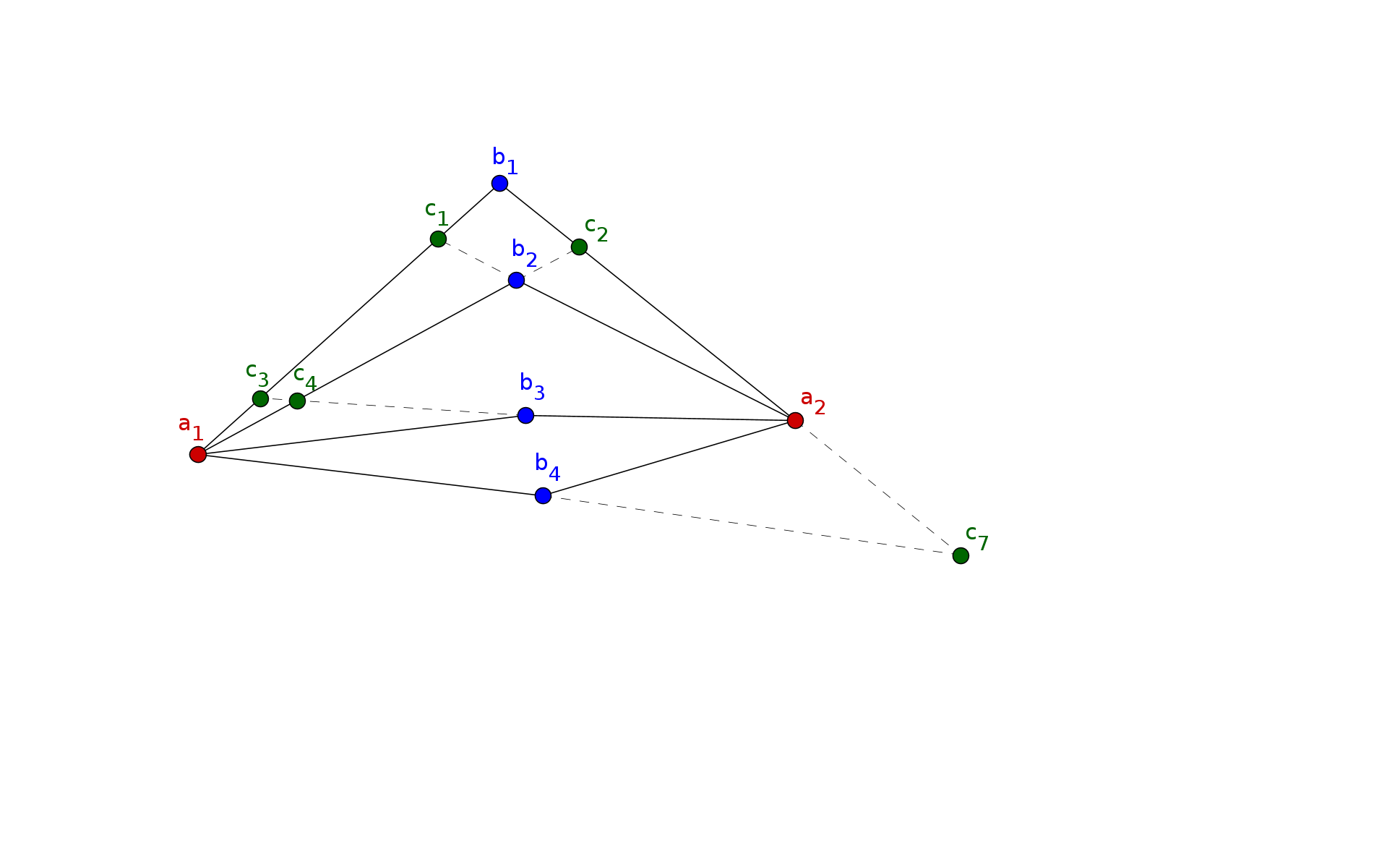}

 However if one assumes some restrictions then a subset of $B$ might be found in $poly(d)$ time. Assume that for some $t\in [d]$:
 \begin{itemize}
 \item $\{a_1,a_2,b_1\}$ are affinely independent.
  \item $fl\{a_2,b_1\} \cap B = \{b_1,\ldots,b_t\}$.
  \item $fl\{a_1,a_2,b_1\}\cap B = \{b_1,\ldots,b_t\}$.
 \end{itemize}

 That is we have a sub configuration that looks like :

\includegraphics[height=3in, width=5in]{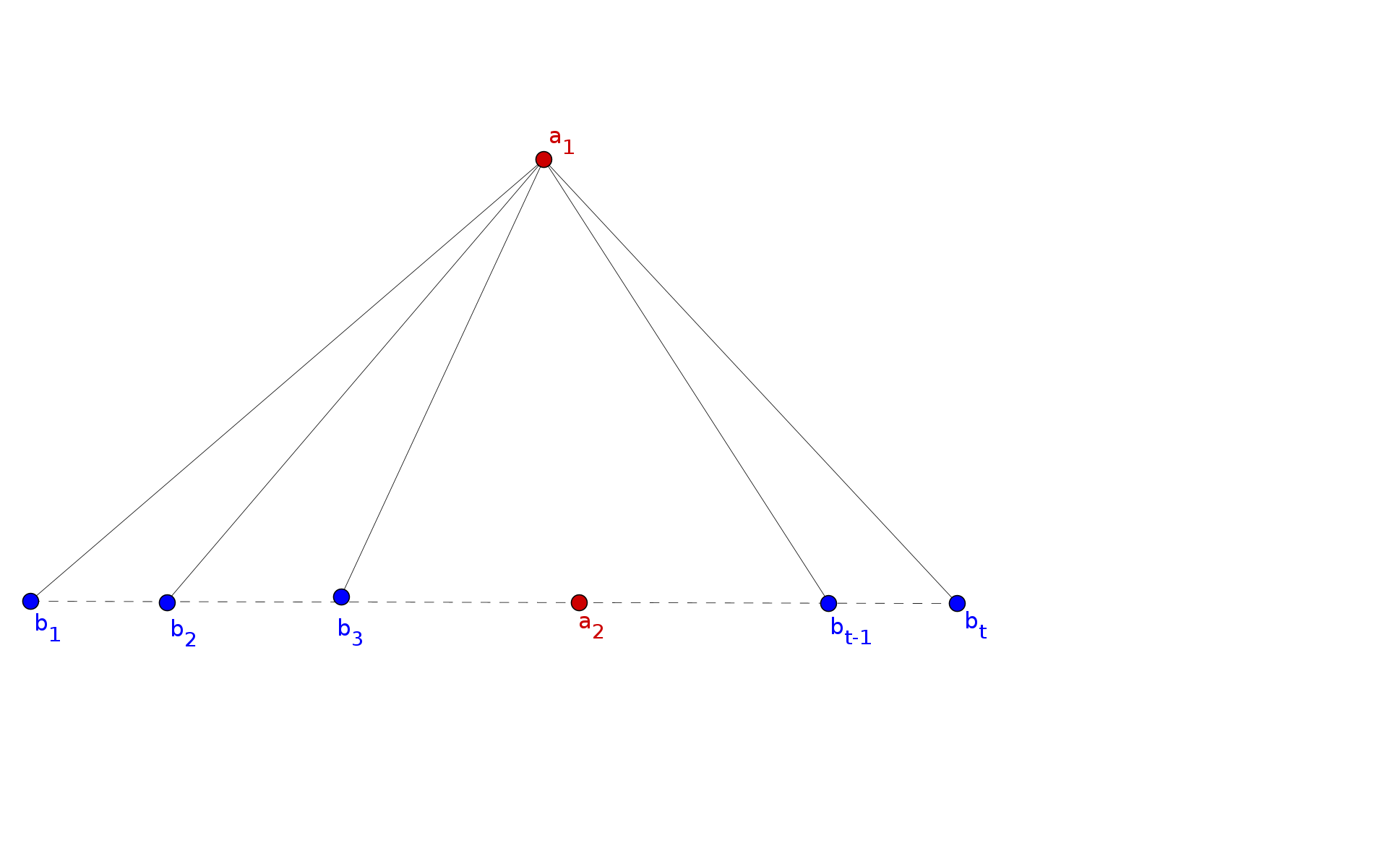}

Here is an algorithm to recover all $\{b_1,\ldots,b_t\}\subset B$ such that the above conditions are satisfied.
\begin{itemize}
 \item We iterate through all lines passing through $a_2$.
 \item For each such line $L$, find the set of lines $S_{L}$ through $a_1$ which intersects $L$. Clearly all lines
 in $S_L$ and $L$ are co-planar.
 \item If this plane does not contain any other line through $a_2$, output the intersections of lines in $S_L$ with $L$.
\end{itemize}

It is more or less straightforward that this algorithm works. The line $L$ we choose has to have some $b_j$ on it.
Now all lines $\tilde L \in S_L$ that intersect $L$ have to intersect it in some $b_i$ otherwise $\tilde{L}$ has
some other $b_s$ on it but then the plane of $S_L,L$ will have another line $\overrightarrow{a_2b_s}$ passing through
$a_2$ on it which is a contradiction. The algorithm actually finds all such configurations $\{b_1,\ldots,b_t\}\subset B$.

\subsubsection{General Overview of the Algorithm}
The broad structure of our algorithm is similar to that of Shpilka in \cite{Shpilka07} however our
techniques are different. We first restrict the blackbox inputs to a low ($O(1)$) dimensional random subspace of
$\F^n$ and interpolate this restricted polynomial. Next we try to recover the $\Sigma\Pi\Sigma(2)$ structure
of this restricted polynomial and finally lift it back to $\F^n$. The random subspace and unique $\Sigma\Pi\Sigma(2)$
structure will ensure that the lifting is unique. Similar to \cite{Shpilka07} we try to answer the following questions.
However our answers (algorithms) are different from theirs

\begin{enumerate}
\item For a $\Sigma\Pi\Sigma(2)$ polynomial $f$ over $r=O(1)$ variables, can one compute a small set of linear forms which contains all factors
from both gates?
 \item Let $V_0$ be a co-dimension $k$ subspace($k=O(1)$) and $V_1,\ldots,V_t$ be co-dimension $1$ subspaces of a linear space $V$.
 Given circuits $C_i$ ($i\in\{0,\ldots,t\}$) computing $f|_{V_i}$(restriction of $f$ to $V_i$) can we reconstruct from them a single
 circuit $C$ for $f|_{V}$?
 \item Given co-dimension $1$ subspaces $V \subset U$ and circuits $f|_{V}$ when is the $\Sigma\Pi\Sigma(2)$ circuit
 representations of lifts of $f|_{V}$ to $f|_{U}$ unique?
\end{enumerate}

Our first question is easily solved using Brill's equations (See Chapter 4 \cite{GKZ94}). These provide a set of polynomials whose simultaneous
solutions completely characterize coefficients of complex $\Pi\Sigma$ polynomials. A linear form $l = x_1-a_2x_2-\ldots-a_rx_r$ divides one of
the gates of $f(x_1,\ldots,x_r)$
$\Rightarrow f(a_2x_2+\ldots+a_rx_r,x_2,\ldots,x_r)$ is a $\Pi\Sigma$ polynomial modulo $l$. When this is applied into Brill's equation
(see Corollary \ref{variety})
we recover possible $l$'s which obviously
include linear factors of gates. We can show that (see Claim \ref{candidate}) the extra linear forms we get are not too many ($poly(d)$) and
also have some special structure. We call
this set $\MC$ of linear forms as Candidate linear forms and non-deterministically guess from this set. It should be noted
that we do all this when our polynomial is over $O(1)$ variables.

We deal with the second question while trying to reconstruct the $\Sigma\Pi\Sigma(2)$ representation of the
interpolated polynomial $f|_{V}$, where $V$ is the random low dimensional subspace. We divide the algorithm into
Easy Case, Medium Case and a Hard Case.
\begin{itemize}
 \item For the Easy Case our algorithm tries to reconstruct one of the
multiplication gates of $f|_{V}$ by first looking at it's restriction to a special co-dimension $1$ subspace $V_1$.
If $f=A+B$ with $A,B$ being $\Pi\Sigma$ polynomials,
the projection of one of the gates (say $A$) with respect to $V_1$ will be $0$ and the other (say $B$) will remain unchanged giving us
$B$ and therefore both gates by factoring $f|_{V}-B$.

\item In the Medium Case we have at least two extra dimensions in one of the gates. This can be used to show that the only linear
factors of $f_{|V}$ are those coming from $G$. Now we can recover $G$ by factoring $f$ and then use Easy Case for the remaining
polynomial. An important consequence of this case is that in the Hard Case we may now assume that both gates are high dimensional
which is very crucial.

\item In the Hard Case we will first need $V_0$, a co-dimension $k$ (where $k=O(1)$) subspace and then iteratively select
co-dimension $1$ subspaces $V_1,\ldots,V_t$. For some gate (say $B$), all pairs $(V_0,V_i)$ ($i\in [t]$) will reconstruct some
linear factors of $B$. This process will either completely reconstruct $B$ or we will fall into the Easy Case. Once $B$ is known
we can factor $f|_{V}-B$ to get $A$.
\end{itemize}

The restrictions that we compute always factor into product of linear forms and can be easily computed since we know $f|_{V}$ explicitly. They can then be factorized into
product of linear forms using the factorization algorithms from \cite{KalTr90}. It is the choice of the subspaces $V_0,V_1,\ldots,V_t$
where our algorithm differs from that in \cite{Shpilka07} significantly. Our algorithm selects $V_0$ and iteratively selects the
$V_i$'s ($i\in [t]$) such that $(V_0,V_i)$ have certain
\emph{"nice"} properties which
help us recover the gates in $f|_{V}$. The existence of subspaces with \emph{"nice"} properties is guaranteed by
Quantitative Sylvester Gallai Theorems given in \cite{BDWY11}. To use the theorems we had to develop more machinery that has been explained later.\\

The third question comes up when we want to lift our solution from the random subspace $V$ to the original space. This
is done in steps. We first consider random spaces $U$ such that $V$ has co-dimension $1$ inside them. Now we reconstruct
the circuits for $f|_{V}$ and $f|_{U}$. The $\Sigma\Pi\Sigma(2)$ circuits for $f|_{V}$ and $f|_{U}$ are unique since
the simple ranks are high enough (because $U,V$ are random subspaces of high enough dimension)
implying that the circuit for $f|_{V}$ lifts to a unique circuit for $f|_{U}$. When this is done for multiple $U$'s we can
find the gates exactly.

\paragraph{Project-Reconstruct-Lift Algorithm :}
Here is a broad outline of the three aspects. This technique is quite common. Details of Project and Lift are in
Section \ref{highdimrecon} and that of Reconstruct is in Section \ref{lowdimrecon}.
\paragraph{Project}
\begin{itemize}
 \item Input:$f \in \F[x_1,\ldots,x_n]$ as blackbox
\item  Choose random basis $\{y_1,\ldots,y_n\}$ of $\F^n$, $V = sp(\{y_1,\ldots,y_s\}), V_i =sp(\{v_1,\ldots,v_s,v_i\})$ for $i\in \{s+1,\ldots,n\}$.
\item Define $f_0(y_1,\ldots,y_s) = f_{|V}, f_i(y_1,\ldots,y_s,y_i) = f_{|V_i}$.
\item Consider sets $H\subset V,H_i\subset V_i$ with $|H|\geq d^s,|H_i|\geq d^{s+1}$ and interpolate to find $f_0,f_i$.
\end{itemize}
\paragraph{Reconstruct}
\begin{itemize}
 \item Reconstruct to get $f_0 = M_0+M_1$ and $f_i = M^i_0 + M^i_1$ with $M_0,M_1\in \Pi\Sigma[y_1,\ldots,y_s], M^i_0,M^i_1\in
 \Pi\Sigma[y_1,\ldots,y_s,y_i]$.
\end{itemize}
\paragraph{Lift}
\begin{itemize}
\item Use $M_0,M_1,M^i_0,M^i_1$ to compute gates $N_0,N_1$ such that $f=N_0+N_1$.
\item If the reconstruction was successful return it, else return failed.
\end{itemize}

\section{An Illustrative Example}
Let $\B{x}$ denote the variables $(x_1,\ldots,x_r)$ where $r$ is a constant (we will fix this constant later).
Consider the following polynomial $f(\B{x})\in \F[x_1,\ldots,x_r]$
\[
 f(\B{x}) = T_0(\B{x}) + T_1(\B{x})
\]
Such that:
\begin{enumerate}
\item $T_0(\B{x})=A_1\ldots A_d, T_1(\B{x}) = B_1\ldots B_d$ with $A_i,B_j$ linear forms
 \item $gcd(A_i,B_j)=1$ for all $1\leq i,j\leq d$.
 \item $dim(\{A_i,B_j : i,j \in [d]\}) =r $ i.e. there are no redundant variables.
\end{enumerate}
Define the sets $A = \{A_1,\ldots,A_d\}$, $B=\{B_1,\ldots,B_d\}$. We are going to view the points in $A$ and
$B$ as points in the space $\F^r$. We also identify (keep only one copy) linear forms which are scalar multiples of each other.

\begin{theorem}
Consider $f(\B{x})$ from above and assume $f(\B{x}) = \sum\limits_{\lambda \in \Lambda}{\bf c}_{\lambda} {\bf x}^{\lambda}$ where $\lambda = (\lambda_1,\ldots,\lambda_r)$ and ${\bf x}^\lambda = x_1^{\lambda_1}\ldots x_r^{\lambda_r}$. Suppose we know all the coefficients ${\bf c}_{\lambda}$ then in time $poly(d)$  we can reconstruct $T_0(\B{x}),T_1(\B{x})$ with high probability.
\end{theorem}

We will describe an algorithm which proves the above theorem. At many points during the algorithm we will need results
that are mentioned later in the paper. For better understanding we encourage the reader to first go through this algorithm assuming
all the claims mentioned.

\subsection{Candidate Linear Forms}
Our job in this algorithm is to reconstruct $T_0(\B{x}),T_1(\B{x})$ i.e. $A_i$'s and $B_j$'s. Let us first observe a property these linear forms
satisfy. One can see that for $l\in \{A_i,B_j : i,j\in [d]\}$ the following holds:
\[
 f_{|_{l=0}} \text{ is a non-zero product of linear forms }
\]
Can we use this to reconstruct $A_i,B_j$? The two questions that pop up are:
\begin{enumerate}
 \item Are there linear forms other than $A_i,B_j$ that satisfy the above condition?
 \item If yes, can we find out some structure of the bad $l$'s ( which are not $A_i,B_j$)?
 \item Can we bound the total number of such $l$'s by a polynomial in $d$?
 \item Can we construct this set efficiently?
\end{enumerate}
The answer to all the above questions is a YES!

\begin{example}
 Consider $f(x_1,\ldots,x_r) = (x_1+x_2)(x_1+x_3)\ldots (x_1+x_{r}) + x_2\ldots x_r$. We can see that $f_{|_{x_1=0}} = x_2\ldots x_r$ but
 $x_1$ is not a factor of any of the gates.
\end{example}

The next claim contains the information structure of the bad $l$'s and their number. Proof will be given
later in the paper in Appendix \ref{findcandidate}.

\begin{claim}\label{structclaim}
 Consider the set $\MC = \{ l : f_{|_{l=0}} \text{ is a non zero product of linear forms }\}$ and let $\{l_1,\ldots,l_{k}\}\subset T_i$
 be a set of LI linear forms where $k=c_\F(3)+2$ (rank bound for $\Sigma\Pi\Sigma(3)$ circuits) then
 \begin{enumerate}
 \item $\{A_i,B_j : i,j\in [d]\}\subseteq \MC$
  \item $|\MC| \leq O(d^4)$
  \item \label{struct}If $l\in \MC \setminus \{A_i,B_j, i,j\in [d]\}$, then there exists $i\in [k]$ and $j\in [d]$ such that $\{l,A_i,B_j\}$
  are linearly dependent i.e. for every $LI$ set $\{A_1,\ldots, A_k\}$, a bad $l$ will match one of these $A_i$ ($i\in [k]$) to some $B_j$.
 \end{enumerate}
\end{claim}

Moreover the above set $\MC$ can be constructed in time $poly(d)$. This is done by solving a set of multivariate polynomial equations of
$poly(d)$ degree in $O(1)$ variables. Please see Appendix \ref{findcandidate} for details.

\subsection{Reconstruction Algorithm}
Before going to the core of the algorithm let's explain an easy case. Recall $A = \{A_1,\ldots,A_d\}$ and $B = \{B_1,\ldots,B_d\}$. Also color the points in $A$ red and the points in $B$ blue.
\subsubsection{Easy Case}
 For this case we assume

\begin{center}
\fbox{$sp(A) \subsetneq sp(B)$}
\end{center}

So let's say $A_1\notin sp(B)$. The main advantage of such an $A_1$ is that on setting $A_1$ to $0$ no linearly independent $\{B_i,B_j\}$ become dependent. Geometrically we have the following picture:

\includegraphics[width=3in, height=2in]{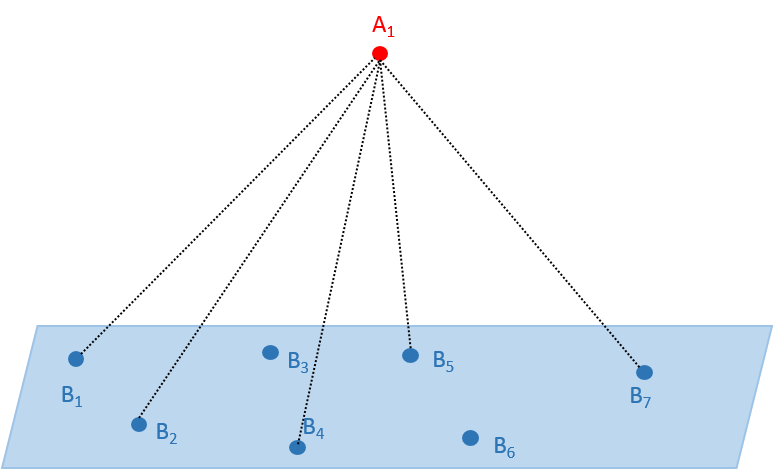}

We guess a basis $\{l_1,\ldots,l_r\}$ of linear forms from the set $\MC$.
While doing this we assume:
\begin{itemize}
 \item $l_1=A_1$
 \item $l_2,\ldots,l_t$ is a basis for $B$
 \item $l_{t+1},\ldots,\ldots,l_r$ are the rest of the basis vectors
\end{itemize}
If our guess was actually a basis we define an invertible linear transformation $T$ sending $l_i$ to $x_i$. We apply $T$ to $f(\B{x})$
by applying it to each variable in the most natural way. If our guess was correct we get
\[
f^\prime(\B{x}) = f(T(\B{x})) = x_1A_2^\prime\ldots A_d^\prime + B_1^\prime\ldots B_d^\prime 													
\]
Note that if our assumption for the basis is correct then none of the $B_i^\prime$'s contain $x_1$. So we can compute
$f^\prime_{|_{x_1=0}} = B_1^\prime\ldots B_d^\prime$. Then we can apply $T^{-1}$ and get back $T_1(\B{x})=B_1,\ldots,B_d$. We remind the
reader that everything is recovered up to a scalar multiple but that is not a problem since that can be merged into one scalar for the gate
$B(\B{x})$ which can be easily recovered.  We then factorize $f-T_1(\B{x})$ and check if it factors into a product of linear forms and recover
$T_0(\B{x})$. Note that during the process we will guess the basis correctly at least once. Also the last step checks if we actually
get a $\Sigma\Pi\Sigma(2)$ circuit and therefore the reconstruction will be complete. The case where $sp(B)\subsetneq sp(A)$ is symmetrical
and is handled in the same way. Next we deal with the hard case.

\subsubsection{Hard Case}

The other case i.e. $sp(A)=sp(B)$ is much harder but high dimensionality enables us to apply the Quantitative version of Sylvester Gallai Theorems
from \cite{BDWY11}. Let's first just give some consequences of the Quantitative Sylvester Gallai theorem (from \cite{BDWY11}) which will
be useful for us. A slightly more general version with proof can be found in Appendix \ref{incidence}.

\begin{corollary} \label{elementary}
Let $S = \{s_1,\ldots,s_n\} \subseteq \C^d$ be a set of points. Assume $dim(S)> \Omega(C^k)$ for some constant $C$,
then there exists a set of linearly
independent points $\{s_1,\ldots,s_k\}$ and a set $T\subset S$ with $|T|\geq 0.99n$, such that $fl(\{s_1,\ldots,s_k,t\})$ is an
elementary $k - flat$ for every $t\in T$. That is:
 \begin{itemize}
  \item $t\notin fl(\{s_1,\ldots,s_k\})$
  \item $fl(\{s_1,\ldots,s_k,t\}) \cap S = \{s_1,\ldots,s_k,t\}$.
 \end{itemize}
\end{corollary}

\begin{lemma}[Bi-chromatic semi-ordinary line]\label{bichromatic}
 Let $X$ and $Y$ be disjoint finite sets in $\C^d$ satisfying the following
conditions.
\begin{enumerate}
\item $dim(Y)>\Omega(C^4)$ where $C$ is the constant in the above corollary.
\item $|Y|\leq  99|X|$
\end{enumerate}
Then there exists a line $l$ such that $|l\cap Y|=1$ and $|l\cap X|\geq 1$
\end{lemma}

At this point we would like to mention that the constants $99,0.99$ and the one hidden in $\Omega(C^k)$ have more general values
given by a parameter $\delta$. For the time being we've fixed them for better exposition. Please see Appendix \ref{incidence}
for more details.

Using high dimensionality of $A,B$ and the above mentioned corollaries we are able to prove the following
theorem which forms the backbone of our algorithm.

\begin{theorem}
For some product gate (say $A$), there exists $k = O(1)$ points {$S =\{A_1,\ldots,A_k\}$} and a large set $D\subset A$ such that
on projecting $D,B$ to the subspace $W$ defined by $\{A_1=0,\ldots,A_k=0\}$ (and throwing away zeros):
\begin{itemize}
\item There exists a  lines $L = \overrightarrow{B_1^\prime D_1^\prime}$ where $B_1 ^\prime$ and $D_1 ^\prime$ are projections of $B_1,D_1$ onto $W$.
Also if $B^\prime$ is the projection of $B$ onto $W$ then $L\cap B^\prime = \{B_1^\prime\}$, so the line is a bi-chromatic semi-ordinary which were discussed in
the lemma above.
\end{itemize}
\end{theorem}
\includegraphics[height=1.75in, width=4in]{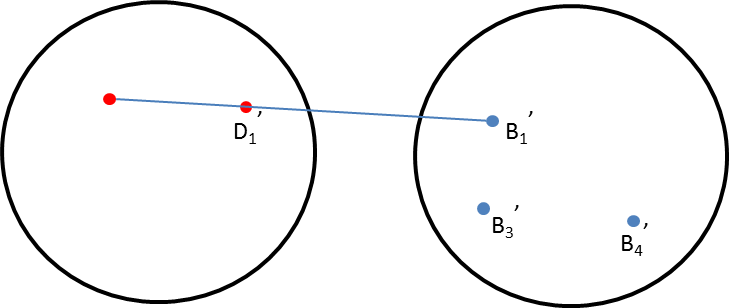}

Let's pick one of these lines and see what would have happened in $\F^r$ which led us to this line in $W$.

\includegraphics[width=4in,height=2in]{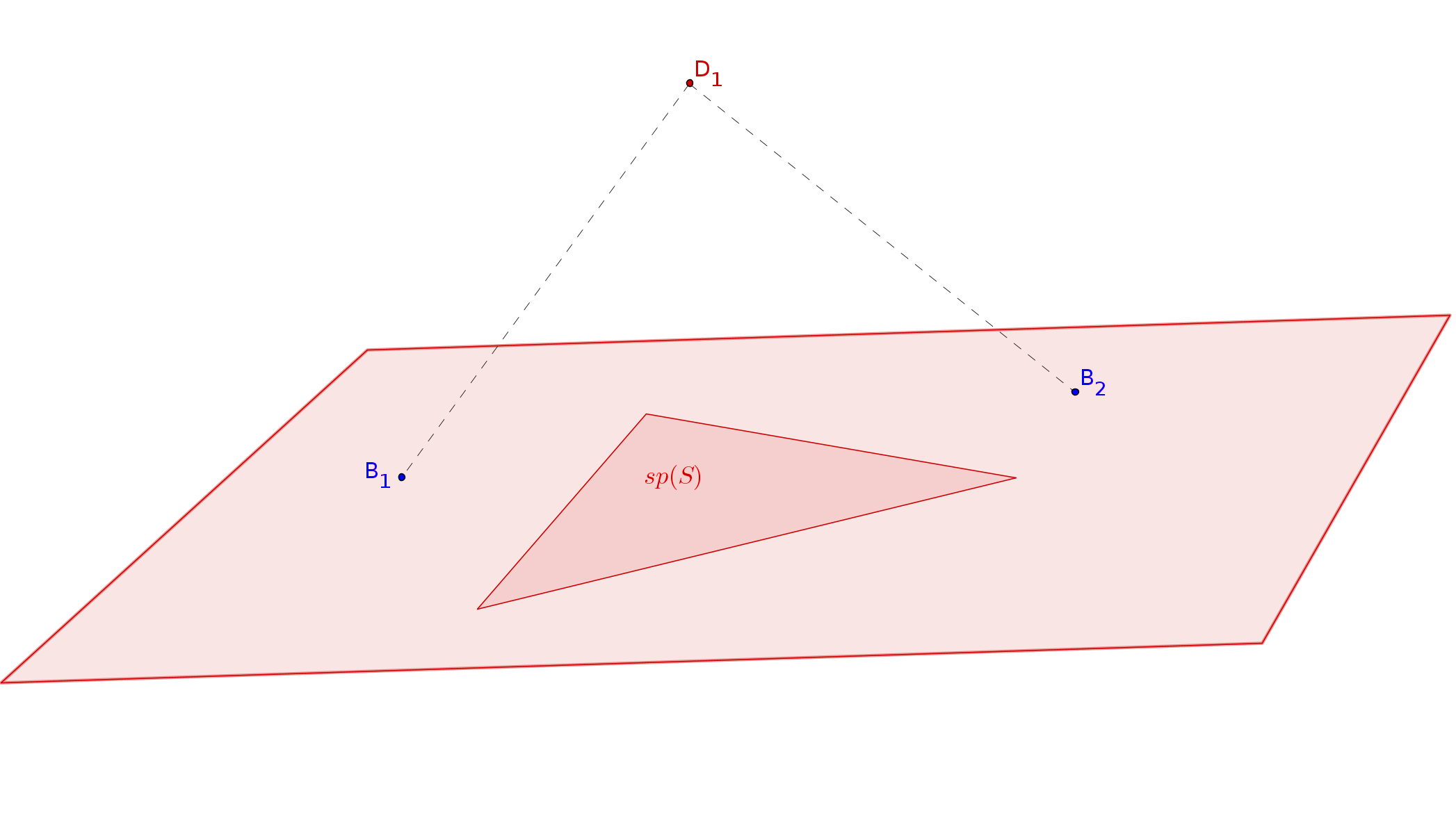}

In the picture above the inner triangle denotes $sp(S)$ and the outer parallelogram denotes $sp(S\cup\{B_1\})$. The line in the previous picture i.e. projecting
the points onto $W$ has only one blue point implying:
\begin{itemize}
 \item $sp(S \cup\{B_1\})\cap B = sp(S\cup \{B_1\})$
 \item $sp(S\cup \{B_1\} \cup \{D_1\}) \cap B = sp(S\cup \{B_1\})$
\end{itemize}

Note that this looks very similar to what we had in Subsection \ref{project}. We used this kind of a configuration to recover
points using their projections. A similar method is implemented here. Given that such a configuration exists
we can come up with the following Algorithm.
\begin{enumerate}
 \item From the set $\MC$ guess the set $S=\{A_1,\ldots,A_k\}$ mentioned in the theorem above.
 \item Using condition \ref{struct} in Claim \ref{structclaim} obtain a set $X$ such that $D\subset X\subset A$.
 This can be done as explained in Algorithm \ref{overestimatedetector}. The reader should just assume this at the moment. We need to
 make sure that $D_1$ comes from $A$ because the algorithm is iterative and we don't want a spurious linear form in $\MC$ give any
 reconstruction. We always want to set some $A_i$'s to $0$ so that we only recover $B_j$'s.
 \item Iterate over this set $X$ and guess $D_1$.
 \item By projecting $f$ to the subspaces $\{A_1=0,\ldots, A_k=0\}$ and $\{D_1=0\}$ we get $B_1^\prime$ and $({B_1})_{|_{D_1=0}}$. Because
 of the diagram above these two projections can be matched and used to reconstruct $B_1$.
 \item If no $D_1\in X$ worked then go to Easy Case since dimension  should have fallen.
\end{enumerate}

Basically the algorithm just exploits the existence of the line mentioned in the previous theorem and reconstructs the corresponding $B_1$ (whose projection lies 
on the line).
This reconstruction was possible because this line had only one blue point. After finding $B_1$ we declare it known so that in the next iteration we can remove it's
 projection when required. We will continue to get such bi-chromatic semi-ordinary lines till the unknown linear forms in the $B$ set have high dimension. If at any 
 stage this reconstruction is not possible then this dimension would have fallen and we can use the Easy Case.

\paragraph{Return Type}
In all our algorithms we wish to return the reconstructed form of $f$. Since $f$ and the two gates $T_0,T_1$ are to be returned we define an
object for it. We call this object Decomposition. We assume having a data type polynomial for general polynomials and pi\_sigma for polynomials
which are product of linear forms. We use C++ syntax to define our structure.
\begin{lstlisting}
struct decomposition {
  bool iscorrect; // iscorrect will be true if f = M_0 + M_1
  polynomial f;
  pi_sigma M_0;
  pi_sigma M_1;

  // Constructor when a reconstruction is found
  decomposition(polynomial g, pi_sigma A, pi_sigma B){
    iscorrect =true;
    f=g;
    M_0=A;
    M_1=B;
  }

  // Constructor when no reconstruction is found
  decomposition(){
    iscorrect=false;
  }
};
\end{lstlisting}

% \subsubsection{Lifting from Low to High dimension}\label{lifting}
% Let's Explain the second block in Picture \ref{blockexplain}. So we have the reconstructions $f_0 = M_0+M_1$ and $f_i = M^i_0+M^i_1$.
% If we set $y_i = 0$ in $f_i$ we should get $f_0$. So ${M^i_0}_{|V} + {M^i_1}_{|V} = M_0+M_1 $. Since the simple rank of $f_0$ is $r$ this
% representation should be unique. The multiplication gates $M^i_0, M^i_1$ should be lifts of $M_0,M_1$. So we can just set $y_i$
% to $0$ and find the correspondence between these gates. Let's say $M^i_0|_{|V} = M_0$, this implies that the linear forms in
% $M^i_0$ are lifts of linear forms in $M_0$. Next notice that with high probability LI linear forms from a gate in circuit of $f$
% remain LI on projecting to $V$. So LD linear forms in $M_0$ cannot have LI lifts in $M^i_0$. Now to find this lift of linear form $l$ dividing
% $M_0$ with multiplicity $k$, find $l_i$ in $M^i_0$ ( with multiplicity $k$) such that on setting $y_i=0$, we get $l$ i.e.
% ${l_i}_{|\{y_i=0\}}=l$. This gives the coefficient of $y_i$ in the lift of $l$. If we do this for all $i$ we get the lift of $l$ to $\F^n$.
% So we can compute lifts of all linear forms in $M_0$ and $M_1$. By uniqueness this will give us the gates $N_0,N_1$ such that $f=N_0+N_1$.

\newpage

\section{Reconstruction for low rank }\label{lowdimrecon}
Let's recall Definition \ref{rvalue} following Theorem \ref{maintheorem} in Section \ref{introduction}.

\begin{definition}
\begin{mdframed}
We fix $s$ to be any constant $> \max (C_{2k-1}+k, c_\F(4) )$ where :
\begin{enumerate}
\item $c_{\F}(l)=3l^2$ is the rank lower bound (see Theorem \ref{rankbound}) that guarantees nonzero-ness of any simple, minimal, $\Sigma\Pi\Sigma(l)$ circuit with 
rank $> c_\F (l)$.
\item $k = c_{\F}(3) +2$.
\item $\delta$ is some fixed number in $(0,\frac{7-\sqrt{37}}{6})$.
 \item $C_k = \frac{C^k}{\delta}$ the constant that appears in Theorem \ref{bdwy}.

\end{enumerate}
\end{mdframed}
\end{definition}

Let $r$ be any constant $\geq s$ (In our application we need $s$ and $s+1$). Our main theorem for this section therefore is:
\begin{theorem}
Let $r$ be as defined above. Consider $f(\B{x})\in
\F[\B{x}]$,
a multivariate homogeneous polynomial of degree $d$ over the variables
$\B{x}=(x_1,\ldots,x_r)$ which can be computed by a $\Sigma\Pi\Sigma(2)$
circuit $C$ over $\F$. Assume that rank of the simplification of $C$ i.e. $Sim(C)=r$. We give a $poly(d)$
time randomized algorithm which computes $C$ given blackbox access to $f(\B{x})$.
\end{theorem}
\par{}\label{randomtransformation}

We assume $f$ has the following
$\Sigma\Pi\Sigma(2)$ representation:
\[
f = \tilde G(\tilde \alpha_0\tilde T_0 + \tilde \alpha_1\tilde T_1)
\]
where $\tilde G,\tilde T_i\in\Pi\Sigma_\F[\B{x}]$ are $normal$ (i.e. leading non-zero coefficient is $1$ in every linear factor)
and $\tilde \alpha_0,\tilde \alpha_1\in \F$ with $gcd(\tilde T_0,\tilde T_1)=1$. The $rank(Sim(C))=r$ condition then becomes
\[
 sp(\ML(\tilde T_0)\cup \ML(\tilde T_1)) = Lin_\F[\B{x}]
\]

Consider the set $T = \ML(\tilde G)\cup\ML(\tilde T_0)\cup \ML(\tilde T_1)$. By abuse of
notation we will treat these linear forms also as points in $\F^r$. Since linear factors of $\tilde G, \tilde T_i$ are normal,
two linear factors of $\tilde G, \tilde T_i$ are LD if and only if they are same.\\

\paragraph{Random Transformation and Assumptions}\label{assumptions}
Let $\Omega,\Lambda$ be two $r\times r$ matrices such that their entries $\Omega_{i,j}$ and $\Lambda_{i,j}$ are picked
independently from the uniform distribution on $[N]$. Here $N = 2^d$.
We begin our algorithm by making a few assumptions. All of these assumptions are
true with very high probability and we assume them in our algorithm. These assumptions make our work easy
by removing redundancy in the co-ordinates. {\bf The idea is to move vectors randomly thereby introducing non-zero coefficients in them}. Consider the standard 
basis of $\F^r$ given
as $\MS = \{e_1,\ldots,e_r\}$. Let $E_j = sp(\{e_1,\ldots,e_j\})$
and $E_j^{\prime} = sp(\{e_{j+1},\ldots,e_r\})$, clearly $\F^r = E_j\oplus E_j^\prime$. Let $\pi_{W_{E_j}}$ be
the orthogonal projection onto $E_j$ w.r.t. this decomposition. Note that $T$ is a finite set of vectors in $\F^r$.
\begin{itemize}
\item {\bf Assumption 0 : } $\Omega$ is invertible. This is just the complement of event $\ME_0$ in Section
\ref{randomtransform} and so occurs with high probability.
 \item {\bf Assumption 1 : } For all $t\in T$,
 $\pi_{W_{E_1}}(\Omega(t))\neq 0$ i.e. $[\Omega(t)]^1_{\MS}\neq 0$ (coefficient of $e_1$ is non-zero) .  This is the complement of event $\ME_1$ in Section 
 \ref{randomtransform}
 and so occurs with high probability.
 \item {\bf Assumption 2 : } For all LI sets $\{t_1,\ldots,t_r\}\subset T$, $\{\Omega(t_1),\ldots,\Omega(t_r)\}$ is LI.
 This essentially means that $\Omega$ is invertible. This is the complement of $\ME_2$ in Section \ref{randomtransform}
 and so occurs with high probability.
 \item {\bf Assumption 3 : } Fix a $k<r$. For all LI sets $\{t_1,\ldots,t_r\} \subset T, \{\Omega(t_1),\ldots,\Omega(t_k),\Lambda\Omega(t_{k+1}),
 \ldots , \Lambda\Omega(t_d) \}$ is LI i.e. is a basis. This is the complement of event $\ME_3$ in Section \ref{randomtransform} and so occurs with
 high probability. It'll be used later in this chapter.
 \item {\bf Assumption 4 : } Fix a $k<r$. For all LI sets $\tilde T = \{t_1,\ldots,t_r\}\subset T$,
 define the set $\MB = \{\Omega(t_1),\ldots,\Omega(t_k),\Lambda\Omega(t_{k+1}),\ldots,\Lambda\Omega(t_r)\}$. By Assumption $3$ this is a basis.
 Consider any $t\in T$ such that $\Omega(t) \notin sp(\{\Omega(t_1),\ldots,\Omega(t_k)\})$.
 Then $[\Omega(t)]^{k+1}_\MB\neq 0$. This event is the complement of $\ME_5$ and so it occurs with high probability. We want
 nonzero-ness of co-ordinates even after projecting to a co-dimension-$k$ subspace. That is where this will be useful.

\end{itemize}
From now onwards we will assume that all the above assumptions are true. Since all of them occur with very high probability,
their complements occur with very low probability and by union bound the union of their complements is a low
probability event. So intersection of the above assumptions occurs with high probability and we assume all of them are true.
{\bf Note that the assumptions will continue to be true if we scale all linear forms ( possibly different scaling for different vectors, but non-zero scalars)
in $T$ i.e. if the assumptions were true for $T$ then they would have been true had we started with a scaling of $T$.}\\

The first step of our algorithm is to apply $\Omega$ to $f$. We have a natural identification between
linear forms and points in $\F^r$. This identification converts $\Omega$ into a linear map on $Lin_\F[\B{x}]$
which can be further converted to a ring homomorphism on polynomials by assuming that it preserves the products and sums of polynomials.
So $\Omega$ gets applied to all linear forms in the $\Sigma\Pi\Sigma(2)$ representation of $f$. Since $f$ is a degree $d$ polynomial in $r$
variables it has at most $poly(d^r)$ coefficients. Applying $\Omega$ to each monomial and expanding it takes $poly(d^r)$ time and gives
$poly(d^r)$ terms. So computing $\Omega(f)$ takes $poly(d^r)$ time and has $poly(d^r)$ monomials.\\

Now we try and reconstruct the circuit for $\Omega(f)$. If this reconstruction can be
done correctly, we can apply $\Omega^{-1}$ and get back
$f$. Note that {\bf Assumption 1} tells us that the coefficient of $x_1$ in $\Omega(l)$ is non-zero for all $l$ in $T$.
Let $X = \{x_1,\ldots,x_r\}$ and $\B{x}$ is used for the tuple $(x_1,\ldots,x_r)$. From this discussion we know that:
\[
 \Omega(f) = \Omega(\tilde G)(\tilde \alpha_0\Omega(\tilde T_0) + \tilde \alpha_1\Omega(\tilde T_1)) = G(\alpha_0 T_0 + \alpha_1 T_1)
\]
where $\alpha_i$ are chosen such that linear factors of $G,T_i$ have their first coefficient( that of $x_1$) equal to $1$. So they are $standard$
$\Pi\Sigma$ polynomials. Note that we've used {\bf Assumption $1$} here. Since we've moved constants to make linear forms standard we can
assume $G = \lambda\Omega(\tilde G), T_i = \lambda_i \Omega(\tilde T_i)$ with $\lambda,\lambda_i \in \F$. Consider some scaling $T_{sc}$ of $T$
such that $\MX = \ML(G)\cup \ML(T_0)\cup \ML(T_1)$ is $ = \Omega(T_{sc})$. All above assumptions are true for $T_{sc}$ and so we may use the conclusions
about $\Omega(T_{sc})$ i.e. $\MX$. Also since $\Omega$ is invertible $gcd(T_0,T_1)=1$.\\

Let
\[
T_i = \prod\limits_{j\in [M]} l_{ij}, i=0,1 \text{ and } G = \prod\limits_{k\in[d-M]}G_k
\]
with $l_{ij},G_k$ linear forms (so $d = deg(f)$ ).\\

For simplicity from now onwards we call $\Omega(f)$ by $f$ and try to reconstruct it's circuit. Once this is done we may apply $\Omega^{-1}$ to
all the linear forms in the gates and get the circuit for $f$. This step clearly takes $poly(d^r)$ time in the same way as applying $\Omega$ took.
Since $r$ is a constant, the steps described above take $poly(d)$ time overall.

\paragraph{Known and Unknown Parts}
We also define some other $\Pi\Sigma$ polynomials $K_i,U_i, i=0,1$
which satisfy
\[
K_i\mid \alpha_iGT_i, U_i=\frac{\alpha_iGT_i}{K_i}.
\]
with the extra condition
\[
gcd(K_i,U_i)=1.
\]
$K_i$ are the known factors of $\alpha_iGT_i$ and $U_i$ the unknown factors. The
$gcd$ condition just means
that that known and unknown parts of $\alpha_iGT_i$ don't have common factors.
In other words linear forms in $\alpha_iGT_i$ are known with full multiplicity.
We initialize $K_i=1$ and during the course of the algorithm update them as and
when we recover more linear forms.
At the end $K_i=\alpha_iGT_i$ and so we know both gates.

\subsection{Outline of the algorithm}

\begin{enumerate}
    \item \label{candidateset}{\bf Set $\MC$ of Candidate Linear Forms :}

    We compute a $poly(d)$ size set $\MC$ of linear forms which contains
$\ML(T_i), i=0,1$. We will non-deterministically guess from this set $\MC$
making only a constant number of guesses every time(thus polynomial work
overall). It is important to note that the
uniqueness of our circuit guarantees that our answer if computed can always be
tested to be right. For more details on this please see Appendix \ref{findcandidate}.
We also give an efficient algorithm to construct this set. See Algorithm \ref{candidatealgo}.\\

    \item \label{partknown} \underline{\bf Easy Case  :}
\centerline{\fbox{$\ML(T_{1-i}) \subsetneq sp(U_{i}),$ for
some $i\in \{0,1\}$ : }\\}\\

        So $T_{1-i}$ has a linear factor $l_{(1-i)1}$ such that
\begin{equation}\label{eq:1}
 sp(\{l_{(1-i)1}\})\cap sp(U_{i})=\{0\}
\end{equation}
Let $W = sp(\{l_{(1-i)1}\})$ and extend
to a basis of $V$ and in the process obtain
        another subspace $W^\prime\subset V$ such that $W\oplus W^\prime = V$.
We can see from Equation \ref{eq:1} that LI linear forms in $U_{i}$ remain LI
    when we project to $W^\prime$. We use this to compute $U_i$ and then
    since $K_iU_i=\alpha_iGT_i$ we know one of the gates. To find the other gate
simply factorize $f-\alpha_iGT_i$.
    If it factors into a product of linear forms we have the reconstruction.\\

       \item \label{dimensiongap} \underline{\bf Medium Case :} -
\centerline{\fbox{ $dim(\sfrac{sp(T_{1-i}) +
sp(T_i)}{sp(T_i)} )\geq 2$ for some $i\in \{0,1\}$ :}\\}\\

This case is just to facilitate the Hard Case.
     We know that $T_{1-i}$ has two linear factors $l_{(1-i)1},l_{(1-i)2}$ such that
$sp(\{l_{(1-i)1},l_{(1-i)2}\})\cap sp(T_i)=\{0\}$. We show that the
only linear factors of $f$ are those which appear in $G$. So we can first
factorize $f$ using Kaltofen's factoring (\cite{KalTr90}) and obtain $G$.
Update $K_j=G, j=0,1$. So $U_j = \alpha_jT_j$ for $j=0,1$. Clearly we also have
$\ML(T_{1-i})\subsetneq sp(T_i) = sp(U_i)$ and we can go to {\bf Easy Case }
above with $K_i=G$.\\

\item \underline{\bf Hard Case :}
  \centerline{\fbox{ $\ML(T_{1-i}) \subseteq sp(U_{i}),$ for $i=0$ and
$1$ :}\\}

    We know that we are not in {\bf Medium Case} and so
$dim(sp(T_0)+
sp(T_1)) - sp(T_i)\leq 1$  for $i=0,1$. Also $dim(sp(T_0)+ sp(T_1)) = r$ by
assumption on the simple rank of our polynomial.
    So this guarantees that $dim(sp(T_{1-i})) \geq r-1 \Rightarrow$ (by the
condition of this hard case) $dim(sp(U_i))
\geq r-1$ for $i=0,1$. This enables us to use the Quantitative
Sylvester Gallai theorems on both sets $\ML(T_i), \ML(U_i)$.
\begin{itemize}
\item Our first step is to identify a certain \emph{"bad"} $\Pi\Sigma$
    factor $I$ of $G$ and get rid of it to get $G = \frac{G}{I}$ and thus $f=\frac{f}{I}$. The factors of $I$ don't satisfy certain properties
    we need later and so we remove them. Thankfully we have an efficient algorithm to recover $I$.
    Our algorithm uses something we call a Detector Pair (See \ref{detectorset}) whose existence is shown
    using the Quantitative Sylvester Gallai Theorems mentioned above.
\item So now our job is to reconstruct $f$
    with known (and unknown resp.) parts as $K_0^\star,K_1^\star$ ($U_0^\star,U_1^\star$ resp.).

\item If $sp(U_{1-i})$ becomes low dimensional we may fall in {\bf Easy Case} and
    recover the circuit for $f$ directly. Otherwise the same detector pairs then provide certain \emph{"nice"}
    subspaces corresponding to linear forms in $T_{i}$.
    Projection of $U_{1-i}$ onto these subspaces can be easily glued together to recover some linear factors(with multiplicities) of
    $U_{1-i}$, which will then be multiplied to $K_{1-i}^\star$.
\item The
process continues as long as $sp(U_{1-i})$ remains high dimensional. As soon
as this condition fails we end up in {\bf Easy Case } and the gates are recovered.
\end{itemize}
    \end{enumerate}

We give algorithms for {\bf Easy} and {\bf Medium} cases. {\bf Hard Case} will require more  preparation and will be done after these subsections. From now onwards
we assume that we have constructed a $poly(d)$ sized set of linear forms $\MC$ which contains $\ML(T_i)$ for $i=0,1$. We have other structural results about linear 
forms
in this set. See Appendix \ref{findcandidate} for more details and algorithms. Algorithm \ref{candidatealgo} constructs this set in $poly(d)$ time.

\subsection{Easy Case} \label{partknowngate}

\begin{center}
\fbox{$\ML(T_{1-i}) \subsetneq sp(U_{i}),$ for some
$i\in \{0,1\}$ }
\end{center}

\begin{claim}
Suppose for some $i\in \{0,1\}$, $\ML(T_{1-i}) \subsetneq sp(U_{i})$ then we can
reconstruct $f$.
\end{claim}

\IncMargin{1em}
\begin{algorithm}[H]

\SetKwData{Left}{left}\SetKwData{This}{this}\SetKwData{Up}{up}
\SetKwInOut{Name}{FunctionName}\SetKwFunction{FindCompress}{FindCompress}
\SetKwInOut{Input}{input}\SetKwInOut{Output}{output}

\Name{EasyCase}
\Input{$f \in \Sigma\Pi\Sigma_\F(2)[\B{x}],K_0 \in \Pi\Sigma_\F[\B{x}], K_1 \in \Pi\Sigma_\F[\B{x}], \MC \subset Lin_{\F}[\B{x}])$}
\Output{An object of type $decomposition$}
\BlankLine
%\emph{special treatment of the first line}\;
\For{$i\leftarrow 0$ \KwTo $1$}{
%\emph{special treatment of the first element of line $i$}\;
\For{ each LI set $\{l_1,l_2,\ldots,l_r\}  \subset \MC $}{\label{guess basis}
Define $K_i^\prime \gets K_i$\;
Find $t$ such that $l_1^t\mid\mid f$ \;
\tcp{i.e. $l_1^t\mid f$ \&\& $l_1^{t+1}\nmid f$}
$W \gets sp(\{l_1\}), W^\prime \gets sp(\{l_2,\ldots,l_r\})$\;
%\This$\leftarrow$ \FindCompress{$Im[i,j]$}\;
\If(){$l_1^t\mid\mid K_i^\prime$}{\label{lt}
$\tilde{f} = \frac{f}{l_1^t}$; $\tilde{K_i} = \frac{K_i^\prime}{l_1^t}$\;
\lIf {$U_i=\frac{\pi_{W^\prime}(\tilde{f})}{\pi_{W^\prime}(\tilde{K_i})}\in
                \Pi\Sigma_\F[\B{x}]$ \&\& $f - K_iU_i \in \Pi\Sigma_\F[\B{x}]$}
            {$K_i = K_iU_i$, $K_{1-i} = f-K_iU_i$}
            \Return {$decomposition(f,K_0,K_1)$\;}
    }
    }
}
\Return {$decomposition()$\;}

\caption{Easy Case Reconstruction}\label{easycaserecon}
\end{algorithm}\DecMargin{1em}

\paragraph{Explanation and Correctness Analysis}

\begin{itemize}
\item The first for loop just guesses the gate with extra dimensions i.e. it's not contained in
span of the unknown part of the other gate.
\item If for some basis $\{l_1,\ldots,l_r\}\subset \MC$ the algorithm actually computes a
$\Sigma\Pi\Sigma(2)$ representation in the end then
it ought to be correct since the last 'if' also checks if it is correct.
 \item If our guess for $i$ is correct, we show that there exists a basis
$\{l_1,\ldots,l_r\} \subset \MC$ for which all conditions will be
 satisfied and we actually arrive at a $\Sigma\Pi\Sigma(2)$ representation in
the end. Since $\ML(T_{1-i})\subsetneq sp(U_i)$ and $\ML(T_{1-i}),\ML(U_i)\subset \MC$ there exists
$l_1\in \ML(T_{1-i})\setminus sp(U_i)\subset \MC$. Choose a basis $\{l_2,\ldots,l_s\}$ of $sp(U_i)$, then $\{l_1,\ldots,l_s\}$ is an LI set.
Now extend this to a basis $\{l_1,\ldots,l_s,l_{s+1},\ldots,l_r\}\subset \MC$ of $V$.  We go over all choices of basis
in $\MC$ and will arrive at the right one.

\item We initialize a dummy polynomial $K_i^\prime$ to represent
$K_i$ since we do not want to update $K_i$ till we actually have a solution.
Let's assume $l_1 ^t \mid\mid f$ i.e. $l_1^t\mid f$ and $l_1^{t+1}\nmid f$. We know $l_1\mid T_{1-i}\Rightarrow l_1\nmid T_i
\Rightarrow l_1\nmid \alpha_iT_i + \alpha_{1-i}T_{1-i}$. Therefore
$l_1^t\mid\mid G \Rightarrow l_1^t\mid\mid \alpha_iGT_i = K_iU_i$. Also $l_1
\notin sp(U_i) \Rightarrow l_1\nmid U_i$ thus $l_1^t \mid\mid K_i \Rightarrow l_1^t\mid\mid K_i^\prime$. We remove
$l_1^t$ from both $f,K_i^\prime$ to get $\tilde{f}, \tilde{K_i}$. Let $W = sp(\{l_1\})$ and $W^\prime =
sp(\{l_2,\ldots,l_r\})$, therefore $V = W\oplus W^\prime$.
 Note that since $l_1\in \ML(T_{1-i})$
 \[
 \pi_{W^\prime}(\tilde{f}) = \pi_{W^\prime}(U_i)\pi_{W^\prime}(\tilde{K_i})
  \]
  Since $\pi_{W^\prime}(\tilde{K_i})\neq 0$, we  get
$\pi_{W^\prime}(U_i)=\frac{\pi_{W^\prime}(\tilde{f})}{\pi_{W^\prime}(\tilde{K_i})}$.
  If $U_i = u_1\ldots u_s$ with $u_j\in W^\prime$, we see that $\pi_{W^\prime}(U_i) =
\pi_{W^\prime}(u_1)\ldots \pi_{W^\prime}(u_s) = u_1\ldots u_s = U_i$.
  So we get $U_i$ and hence $\alpha_iGT_i =K_iU_i$ . Once $\alpha_iGT_i$ is
known we factorize $f-\alpha_i GT_i$ to get $\alpha_{1-i} GT_{1-i}$. For the correct choice of our basis this will
factorize completely into a $\Pi\Sigma$ polynomial. Now we update $K_i = K_iU_i$ and $K_{1-i}=f-K_iU_i$ and an object $decomposition(f,K_0,K_1)$. Throughout
the algorithm we use Kaltofen's factoring \cite{KalTr90} wherever necessary.
\item If we were not able to find the $\Sigma\Pi\Sigma(2)$ representation then we return an object $decomposition()$.
\end{itemize}

\paragraph{Time Complexity - }
We can see above all loops run only $poly(d)$ many times. The most expensive step is choosing
$~ r$ vectors from $\MC$. But recall that $r$ is a constant and so this also takes only polynomial time in $d$.
Other steps like factoring polynomials (using Kaltofen's factoring algorithm from \cite{KalTr90}), taking projection onto known subspaces, divding
by polynomials require $poly(d)$ time ($r$ is a constant) as has been explained multiple times before.\\

\subsection{Medium Case} \label{mediumcase}

\begin{center}
\fbox{$dim(\sfrac{sp(T_{1-i}) +
sp(T_i)}{sp(T_i)} )\geq 2$ for
some $i\in \{0,1\}$ }
\end{center}

\begin{claim}
If $dim(\sfrac{sp(T_{1-i}) + sp(T_i)}{sp(T_i)} )\geq 2$ then
$\ML(\alpha_iT_i+\alpha_{1-i}T_{1-i})=\phi$.
\end{claim}
\emph{Proof.}
 $dim(\sfrac{sp(T_{1-i}) + sp(T_i)}{sp(T_i)} )\geq 2\Rightarrow$, there exists $l_1^\prime,l_2^\prime\in \ML(T_{1-i})\setminus sp(T_{i})$ be
such that $dim(\{l_1^\prime,l_2^\prime\}\cup \ML(T_{i}))= dim(\ML(T_{i}))+2$.
Assume there
 exist $l\in \ML(\alpha_iT_i+ \alpha_{1-i}T_{1-i})$.

 \[
 l \mid \alpha_iT_i+ \alpha_{1-i}T_{1-i} \Rightarrow  l \nmid T_i \text{ and }
l\nmid T_{1-i} \text{ (since they are coprime) }
 \]

\[
0\neq \alpha_i\prod\limits_{j\in[M]}l_{ij} =
-\alpha_{1-i}\prod\limits_{j\in[M]}l_{(1-i)j} \pmod{\{l\}}.
 \]

 Thus there exist $l_1,l_2\in \ML(T_i)$ and scalars $\gamma_j,\delta_j, j\in [2]$ such that $l = \gamma_j l_{j} + \delta_j l_{j}^\prime$.
 Since $l\nmid T_0,l\nmid T_1$ we get $\gamma_j,\delta_j$ are non zero.

$\delta_1,\delta_2\neq 0 \Rightarrow$,
\[
l_1^\prime,l_2^\prime\in sp(\{l\}\cup \ML(T_{i}))\Rightarrow
dim(\{l_1^\prime,l_2^\prime\}\cup \ML(T_{i})) \leq dim(\ML(T_{i}))+1
 \]
 which is a contradiction.
So $\ML(\alpha_iT_i+ \alpha_{1-i}T_{1-i})=\phi$. \

Therefore the only linear factors of $f$ are
present in $G$, which can now be correctly found by using Kaltofen's algorithm
\cite{KalTr90} and identifying the linear factors.
Update $K_j=G$ for $j=0,1$, therefore $U_j=T_j$. Also this case implies that
$\ML(T_{1-i})\subsetneq sp(T_i)=sp(U_i)$.
and so we can use Easy Case.

So we have the following claim: 	 	
\begin{claim}
If the condition in Medium Case is true, the following algorithm reconstructs $f$, if there is a reconstruction.
\end{claim}

\IncMargin{1em}
\begin{algorithm}[H]
\SetKwInOut{Name}{FunctionName}
\SetKwData{Left}{left}\SetKwData{This}{this}\SetKwData{Up}{up}
\SetKwFunction{Union}{Union}\SetKwFunction{FindCompress}{FindCompress}
\SetKwInOut{Input}{input}\SetKwInOut{Output}{output}
\Name {MediumCase}
\Input{$f \in \Sigma\Pi\Sigma_\F(2)[\B{x}], \MC \subset Lin_{\F}[\B{x}])$}
\Output{An object of type $decomposition$}
\BlankLine
$L\gets Lin(f)$\;
\tcp{Use Kaltofen's factoring from \cite{KalTr90} to compute $Lin(f)\eqdef$ product of all linear factors of $f$}
\If(){EasyCase$(f,L,L,\MC)\rightarrow$ iscorrect}{\label{lt}
            \Return { EasyCase $(f,L,L,\MC)$\;}
    }
\Return {$decomposition()$\;}

\caption{Medium Case Reconstruction}\label{easycaserecon}
\end{algorithm}\DecMargin{1em}

The above algorithm does exactly what has been explained in the preceding paragraph. It works in $poly(d)$
time if EasyCase$(f,K_0,K_1,\MC)$ works in $poly(d)$ time. Kaltofen's factoring and all other steps
are $poly(d)$ time.\\

Now we need to handle the {\bf Hard Case}. This is quite technical and so we do some more preparation. We devise a technique
to get rid of some factors of $f$ to get a new polynomial $f$ without destroying the $\Sigma\Pi\Sigma(2)$ structure.
If Easy Case holds for $f$ we stop there itself. Otherwise we will use
combination of different subspaces of $V$, project $f$ onto them and glue projections to get gates for $f$.

\subsection{Detector Pair, Reducing Factors, Hard Case Preparation}\label{reducefactors}

Let's recall:
\[
g = \frac{f}{G} = \alpha_0T_0+\alpha_1T_1
\]

We outline an approach to identify some factors of $f$. These factors will
divide $G$ but won't divide $g$. This is going to be useful in the Hard Case.
The linear factors left after removing these identified factors will have very
strong structural properties and so will be instrumental in reconstruction. The
main tool in this identification is a pair $(S,D)$ (defined below) inside one
of the $\ML(T_i)$'s. This pair will be called a \emph{``Detector Pair''}. It will
also decide the subspaces on which we take projections of $f$ and glue back to
get the gates.
\paragraph{Detector Pairs $(S,D)$ }\label{detectorset}
Fix $k=c_{\F}(3)+2$ (See Theorem \ref{rankbound} for definition of $c_\F(m)$). Let $S = \{l_{1},\ldots,l_{k}\} \subset \ML(T_i)$ be an LI
set of linear forms. Let $D(\neq \phi)\subseteq \ML(T_i)$ .We say that $(S,D)$
is a
\emph{"Detector Pair"} in $\ML(T_i)$ if the following are satisfied for all $l_{k+1}
\in D$:
\begin{itemize}
\item $\{l_{1},\ldots,l_{k},l_{k+1}\}$ is an LI set. Let $\mathcal{F} =
fl(\{l_{1},\ldots,l_{k},l_{k+1}\})$. $\mathcal{F}$ is elementary in $\ML(T_i)$
i.e.
$\mathcal{F}\cap \ML(T_i) = \{l_{1},\ldots,l_{k},l_{k+1}\}$. See
Definition \ref{elementaryset}.
\item $\mathcal{F}\cap \ML(T_{1-i}) \subseteq fl(\{l_{1},\ldots,l_{k}\})$ i.e.
$\mathcal{F}$  contains only those points from $\ML(T_{1-i})$ which lie inside
$fl(\{l_{1},\ldots,l_{k}\})$.
\end{itemize}

\subsubsection{Identifying Some Factors Which Don't Divide $g$}\label{identification}

The two claims below give results about structure of linear forms which
divide $g$. The proofs are easy but technical and so we move them to the
appendix.
\begin{claim} \label{spuriousli}
Let $(S = \{l_{1}\ldots, l_{k}\},D)$ be a Detector set in $\ML(T_i)$. Let
$l_{k+1}\in D$. For a $standard$ linear form $l\in V$, if $l\mid g$ then
$l\notin sp(\{l_{1},\ldots,l_{k}\})$ .
\end{claim}
\emph{Proof.}
See \ref{spuriousliproof} in Appendix

\begin{claim}\label{spuriousextra}
Let $l \in Lin_\F[\B{x}]$ be $standard$ such that $l \mid g$ and
$\MC$ be the candidate set. Assume $(S = \{l_{1},\ldots,l_{k}\}, D(\neq \phi))$
is a Detector pair in $\ML(T_i)$. Then $|\ML(T_{1-i}) \cap (fl(S\cup\{l\}) \setminus
fl(S))|\geq 2$. That is the flat $fl(\{l_{1},\ldots,l_{k},l\})$ contains
atleast two distinct points from $\ML(T_{1-i})(\subseteq \MC)$ outside $fl(\{l_1,\ldots,l_k\})$.
\end{claim}

\emph{Proof.}
See \ref{spuriousextraproof} in Appendix

\begin{claim}\label{identifylinform}
Suppose $(S=\{l_{1},\ldots,l_{k}\},D(\neq \phi))$ is a Detector Pair in $\ML(T_i)$.
The following algorithm identifies some factors in $\ML(G)\setminus \ML(g)$.
It returns the product of all linear forms identified.
\end{claim}

\IncMargin{1em}
\begin{algorithm}[H]
\SetKwInOut{Name}{FunctionName}
\SetKwData{Left}{left}\SetKwData{This}{this}\SetKwData{Up}{up}
\SetKwFunction{Union}{Union}\SetKwFunction{FindCompress}{FindCompress}
\SetKwInOut{Input}{input}\SetKwInOut{Output}{output}
\Name{IdentifyFactors}
\Input{$f\in\Sigma\Pi\Sigma_\F(2)[\B{x}], \MC \subset Lin_\F[\B{x}], S =\{l_{1},\ldots,l_{k} \} \subset Lin_\F[\B{x}])$}
\Output{a $\Pi\Sigma_\F[\B{x}]$ polynomial }
\BlankLine
{\bf I} $= 1$, $bool$ $flag$\;
\For{ each factor $l$ of $f$ } {
$flag=false$\;
\If{$l, l_{1}, \ldots,l_{k}$ are LI}{
\For{ $l_1^\prime \neq l_2^\prime \in \MC\setminus fl(\{l_{1},\ldots,l_{k}\})$} {
\lIf {$l_1^\prime,l_2^\prime\in sp(\{l,l_{1},\ldots,l_{k}\})$}{$flag=true$\; $break$}}}
\If { $!flag$ } {{\bf I} = {\bf I}$\times l$\;}
}

\Return {{\bf I}\;}

\caption{Identify Factors}\label{easycaserecon}
\end{algorithm}\DecMargin{1em}

\emph{Proof.}
 The proof of the claim is a part of Lemma \ref{filteredfactor} below.

\paragraph{Time Complexity - }
Since $\MC$ has size $poly(d)$ and $deg(f)=d$, the nested loops run $poly(d)$ times. $k,r$ are constants
so checking linear independence of $k+1$ linear forms in $r$ variables takes constant time. Checking if some
vectors belong to a $k+1$ dimensional space also takes constant time. Multiplying linear forms to {\bf I} takes
$poly(d)$ time. So overall the algorithm runs in $poly(d)$ time.\\

So the above algorithm identified a factor {\bf I} of $G$  for us. Let us define new polynomials
\[
 G = \frac{G}{{\bf I}} = \prod\limits_{t\in [N_1]}G_t
\]
and
\[
f = \frac{f}{{\bf I}} = G (\alpha_0T_0 + \alpha_1T_1)
\]

\begin{lemma}\label{filteredfactor}
The following are true:
\begin{enumerate}
\item If $l\mid I$ (i.e. $l$ was identified) then $l\in \ML(G)\setminus
\ML(g) $.
\item \label{retainedfactor}If $l\mid G$ (i.e. $l$ was retained) then
$(fl(\{l_{1},\ldots,l_{k},l\})\setminus fl(\{l_{1},\ldots,l_{k}\})) \cap
(\ML(T_{1-i})\cup (\ML(T_i)\setminus D)) \neq \phi $ that is:

$(fl(\{l_{1},\ldots,l_{k},l\})\setminus fl(\{l_{1},\ldots,l_{k}\}))$ contains
a point from $\ML(T_i)\setminus D$ or $\ML(T_{1-i})$.

\item \label{retaineddetector} If $l\mid G$ and $l_{k+1}\in D$ then
$l \notin sp(\{l_{1},\ldots,l_{k},l_{k+1}\})$.
    \end{enumerate}
\end{lemma}

\emph{Proof.}
See \ref{prooffilteredfactor} in Appendix.

\subsubsection{Overestimating the set $D$ of the detector pair $(S,D)$}

Lemma \ref{filteredfactor} is
going to help us actually find an overestimate of $D$ corresponding to
$S=\{l_{1},\ldots,l_{k}\}$ in the detector pair $(S,D)$ as described in the lemma
below. This will be important since we need $D$ during our algorithm for the Hard Case.

\begin{lemma} \label{detectorexpansion}
Let $(S=\{l_{1},\ldots,l_{k}\},D)$ be a detector in $\ML(T_i)$. For each
$(l,l_j) \in \MC \times S$ define the space $U_{\{l,l_j\}} = sp(\{l,l_j\})$.
Extend $\{l,l_j\}$ to a basis and in the process obtain $U_{\{l,l_j\}}^\prime$
such that $V = U_{\{l,l_j\}}\oplus U_{\{l,l_j\}}^\prime$. Define the set:
\[
 X = \{l\in \MC : \pi_{U^\prime_{\{l,l_j\}}}(f) \neq 0, \text{  for all }
l_j\in S\}
\]
Then $D\subset X\subset \ML(T_i)$.
\end{lemma}

\emph{Proof.}
See \ref{detectorexpansionproof} in Appendix.

This set $X$ is an overestimate of $D$ inside $\ML(T_i)$ and also easy to
compute. Given $S$ we may
easily construct $X$ in time $poly(d)$ because of it's
simple description. Let's give an algorithm to compute $X$ given
$f,S,\MC$.

\begin{claim}
The following algorithm computes the overestimate $X$ of $D$ as discussed above
\end{claim}

\IncMargin{1em}
\begin{algorithm}[H]\label{overestimatedetector}
\SetKwInOut{Name}{FunctionName}
\SetKwData{Left}{left}\SetKwData{This}{this}\SetKwData{Up}{up}
\SetKwFunction{Union}{Union}\SetKwFunction{FindCompress}{FindCompress}
\SetKwInOut{Input}{input}\SetKwInOut{Output}{output}
%\Function{IdentifyFactors}
\Name{OverestimateDetector}
\Input{$f \in \Sigma\Pi\Sigma_\F(2)[\B{x}], S =\{l_1,\ldots,l_k\} \subset Lin_\F[\B{x}], \MC \subset Lin_\F[\B{x}])$}
\Output{ Set of linear forms }
\BlankLine
$bool$ $flag$\;
Define $X\gets \phi$\;
\For { each $l\in \MC$ } { $flag=true$\;
  \For{ each $l_j\in S$ with $\{l,l_j\}$ LI }{
  Find $\{l_1^\prime,\ldots,l_{r-2}^\prime\}\subset \MC$ such that $\{l,l_j,l_1^\prime,\ldots,l_{r-2}^\prime\}$ is LI\;
  $U \gets \F l\oplus \F l_j;  U^\prime \gets \F l_1^\prime \oplus\ldots\oplus \F l_{r-2}^\prime$\;
\If {$\pi_{U^\prime}(f) == 0$ } {$flag=false$\; $break$\;}
}
\If { $flag$ } {$X\gets X\cup \{l\}$\;}
}

\Return {$X$\;}

\caption{Overestimate Detector}\label{easycaserecon}
\end{algorithm}\DecMargin{1em}

\paragraph{Time Complexity - } Inside the inner for loop we look for $(r-2)$ linear forms from $\MC$. $|\MC| = poly(d)$
and $r$ is a constant and so this step only needs $poly(d)$ time. The nested loops run polynomially many times. Checking linear independence
of $r$ linear forms and projecting to known constant dimensional subspaces also take $poly(d)$ time as has been
discussed before. So the algorithm runs in $poly(d)$ time.

\subsection{Hard Case } \label{hardcase}

\begin{center}
\fbox{$\ML(T_{1-i}) \subseteq sp(U_{i}),$ for $i=0$ and
$1$
}
\end{center}
\paragraph{}\label{hardcasediss}
This Subsection will involve the most non trivial ideas. We handled
$dim(\sfrac{sp(T_{1-i}) + sp(T_i)}{sp(T_i)} )\geq 2$ in the Medium Case (see Subsection \ref{mediumcase})
completely, so let's assume $dim(\sfrac{sp(T_{1-i}) + sp(T_i)}{sp(T_i)} )\leq
1\Rightarrow dim(\ML(T_{1-i})\cup \ML(T_i))\leq dim(\ML(T_i))+1$ for both
$i=0,1$. We already know that $rank(f)=r,$ implying $dim(\ML(T_i)\cup
\ML(T_{1-i}))=r$. Thus for $i=0,1$; $dim(\ML(T_i))\geq r-1$.
This works in our favor for applying the quantitative
version of the Sylvester Gallai theorems given in \cite{BDWY11}. To be precise
we will use
Corollary \ref{bichromatic} from Appendix \ref{incidence} in this paper.

\begin{enumerate}

\item Our first application (See Lemma \ref{largedetector}) of Quantitative Sylvester
Gallai will help us prove the existence of
a Detector pair $(S=\{l_{1},\ldots,l_{k}\},D)$ in $\ML(T_i)$ with $k=c_\F(3)+2$ (See definition of $c_\F(.)$ in Theorem \ref{rankbound})
and large size of $D$. For this we will
only need $dim(\ML(T_i))\geq C_{2k-1}$ for $i=0,1$(See Appendix \ref{incidence}
for definition of $C_{2k-1}$). From Definition \ref{rvalue} we know that this is true with $k =c_\F(3) +2$.

\item The above point shows the existence of a detector pair $(S,D)$ in $\ML(T_i)$ with large $|D|$. So now we go back to
Subsection \ref{reducefactors} and remove some factors of $f$ to get $f = G(\alpha_0T_0 + \alpha_1T_1)$
such that linear factors of $G$ satisfy properties given in Lemma \ref{filteredfactor}. We also compute the overestimate
$X$ of $D$ using Algorithm \ref{overestimatedetector}. Let the known and unknown parts of $f$ be $K_0^\star,K_1^\star$
and $U_0^\star, U_1^\star$ respectively. If for some $i\in \{0,1\}$, $\ML(T_i)\subsetneq sp(U_{1-i})$ then we are in
Easy Case for $f$ and can recover the gates for $f$. Otherwise for both $i=0,1;$ $\ML(T_i)\subseteq sp(U_{1-i})
\Rightarrow dim(\ML(U_{1-i}))\geq r-1$ and we continue with reconstruction below.

\item Next to actually reconstruct linear forms in $U_{1-i}$, we will use it's high-dimensionality ($\geq r-1\geq C_{2k-1}$) discussed above.
Corollary \ref{bichromatic} from Section \ref{incidence} will enable us to prove
the existence
of a $d_1\in D$ which together with the set $S$ found above will give the existence of a \emph{"Reconstructor"}( See
Claim \ref{reconalgoclaim} and Algorithm \ref{reconalgo}) which recovers
some linear factors of $U_{1-i}$ with multiplicity (See Theorem \ref{foundreconstructor}) .

\end{enumerate}

\subsubsection{Large Size of Detector Sets}

w.l.o.g. we assume $|\ML(T_0)|\leq |\ML(T_1)|$.
First we point out a simple calculation that will be needed later.
For $\delta \in (0, \frac{7-\sqrt{37}}{6})$ and $\theta\in (
\frac{3\delta}{1-\delta}, 1-3\delta )$,
let $v(\delta ,\theta)$ be defined as follows:
\[
 v(\delta,\theta) =
  \begin{cases}
      \hfill 1-\delta-\theta    \hfill & \text{ if $|\ML(T_{0})|\leq \theta
|\ML(T_1)|$} \\
      \hfill (1-\delta)(1+\theta)-1 \hfill & \text{ if $\theta |\ML(T_1)| <
|\ML(T_0)| \leq |\ML(T_1)|$} \\
  \end{cases}
\]
\begin{claim}\label{calculation}
The following is true

\[
  \frac{(2-v(\delta,\theta))}{v(\delta,\theta)}\leq \frac{1-\delta}{\delta}
\]

\end{claim}

\emph{Proof.}
See \ref{calculationproof} in Appendix.

%For convenience we will fix $\delta=\frac{1}{12}$ and $\Theta = \frac{1}{2}$ from now onwards. Both belong to our range
%defined in the Claim above.

\begin{lemma}\label{largedetector}
Let $k=c_{\F}(3)+2$ (see definition of $c_\F(m)$ in Theorem \ref{rankbound}). Fix $\delta, \theta$ in range given in Claim \ref{calculation} above .
Then for some $i\in \{0,1\}$ there exists a Detector
$(S=\{l_{1},\ldots,l_{k}\},D)$ in $\ML(T_i)$  with
$|D|\geq v(\delta,\theta) \max(|\ML(T_{0})|,|\ML(T_{1})|)$.

\end{lemma}

\emph{Proof.}
 See \ref{largedetectorproof} in Appendix.

\subsubsection{Assuming $\ML(T_i)\subseteq sp(\ML(U_{1-i}))$ and reconstructing factors of $U_{1-i}$}\label{getreconstructor}

Let's begin by stating our main reconstruction theorem for this Sub-subsection. We will go through several steps to prove it:

\begin{theorem}\label{foundreconstructor}
There exist pairwise disjoint LI sets $S_0,S_1,S_2$ with $S_0\cup S_1\cup
S_2$ being a basis of $V =  Lin_\F[x_1,\ldots,x_r] \simeq \F^r$, and non constant polynomials $P,Q$ dividing $ U_{1-i}$
such that $P\mid Q$ and
$(Q,P,S_0,S_1,S_2)$ is a Reconstructor.

\end{theorem}

Once we know this result we actually recover $P$ by computing $\pi_{W_0^\prime}(Q)$ and $\pi_{W_1^\prime}(Q)$ and then using
Algorithm \ref{reconalgo}. We state this in the following corollary. Proof is given as Algorithm \ref{hardcasealgo}

\begin{corollary}\label{findprojections}
 Using $f, K_{1-i}, S_0,S_1,S_2$ from above we can compute
$\pi_{W_0^\prime}(Q),
\pi_{W_1^\prime}(Q)$ for $Q$ defined in the proof above.
\end{corollary}

Before going to the proof let's do some more more preparation.

\paragraph{}\label{lambda}
Consider the set of linear forms $\MX = \ML(G)\cup \ML(T_0)\cup \ML(T_1)$. We know
that $sp(\MX) = V = Lin_\F[\B{x}]\simeq \F^r$ (By abuse of notation we will use linear forms as points in $\F^r$ wherever required).
Let $(S_0 = \{l_{1},\ldots,l_{k}\},D)$ be a detector in $\ML(T_i)$ with $|D|\geq
v(\delta,\theta)\max(|\ML(T_{0})|, |\ML(T_{1})|)$ as obtained in the preceding discussion. \\

Define $W_0 = sp(S_0)$ and extend $S_0$ to a basis
$\{l_1,\ldots,l_k,l_{k+1}^\prime,\ldots,l_r^\prime\}$. Now it's time to use the other random matrix $\Lambda$. Since we had
applied $\Omega$ in the beginning, $\{\Omega^{-1}(l_1),\ldots,\Omega^{-1}(l_k)\}$ are linear forms in our input
polynomial for this section. By {\bf Assumption 3 }
we know that the set
\[
\{\Omega (\Omega^{-1}l_1),\ldots , \Omega (\Omega^{-1}l_k), \Lambda \Omega(\Omega^{-1}l_{k+1}^\prime), \ldots, \Lambda \Omega(\Omega^{-1}l_{r}^\prime)
\}
\]
is LI.  Let $l_j = \Lambda l_j^\prime, j\in \{k+1,\ldots,r\}$. So $\MB = \{l_1,\ldots,l_r\}$ is a basis. and define
$W_0^\perp = sp(\{l_{k+1},\ldots,l_r\})$.
Clearly $V = W_0\oplus W_0^\perp$.\\

By {\bf Assumption 4} for any $l\in \MX \setminus W_0$, $[l]^{k+1}_{\MB} \neq 0$. We re-normalize all linear forms in
$\MX \setminus W_0$ making sure that the coefficient of $l_{k+1}$ is $1$ in them. From now onwards this will be assumed.\\

With this notation we proceed towards detecting linear factors of the unknown parts. But first let's show that even after
projecting onto $W_0^\perp$, the detector is larger in size (up to a function of $\delta$) compared to one of the unknown parts.

\begin{lemma}\label{findreconstructor}
The following are true:
\begin{enumerate}
 \item $dim(\pi_{W_0^\perp}({\ML(U_{1-i})}))> C_4$
 \item $\pi_{W_0^\perp}({\ML(U_{1-i})})\cap \pi_{W_0^\perp}({D}) = \phi$
 \item $|\pi_{W_0^\perp}({\ML(U_{1-i})})\setminus \{0\}| \leq
\frac{1-\delta}{\delta}|\pi_{W_0^\perp}({D})|$
\end{enumerate}
\end{lemma}

\emph{Proof.}
See \ref{findreconstructorproof} Appendix.\\

This Lemma enables us to apply Lemma \ref{bichromatic} from Section \ref{incidence}.
Consider the sets $Y=\pi_{W_0^\perp}({\ML(U_{1-i})})\setminus \{0\}$ and
$X=\pi_{W_0^\perp}({D})$.
We've shown all conditions in Lemma \ref{bichromatic}, so there exists a line $\vec L$ (called a \emph{"semi-ordinary bi-chromatic"} line) in
$W_0^\perp$ such that
$|\vec L\cap Y|=1$ and $|\vec L\cap X|\geq 1$.\\

Let's prove another short lemma which is useful for technical reasons.

\begin{lemma}\label{technical}
For any subspace $W_0^\prime$ such that $V = W_0\oplus W_0^\prime = W_0 \oplus W_0^\perp$ there is a line
$\vec L \subset W_0^\prime$ such that
\begin{enumerate}
\item $|\vec{L} \cap \pi_{W_0^\prime}({D})|\geq 1$
 \item $|\vec L \cap (\pi_{W_0^\prime}({\ML(U_{1-i})})\setminus \{0\})|=1$
\end{enumerate}
\end{lemma}

\emph{Proof.}
We have the following commutative diagram :

\begin{tikzpicture}[every node/.style={midway}]
\centering
  \matrix[column sep={4em,between origins}, row sep={2em}] at (0,0) {
    \node(V) {$V$} ; \\
    \node(prime){$W_0^\prime$} ; & \node (perp) {$W_0^\perp$};\\
  };
  \draw[->] (V) -- (prime) node[anchor=east]  {$\pi_{W_0 ^\prime}$};
  \draw[->] (V) -- (perp) node[anchor=west]  {$\pi_{W_0 ^\perp}$};
  \draw[->] (perp) -- (prime) node[anchor=north] {$\pi_{W_0 ^\prime}$};
\end{tikzpicture}

Let $v = w + w^\perp \in V$ where
$w\in W_0, w^\perp \in W_0^\perp$, then
\[
\pi_{W_0^\prime}(\pi_{W_0^\perp}(v)) =
\pi_{W_0^\prime}(w^\perp) = \pi_{W_0^\prime}(w^\perp) + \pi_{W_0^\prime}(w)
=\pi_{W_0^\prime}(v)
\]
So $\pi_{W_0^\prime} = \pi_{W_0^\prime} \circ \pi_{W_0^\perp}$\\

Next let $T : V \rightarrow V$ be any bijection then $T(A\cap B) = T(A) \cap T(B)$ and therefore
$|A\cap B| = |T(A)\cap T(B)|$. Since the maps above are projections one can easily see that
$\pi_{W_0^\prime} : W_0^\perp \rightarrow W_0^\prime$
is an isomorphism where the inverse of any $w^\prime \in W_0^\prime$ is given as $\pi_{W_0^\perp}(w^\prime)$.
Call this map $T$. Now any
linear isomorphism between vector spaces also preserves affine dependence since:
\[
 T(\lambda u + (1-\lambda) v) = \lambda T(u) + (1-\lambda) T(v)
\]
So image of a line is a line. Let $\vec L$ be the line obtained in Lemma \ref{findreconstructor}.
\begin{itemize}
 \item $T(\vec L)$ is a line in $W_0^\prime$.
 \item $|T(\vec L) \cap \pi_{W_0^\prime}({D})| = |T(\vec L) \cap T(\pi_{W_0^\perp}({D}))| = |\vec{L} \cap \pi_{W_0^\perp}({D})| \geq 1$
 \item $ |T(\vec L) \cap \pi_{W_0^\prime}({\ML(U_{1-i})})| = |T(\vec L) \cap T(\pi_{W_0^\perp}({\ML(U_{1-i})}))| = |\vec L \cap \pi_{W_0^\perp}({\ML(U_{1-i})})|$
\end{itemize}
Since $T$ is a linear isomorphism
$0 \in \pi_{W_0^\perp}({\ML(U_{1-i})}) \Leftrightarrow 0 \in T(\pi_{W_0^\perp}({\ML(U_{1-i})})) = \pi_{W_0^\prime}({\ML(U_{1-i})})$ and
$0\in \vec L \Leftrightarrow 0\in T(\vec L)$, therefore
the third condition above is same as
\[
|T(\vec L) \cap (\pi_{W_0^\prime}({\ML(U_{1-i})})\setminus \{0\})| =
|\vec L \cap (\pi_{W_0^\perp}({\ML(U_{1-i})})\setminus \{0\})| =1
\]
So the lemma is true with $\vec L$ being the line $T(\vec L)$ obtained in this proof. \\

Finally it's time to give the proof of Theorem \ref{foundreconstructor}. Let $d_1\in D$ such that
$\pi_{W_0^\perp}(d_1) \in \vec L$ where $\vec L$ was the line obtained right after Lemma \ref{findreconstructor}.
Since coefficient of $l_{k+1}$ is non-zero in $d_1$, $\{l_1,\ldots,l_k,d_1,l_{k+2},\ldots,l_r\}$ is also a basis. Define
$S_0=\{l_1,\ldots,l_k\}, S_1 = \{d_1\}, S_2 = \{l_{k+2},\ldots,l_r\}, W_i = sp(S_i), W_i^\prime = \bigoplus\limits_{j\neq i}
W_j$. Note this implies $V = W_0 \oplus W_0^\prime$ and so Lemma \ref{technical} above can be used. Let $\vec L$
be the line the Lemma \ref{technical} gives. {\bf By re-normalization we also assume that all linear forms in $
\MX \setminus W_0^\prime$ have coefficient of $d_1$ equal to $1$}.\\

\emph{Proof of Theorem \ref{foundreconstructor}.}
We show this in steps:
\begin{itemize}
 \item Let $S_0,S_1,S_2$ be as defined in the discussion above.

\item Let $Q$ be the largest
factor of $U_{1-i}$ such that for all linear forms $q\mid Q$, $\pi_{W_2}(q)\neq
0$.
So $\pi_{W_2}(Q)\neq 0$ and if $u^\star\mid\frac{U_{1-i}}{Q}$ is a linear form
then $\pi_{W_2}(u^\star)=0$.
Let $P$ be the $\Pi\Sigma$ polynomial with the largest
possible degree such that
for all linear factors $p$ of $P$, $\pi_{W_0^\prime}( p) \in
\vec{L} \cap (\pi_{W_0^\prime}(\ML(U_{1-i}))\setminus \{0\})$.
Clearly $P$ is non constant since $|\vec{L} \cap (\pi_{W_0^\prime}(\ML(U_{1-i}))\setminus \{0\})|=1$.
Clearly $\pi_{W_0^\prime}(P)\neq 0 \Rightarrow P\mid Q$.
Then $(Q,P,S_0,S_1,S_2)$ is a \emph{Reconstructor} (See Subsection \ref{Identifier} for definition) for $P$. Let's check
this is true:

\begin{itemize}
\item $\pi_{W_2}(Q) \neq 0$ - By definition of $Q$ we know this for all it's
factors and therefore for $Q$ itself.
\item $\pi_{W_0^\prime}(P) =  \pi_{W_0^\prime}( p)^t$, for some
linear form $p\mid P$ (since $|\vec{L} \cap (\pi_{W_0^\prime}(\ML(U_{1-i}))\setminus \{0\})|=1$).
\item Let $q\mid \frac{Q}{P}$ such that
$gcd(\pi_{W_2}(P), \pi_{W_2}(q)) \neq 1 \Rightarrow$ there exists some
linear factor
$p\mid P$ such that $\pi_{W_2}(p), \pi_{W_2}(q)$ are LD.
 $\{\pi_{W_2}(p), \pi_{W_2}(q)\}$ are LD and non-zero implies
$q\in sp(\{l_1,\ldots,l_k,d_1,p\})
\Rightarrow
\pi_{W_0^\prime}(q)\in
sp(\{\pi_{W_0^\prime}(d_1),\pi_{W_0^\prime}(p)\})=
sp(\{d_1,\pi_{W_0^\prime}( p)\})$.
So clearly $\pi_{W_0^\prime}( q)\in sp(\{d_1,\pi_{W_0^\prime}( p)\})$.
Since coefficient of
 $d_1$ in $\pi_{W_0^\prime}( q),d_1,$ and $\pi_{W_0^\prime}( p)$ is
$1$, and therefore using Lemma \ref{spantoflat} it's
easy to see that
  $\pi_{W_0^\prime}( q)\in fl(\{d_1,\pi_{W_0^\prime}( p)\}) =
\vec L$.
Since $Q\mid U_{1-i}$ we have $\pi_{W_0^\prime}( q) \in
\pi_{W_0^\prime}( {\ML(U_{1-i})}) \setminus \{0\}\Rightarrow \pi_{W_0^\prime}( q) \in
\vec L\cap (\pi_{W_0^\prime}({\ML(U_{1-i})})\setminus \{0\}) =
\{\pi_{W_0^\prime}( p)\}$ which can't be true
since $P$ is the largest polynomial dividing $Q$ where linear factors have this
property and $q\nmid P$. So such a $q$ does not exist.
\end{itemize}
\end{itemize}

Now we give the algorithm for reconstruction in this case.

\begin{algorithm}[H]
\SetKwInOut{Name}{FunctionName}
\SetKwData{Left}{left}\SetKwData{This}{this}\SetKwData{Up}{up}
\SetKwFunction{Union}{Union}\SetKwFunction{FindCompress}{FindCompress}
\SetKwInOut{Input}{input}\SetKwInOut{Output}{output}
\SetKwInOut{Init}{Fix}
\Name{HardCase}
\Init{$k = c_\F(3) +2$}
\Input{$f\in \Sigma\Pi\Sigma_\F(2)[\B{x}], \MC \subset Lin_\F[\B{x}], \Lambda \in \F^{r\times r}$}
\Output{An object of type $decomposition$}
\BlankLine
 \For{$i\leftarrow 0$ \KwTo $1$}
   {
   \For{each LI $\MB^\prime= \{l_1,\ldots,l_k,l_{k+1}^\prime,\ldots,l_r^\prime\} \subset \MC$}
       {$S_0 = \{l_1,\ldots,l_k\}$\;
 	\For{$j \leftarrow k+1$ \KwTo $r$}{$l_j\gets\Lambda(l_j^\prime)$\;}
 	  \If {$\MB = \{l_1,\ldots,l_r\}$ is LI}
 	    {
		$I \gets IdentifyFactors(f,\MC, S_0)$\;
 		\If{$I\mid f$}
 		{
 		  $f \gets \frac{f}{I}$, $ K_0^{\star}=1,K_1^\star=1$, $X \gets$ OverestDetector$(f^{\star}, \MC, S_0)$\;
 		  \While {$ deg(K_{1-i}^\star) < deg(f)$}
 		    {
 		      \If{EasyCase$(f,  K_0^\star,  K_1^\star, \MC)\rightarrow iscorrect$}
 			{
 			  \Return object $decomposition(f,IK_0^\star,IK_1^\star)$\;
 			}
 			
 			\Else
 			{
 			  \For{each $d_1\in X$}
			    {
 			      \If{$\MB_2=\{l_1,\ldots,l_k,d_1,l_{k+2},\ldots,l_r\}$ is LI}
 			      {
  				$V_j=\F l_j, j\in [r]\setminus \{k+1\}$, $V_{k+1} = \F d_1$, $V_j^\prime =\bigoplus\limits_{t\in [r]\setminus \{j\}}V_t$\;
  				$S_0 = \{l_1,\ldots,l_k\}, S_1=\{u_{k+1}\}, S_2 = \{l_{k+2},\ldots,l_r\}$\;
  				$W_j = sp(S_j), W_{j}^\prime =\bigoplus\limits_{j_1\neq j} W_{j_1}$ for $j\in\{0,1,2\}$\;
  				$Q_0 = \frac{\pi_{V_1^\prime}(f)}{\pi_{V_1^\prime}(K_{1-i}^\star)}$, $Q_1 = \frac{\pi_{W_1^\prime}(f)}{\pi_{W_1^\prime}(K_{1-i}^\star)}$\;
 				\If{$Q_0,Q_1 \in \Pi\Sigma[\B{x}]$ and non-zero }
 				  {
 				      \For{ $q_0 \mid  Q_0$ \&\& $q_0\in W_2^\prime$, $q_1 \mid  Q_1$ \&\& $q_1\in W_2^\prime$}
 					{
 					  $Q_0 = \frac{ Q_0}{q_0}, Q_1 = \frac{ Q_1}{q_1}$\;
 					}
 					$Q_0 = \pi_{W_0^\prime}(Q_0)$\;
 					 \If{$deg(Reconstructor(Q_0,Q_1,S_0,S_1,S_2))\geq1$ }
 					  {
 					    $K_{1-i}^\star \gets K_{1-i}^\star \times Reconstructor(Q_0,Q_1,S_0,S_1,S_2)$\;
 					  }
 				  }
 			      }
 			
			    }
 			}
 		
 		    }
		      \If{ $f-IK_{1-i}^\star \in \Pi\Sigma[{\B{x}}]$}
 		      {
 		      $M_0=IK_{1-i}^\star$, $M_1 = f-M_0$, \Return {new object $decomposition(f,M_0,M_1)$\;}
 		      }
 		}
 	    }
      }
 }
\Return {$decomposition()$\;}

\caption{Hard Case Reconstruction}\label{hardcasealgo}
\end{algorithm}

%	 \item To take projections define the following spaces :
%
%	\item Consider the largest $Q\mid U_{1-i}$ such that for all $q\mid Q, q\notin W_2^\prime=W_0\oplus W_1$(defined above).
%	Next we try to compute $\pi_{W_0^\prime}(Q), \pi_{W_1^\prime}(Q)$.
%	  \item To compute $\pi_{W_0^\prime}(Q)$ first compute

%\item Define .
%\end{longenum}
%\end{longenum}
%\item Outside all the loops return new object $decomposition()$.
%\end{longenum}
%
%\EndProcedure
%\end{algorithmic}
%\end{algorithm}

\paragraph{\bf Correctness }

Let's assume we returned an object $obj$ of type decomposition.
\begin{enumerate}
 \item {\bf If $obj\rightarrow iscorrect ==true$ :} then we ought to be right since we check if
 $obj\rightarrow f=obj\rightarrow M_0 + obj\rightarrow M_1$. Since the representation is unique
this will be the correct answer.

 \item {\bf If $obj\rightarrow iscorrect ==false$:} Let's assume $f$ actually has a $\Sigma\Pi\Sigma(2)$
 representation. If we were in Easy Case or Medium Case we would have already found the circuit using their algorithms.
 So we are in the Hard Case. So by Lemma \ref{largedetector} there exists $i$ such that $\ML(T_i)$ has a detector pair $(S_0,D)$
 with $|D|$ large. For this $i$ there exists such an $S_0$, so sometime during the algorithm we would have guessed the correct $i$ and
 the correct $S_0$. {\bf Now let's analyze what happens inside the while and the third for loop when the first two guesses are correct.}
 Note that
 this also implies that the $I$ we have identified is correct and now we need to solve for
 \[
  f = G(\alpha_0 T_0 + \alpha_1 T_1)
 \]
  Let $K_0^\star,K_1^\star$ (initialized to $1$) be the known parts of the gates for this polynomial $f$ and
  $U_0^\star,U_1^\star$ be the unknown parts.
  Note that $T_0,T_1$ are same for both polynomials so $rank(f) = rank(f)$ and for $j=0,1;$ $K_j\mid G T_j$.\\

  {\bf Assume till the $m^{th}$ iteration of the while loop $K_{t}^\star \mid G T_{t}$ for $t\in \{0,1\}$, we show that after the $(m+1)^{th}$
  iteration, this property continues to hold and $deg(K_{1-i}^\star)$ increases.}

 \begin{itemize}
  \item  If after the $m^{th}$ iteration of the while loop for some $j\in \{0,1\}$, $\ML(T_j)\subsetneq sp(\ML(U_{1-j}^\star))$
  we are in Easy Case for $f$ . The first step in while loop is to call EasyCase$(f,\MC,K_0^\star,K_1^\star)$.
  This will clearly recover the circuit for $f$ and return true since $K_{t}^\star \mid G T_{t}$ for $t\in \{0,1\}$.
  However this does not happen so for both $j=0,1$, we have $\ML(T_i)\subsetneq \ML(U_{1-i})$. This means that we can use
  the ideas in Subsection \ref{getreconstructor}, specifically Theorem \ref{foundreconstructor}.
  \item The first two guesses are correct imply that $D\subseteq X\subseteq \ML(T_i)$.
  \item If $d$ gets rejected then $K_t, t\in\{0,1\}$ remain unchanged. If some $d_1$ does not get rejected then since
  $d_1\in \ML(T_i)$, $Q_0 = \pi_{V_1^\prime}(U_{1-i})$ is a non zero $\Pi\Sigma$ polynomial. Then some factors (the ones $\in W_2^\prime$) are
  removed from $Q_0$. Also on projecting to $W_0^\prime$ this
  still remains non-zero (as $d_1$ was not rejected).

  \item We know that $d_1\in \ML(T_i)$ and $d_1$ not getting rejected implies that $Q_1 =\pi_{W_1^\prime}(U_{1-i}) $ is a non-zero
  $\Pi\Sigma$ polynomial. We again remove some factors (i.e. the ones in $W_2^\prime$) from $Q_1$. The non-zeroness of
  $Q_0,Q_1$ imply that $Q_0= \pi_{W_1^\prime}(Q), Q_1= \pi_{W_1^\prime}(Q)$ i.e. they are projections of the same polynomial $Q$
  which is the largest factor of $U_{1-i}$ with the property that any linear form $q\mid Q$ is not in $W_2^\prime=W_0\oplus W_1$.
  \item $d_1$ was not rejected implies that $Reconstructor(Q_0,Q_1,S_0,S_1,S_2)$ returned a non-trivial polynomial $P$.
  This has to be a factor of $Q$ by Claim \ref{reconalgoconverse} following Algorithm \ref{reconalgo} and therefore a factor of $U_{1-i}$.
  \item Proof of Theorem \ref{foundreconstructor} implies that in every iteration atleast some $d_1$ will not be rejected.
  \item So clearly the new $K_{1-i}^\star = K_{1-i}^\star \times P$ divides $G T_{1-i}$. $K_i$ remains unchanged. Therefore
  even after the $(m+1)^{th}$ iteration $K_t\mid G T_t$ for both $j=0,1$ but degree of $K_{1-i}^\star$ increases.
 \end{itemize}
 So the while loop cannot run more than $deg(f)$ times and in the end $G T_{1-i}$ will be reconstructed completely
 and correctly and we should have returned $obj$ with $obj\rightarrow iscorrect=true$.
 Therefore we have a contradiction and so $f$ did not have a $\Sigma\Pi\Sigma(2)$ circuit and we correctly returned false.

\end{enumerate}

\paragraph{Running Time}
\begin{itemize}
 \item First for loop runs twice.
 \item Inside it choosing $r$ linear forms from $\MC$ ($|\MC| = poly(d)$) takes $poly(d)$ time.
 \item Applying $\Lambda$ to $r-k$ vectors takes $poly(r) = O(1)$ time.
 \item Checking if a set of size $r$ inside $\F^r$ is LI takes $poly(r)=O(1)$ time since it is equivalent to
 computing determinant.
 \item $IdentifyFactors()$ takes $poly(d)$ time and computing $f$ also takes $poly(d)$ time.
 \item $OverestDetector()$ runs in $poly(d)$ time.
 \item while loop runs at most $d$ times
 \item $EasyCase$ runs in $poly(d)$ time and so does polynomial multiplication.
 \item $X\subseteq \ML(T_i)$ and $|\ML(T_i)|\leq deg(f)$ and so for loop runs $d$ times.
 \item Change of bases in $\F^r$ and application to a polynomial of degree $d$ takes $poly(d)$ time.
 \item Therefore projecting to subspaces also takes $poly(d)$ time.
 \item $Reconstructor()$ runs in $poly(d)$ time (since $r$ is a constant) and so does polynomial
 multiplication and factoring by \cite{KalTr90}.
\end{itemize}
Since all of the above steps run in $poly(d)$ time, nesting them a constant number of
times also takes $poly(d)$ time. Therefore the running time of our algorithm is $poly(d)$.

\subsection{Algorithm including all cases : }

The algorithm we give here will be the final algorithm for rank $r$
$\Sigma\Pi\Sigma$ polynomials. It will use the previous three cases.
Our input will be a $\Sigma\Pi\Sigma(2)$ polynomial $f(x_1,\ldots,x_r)$ and
output will be a circuit computing the same.

\IncMargin{0.5em}
\begin{algorithm}[H]
\SetKwInOut{Name}{FunctionName}
\SetKwData{Left}{left}\SetKwData{This}{this}\SetKwData{Up}{up}
\SetKwFunction{Union}{Union}\SetKwFunction{FindCompress}{FindCompress}
\SetKwInOut{Input}{input}\SetKwInOut{Output}{output}
\Name{RECONSTRUCT}
\Input{$f\in \Sigma\Pi\Sigma_\F(2)[\B{x}]$}
\Output{An object of type $decomposition$}
\BlankLine
$decomposition$ obj\;
$(\Omega_{i,j}),(\Lambda_{i,j})$, $r\times r$ matrices with entries chosen uniformly randomly from $[N]$\;
$L_i(\B{x}) \gets \sum\limits_{k=1}^r \Omega_{i,k}x_k$\;
$f(x_1,\ldots,x_r) \gets f(L_1(\B{x}),\ldots,L_r(\B{x}))$\;
$\MC \gets Candidates(f(x_1,\ldots,x_r))$\;
\If { MediumCase$(f, \MC)) \rightarrow iscorrect$}
  {
    obj $\gets$ MediumCase$(f, \MC)$\;
  }
 \ElseIf{EasyCase$(f, K_0, K_1, \MC) \rightarrow iscorrect$ }
 {
    obj $\gets$ EasyCase$(f, K_0, K_1, \MC)$\;
 }
 \Else
 {
   obj $\gets$ HardCase$(f,\MC,\Lambda)$\;
 }
 Apply $\Omega^{-1}$ to obj$\rightarrow f,$ obj$\rightarrow M_0, $ obj$\rightarrow M_1$\;
\Return {obj\;}

\caption{Reconstruction in low rank}\label{overallalgo}
\end{algorithm}\DecMargin{0.5em}

\paragraph{Explanation : }

Here we explain every step of the given algorithm:
\begin{itemize}
 \item The function RECONSTRUCT$(f)$ takes as input a polynomial $f\in \Sigma\Pi\Sigma_\F (2)[\B{x}]$
 of $rank=r$ and outputs two polynomials $K_0,K_1\in \Pi\Sigma_\F[\B{x}]$ which are the two gates in it's
 circuit representation.
 \item Steps $2,3$ picks a random matrix $\Omega$ and transforms each variable
using the linear transformation this matrix defines. With high probability this
will be an invertible transformation. We do the reconstruction for this new
polynomial since the linear factors of it's gates satisfy some non-degenerate
conditions(because they have been randomly transformed) our algorithm needs. We
apply $\Omega^{-1}$ after the reconstruction and get back our original $f$.

 \item The next step constructs the set of candidate linear forms $\MC$. We've
talked about the size, construction and
 structure of this set in Section \ref{findcandidate}.
\item We first assume Medium Case. It that was not the case we check for Easy
Case . If both did not occur we can be sure we are in the Hard case.
\item We apply $\Omega^{-1}$ to polynomials in obj and return it.
\end{itemize}

\section{Reconstruction for arbitrary $rank$}\label{highdimrecon}
This section reduces the problem from $\Sigma\Pi\Sigma(2)$ Circuits with
arbitrary rank $n$ ($> s$) to one with
constant rank $r$ (still $> s$).  Also once the problem has been solved
efficiently in the low rank case
we use multiple instances of such solutions to lift to the general
$\Sigma\Pi\Sigma(2)$ circuit. The idea is to project the polynomial to a small (polynomial) number of
random subspaces of dimension $r$, reconstruct these low rank polynomials and then lift back to the original
polynomial. The uniqueness of
our circuit's representation plays a major role in both the projection and
lifting steps. Let
\[
 f = G(\alpha_0 T_0 + \alpha_1 T_1)
\]
$G,T_i$ are normal $\Pi\Sigma$ polynomials. All notations are borrowed from the previous section.
It is almost identical to the restriction done in \cite{Shpilka07} except that the dimension of random subspaces is different.
For more details see Section 4.2.1 and 4.2.3. in \cite{Shpilka07}. Since all proofs have been done in detail in
\cite{Shpilka07} we do not spend much time here. A clear sketch with some proofs is however given.

\subsection{Projection to a Random Low Dimensional Subspace}\label{projectrandom}
We explain the procedure of projecting to the random subspace below. In this low
dimensional setup we can get coefficient
representation of $\pi_V(f)$, also some important properties of $f$ are retained by
$\pi_V(f)$. Proofs are simple and standard so we discuss them in the appendix at
end.\\

Pick $n$ vectors $v_i, i\in [n]$ with each co-ordinate chosen independently from the uniform distribution on $[N]$.
Let $V = sp(\{v_i : i\in [r]\})$ and $V^\prime = sp\{v_i : i\in \{r+1,\ldots,n\}\}$. Then $V\oplus V^\prime = \F^n$ Let $\pi_V$
denote the orthogonal projection onto $V$.
With high probability the following hold :
\begin{enumerate}
\item $\{v_i : i\in [n]\}$ is linearly independent (See Appendix \ref{linindrandom} for proof).
\item Let $\{l_1,\ldots,l_r\}$ be a set of $r$ linearly independent linear forms in $\ML(T_0)\cup
\ML(T_1)$. Then $\pi_{V}(\{l_1,\ldots,l_r\})$ is linearly independent with high probability. So
 $rank(\pi_V(f))=r$ (See Appendix \ref{linindproj} for Proof).
\item Let $l_{01}\in \ML(T_0), l_{11}\in \ML(T_1)$, then
$\pi_V(l_{01}), \pi_V(l_{11})$ are linearly independent with high probability and so $gcd(\pi_V(T_0), \pi_V(T_1))=1$.
\end{enumerate}
Pick large number of ($\geq d^{r}$) random points $p_i, i=1,\ldots,d^{r}$
in the space $V$. Use the values $\{f(p_i)\}$ and get a coefficient representation for
$\pi_V(f)$. With high probability over
the choice of points interpolation will work (See Appendix
\ref{lagrangeinterp} for Proof). We will effectively be solving a linear system. Note that the number of coefficients in $f|_{V} = O(d^r)$.
Now this coefficient representation of $\pi_V(f)$ is reconstructed using the
algorithm in Chapter \ref{lowdimrecon}. A number of such reconstructions are then glued to reconstruct the original polynomial.
\subsection{Lifting from the Random Low Dimensional Subspace}\label{liftingback}
\begin{enumerate}
\item Consider spaces $V_i = V \oplus \F v_{i}$ for $i=r+1,\ldots,n$.
\item Reconstruct $\pi_{V_i}(f)$ and $\pi_V(f)$ for each $i\in \{r+1,\ldots,n\}$.
\item Let $l = \sum\limits_{i=1}^n a_iv_i $ be a linear form dividing one of the
gates of $f$ say $T_0$. $\pi_V(l) = \sum\limits_{i=1}^r a_iv_i $ and $\pi_{V_i}(l) =
\sum\limits_{j=1}^{r} a_jv_j + a_iv_i$. Using our algorithm discussed in Chapter
\ref{lowdimrecon} we would have reconstructed $\pi_V(f)$ and $\pi_{V_i}(f)$. So
we know the triples
    $(\pi_V(G), \pi_V(T_0),\pi_V(T_1))$ and
$(\pi_{V_i}(G), \pi_{V_i}(T_0),\pi_{V_i}(T_1))$

     On restricting $V_i$ to $V$ :

     a) {\bf Only Factors become factors} with high probability so we can easily
find the correspondence between $\pi_V(G)$ and $\pi_{V_i}(G)$.

     b)  $\pi_V(\pi_{V_i}(T_0)) = \pi_V(T_0)$ and $\neq \pi_V (T_1)$
because of uniqueness of representation and therefore we get the correspondence between gates.

     c) Now to get correspondence between linear forms. Let $\pi_V(l)$
have multiplicity $k$ in $\pi_V(T_0)$. Then with high probability
$l$ has multiplicity $k$ in $T_0$
    Since two LI vectors remain LI on projecting to a random subspace of
dimension $\geq 2$ (again See Appendix \ref{linindproj} for proof). Therefore
$\pi_{V_i}(l)$ has multiplicity $k$ and is the unique lift of
    $\pi_V(l)$ for all $i$. Let $\pi_{V_i}(l) = \pi_V(l) + a_iv_i$.
Then $l = \pi_V(l) + \sum_{i=r+1}^n a_iv_j$. This finds $G,T_0,T_1$ for us

\subsection{Time Complexity}
\begin{itemize}
 \item Interpolation to find coefficient representation $\pi_V(f)$ which is a degree $d$ polynomial over $r$ variables
 clearly takes $poly(d^r)$ time (accounts to solving a linear system of size $poly(d^r$)).
 \item Solving $n-r$ instances of the low rank problem (simple ranks $r$ and $r+1$) takes $npoly(d^r)$ time.
 \item The above mentioned approach to glue the linear forms in the gates clearly takes $poly(n,d)$ time.
 \item Overall the algorithm takes $poly(n,d)$ time since $r$ is a constant.
\end{itemize}

\end{enumerate}

\section{Conclusion and Future Work}
We described an efficient randomized algorithm to reconstruct circuit
representation of multivariate polynomials which exhibit a $\Sigma\Pi\Sigma(2)$
representation. Our algorithm works for all polynomials with rank(number of
independent variables greater than a constant $r$). In future we would like to
address the following:
\begin{itemize}
\item {\bf Reconstruction for Lower Ranks - } As can be seen in the paper, rank
of the polynomial for uniqueness (i.e. $c_{\F}(4)$) and the rank we've assumed
in the low rank reconstruction (i.e. $r$) are both $O(1)$ but $c_{\F}(4)$ is smaller than $r$. Since one would
expect a reconstruction
    algorithm whenever the circuit is unique we would like to close this gap.
    \item {\bf $\Sigma\Pi\Sigma(k)$ circuits - } The obvious next step would be to
consider more general top fan-in. In particular we could consider
$\Sigma\Pi\Sigma(k)$ circuits with $k=O(1)$.
        \item {\bf De-randomization - } We would like to de-randomize the
algorithm as it was done in the finite field case in \cite{KarShp09}.
\end{itemize}

\section{Acknowledgements}
I am extremely thankful to Neeraj Kayal for introducing me to this problem.
Sukhada Fadnavis, Neeraj Kayal and myself started working on the problem
together during my summer internship at Microsoft Research India Labs in 2011.
We solved the first important case together. A lot of intuition I developed was during my collaboration with them. A version of the construction
of our \emph{"Candidate Set"} was adopted from a write-up of Neeraj Kayal and Shubhangi Saraf which Neeraj shared with me. 
I'm grateful to them for all helpful discussions, constant guidance and encouragement.\\

I would like to thank Zeev Dvir for communicating the most recent rank bounds
on $\delta-SG_k$ configurations from \cite{DSW12} and his feedback on the work.
This reduces the gap in the first problem we mentioned above. \\

I would also like to thank Vinamra Agrawal, Pravesh Kothari
and Piyush Srivastava for helpful discussions. I would also like to thank the anonymous reviewers
for their suggestions.
Lastly I would like to thank Microsoft Research for giving me the
opportunity to intern at their Bangalore Labs with the Applied Mathematics
Group.

\newpage

\appendix

\section{Characterizing $\Pi\Sigma$ polynomials (Brill's Equations)}\label{brills}
In this section we will explicitly compute a set of polynomials whose common solutions characterize the coefficients
of all homogeneous $\Pi\Sigma_\C[x_1,\ldots,x_r]$ polynomials of degree $d$. A clean mathematical construction is given by Brill's Equations given in Chapter $4$,
\cite{GKZ94}.
However we still need to calculate the time complexity. But before that we define some operations on
polynomials and calculate the time taken by the operation along with the size of the output. Note that all polynomials are
over the field of complex numbers $\C$ and all computations are also done for the complex polynomial rings.\\

Let $\B{x} = (x_1,\ldots,x_r)$ and $\B{y} = (y_1,\ldots,y_r)$ be variables. For any homogeneous polynomial $f(\B{x})$
of degree $d$, define
\[
 f_{\B{x}^k}(\B{x},\B{y}) = \frac{(d-k)!}{d!}(\sum\limits_{i}x_i\frac{\partial}{\partial y_i})^k f(\B{y})
\]

Expanding $(\sum\limits_{i}x_i\frac{\partial}{\partial y_i})^k$ as a polynomial of differentials takes
$O((r+k)^r)$ time and has the same order of terms in it. $f(\B{y})$ has $O((r+k)^r)$ terms. Taking
partial derivatives of each term takes constant time and therefore overall computing
$(\sum\limits_{i}x_i\frac{\partial}{\partial y_i})^k f(\B{y})$ takes $O((r+k)^{2r})$ time. Also
the expression obtained will have at most $O((r+k)^{2r})$ terms. Computing the external
factor takes $poly(d)$ time and so for an arbitrary $f(\B{x})$ computing all $f_{\B{x}^k}(\B{x},\B{y})$
for $0\leq k\leq d$ takes $poly((r+d)^r)$ time and has $poly((r+d)^r)$ terms in it. From Section E., Chapter $4$ in \cite{GKZ94}
we also know that $f_{\B{x}^k}(\B{x},\B{y})$ is a bi-homogeneous form of degree $k$ in $\B{x}$ and degree $d-k$ in $\B{y}$. It
is called the $k^{th}$ polar of $f$.\\

Next we define an $\odot$ operation between homogeneous forms. Let $f(\B{x})$ and $g(\B{x})$ be homogeneous polynomials
of degrees $d$, define
\[
(f\odot g)(\B{x},\B{y}) = \frac{1}{d+1}\sum\limits_{k=0}^d(-1)^k{d\choose k}f_{\B{y}^k}(\B{y},\B{x})g_{\B{x}^k}(\B{x},\B{y})
\]

From the discussion above we know that computing $f_{\B{y}^k}(\B{y},\B{x})g_{\B{x}^k}(\B{x},\B{y})$ takes $poly((r+d)^r)$
time and it is obvious that this product has $poly((r+d)^r)$ terms. Rest of the operations take $poly(d)$ time and therefore
computing $(f\odot g)(\B{x},\B{y})$ takes $poly((r+d)^r)$ time and has $poly((r+d)^r)$ terms. From the discussion before we may also
easily conclude that the degrees of $\B{x},\B{y}$ in $(f\odot g)(\B{x},\B{y})$ are $poly(d)$. The form $(f\odot g)$ is called the
vertical(Young) product of $f$ and $g$. See Section G., Chapter $4$ in \cite{GKZ94}.\\

Next for $k\in \{0,\ldots,d\}$ and $\B{z} = (z_1,\ldots,z_r)$ consider homogeneous forms:
\[
 e_k = {d\choose k}f_{\B{x}^k}(\B{x},\B{z})f(\B{z})^{k-1}
\]
Following arguments from above, it's straightforward to see that computing $e_k$ takes $poly((r+d)^r)$ time and
has $poly((r+d)^r)$ terms.
Each $e_k$ is a homogeneous form in $\B{x},\B{z}$ and $f$. It has degree $k$ in $\B{x}$, degree $k(d-1)$ in $z$, and $k$ in
coefficients of $f$. See Section H. of Chapter $4$ in \cite{GKZ94}. Let's define the following function of $\B{x}$ with parameters $f,z$
\[
 P_{f,z}(\B{x}) = (-1)^dd\sum\limits_{i_1+2i_2+\ldots+ri_r=d}(-1)^{(i_1+\ldots+i_r)}\frac{(i_1+\ldots+i_r-1)!}{i_1!\ldots i_r!}e_1^{i_1}\ldots e_r^{i_r
  }
\]
Note that $\{(i_1,\ldots,i_r) :i_1+2i_2+\ldots+ri_r=d\}\subseteq \{(i_1,\ldots,i_r) : i_1+i_2+\ldots+i_r\leq d\}$ and therefore
the number of additions in the above summand is $O(poly(r+d)^r)$. For every fixed $(i_1,\ldots,i_r)$ computing the
coefficient $\frac{(i_1+\ldots+i_r-1)!}{i_1!\ldots i_r!}$ takes $O(poly((r+d)^r))$ time using multinomial coefficients.
Each $e_k$ takes $poly((r+d)^r)$ time to compute. There are $r$ of them in each summand and so overall we take
$O(poly((r+d)^r))$ time. A similar argument shows that number of terms in this polynomial is $O(poly((r+d)^r))$. Some more
analysis shows that $P_{f,z}(\B{x})$ is a form of degree $d$ in $\B{x}$ whose coefficients are homogeneous polynomials
of degree $d$ in $f$ and degree $d(d-1)$ in $\B{z}$. Let
\[
 B_f({\B{x},\B{y},\B{z}}) = (f\odot P_{f,z})(\B{x},\B{y})
\]

By the arguments given above calculating this form also takes time $poly((r+d)^r)$ and it has $poly((r+d)^r)$ terms.
This is a homogeneous form in $(\B{x},\B{y},\B{z})$ of multi-degree $(d,d,d(d-1))$ and it's coefficients are forms
of degree $(d+1)$ in the coefficients of $f$. See Section H., Chapter $4$ in \cite{GKZ94}.
So in time $poly((r+d)^r)$ we can compute $B_f({\B{x},\B{y},\B{z}})$ explicitly.\\

Now we arrive at the main theorem
\begin{theorem}[{\bf Brill's Equation}, See 4.H, \cite{GKZ94}]
 A form $f(\B{x}$) is a product of linear forms if and only if the polynomial $B_f(\B{x},\B{y},\B{z})$ is identically $0$.
\end{theorem}

We argued above that computing $B_f(\B{x},\B{y},\B{z})$ takes $O(poly((r+d)^r))$ time. It's degrees
in $\B{x},\B{y},\B{z}$ are all $poly(d)$ and so the number of coefficients when written as a polynomial over the $3r$ variables

$(x_1,\ldots,x_r,y_1,\ldots,y_r,z_,\ldots,z_r)$ is $poly((r+d)^r)$. We mentioned that each coefficient is a polynomial
of degree $(d+1)$ in the coefficients of $f$. Therefore we have the following corollary.

\begin{corollary}\label{variety}
 Let
\[
 I\eqdef \{(\alpha_1,\ldots,\alpha_n) : \forall i : \alpha_i\geq 0,
 \sum\limits_{i\in[r]}\alpha_i=d\}
 \]
be the set capturing the indices of all possible monomials of degree exactly $d$
in $r$ variables $(x_1,\ldots,x_r)$. Let
$f_{\bf a}(y_1,\ldots,y_r) = \sum_{\alpha\in I}a_{\alpha}{\bf y}^{\alpha}$
denote an arbitrary homogeneous polynomial. The coefficient vector then
becomes ${\bf a }= (a_\alpha)_{\alpha\in I}$.
Then there exists an explicit set of polynomials $F_1({\bf a}),\ldots,F_m({\bf
a})$ on $poly((r+d)^r)$ variables (${\bf a} = (a_\alpha)_{\alpha\in I}$), with $m=poly((r+d)^r)$,  $deg(F_i)\leq poly(d)$ such that for any
particular value of ${\bf a}$, the corresponding polynomial $f_{\bf a}({\bf
y})\in \Pi\Sigma_\F^{d}[\B{y}]$  if and only if $F_1({\bf a})=\ldots=F_m({\bf
a})=0$. Also this set $\{F_i, i\in [m]\}$ can be computed in time $poly((r+d)^r)$ time.
\end{corollary}
\emph{Proof.}
 Clear from the theorem and discussion above.

Note that in our application $r=O(1)$ and so $poly((d+r)^r) = poly(d)$.

\section{Tools from Incidence Geometry}\label{incidence}

Later in the paper we will use the quantitative version of Sylvester-Gallai Theorem from \cite{BDWY11}. In this subsection we do preparation
for the same. Our main application will also involve a corollary we prove towards the end of this subsection.

\begin{definition}[\cite{BDWY11}]\label{elementaryset}
Let $S$ be a set of $n$ distinct points in complex space $\C ^r$. A $k - flat$
is elementary if its intersection with $S$ has exactly $k+1$ points.
\end{definition}

\begin{definition}[\cite{BDWY11}]
Let $S$ be a set of $n$ distinct points in $\C^r$. $S$ is called a $\delta -
SG_k$ configuration if for every
independent $s_1,\ldots,s_k \in S$ there are at least $\delta n$ points $t\in S$
such that either $t\in fl(\{s_1,\ldots,s_k\})$ or the $k-$flat
$fl(\{s_1,\ldots,s_k,t\})$ contains a point in $S\setminus
\{s_1,\ldots,s_k,t\}$.
\end{definition}

\begin{theorem}[\cite{BDWY11}]
Let $S$ be a $\delta-SG_k$ configuration then
$dim(S) \leq \frac{2^{C^k}}{\delta^2}$. Where $C>1$ is a universal constant.
\end{theorem}

This bound on the dimension of $S$ was further improved by Dvir et. al. in \cite{DSW12}. The latest version
now states
\begin{theorem}[\cite{DSW12}]\label{bdwy}
 Let $S$ be a $\delta-SG_k$ configuration then
$dim(S) \leq C_k = \frac{C^k}{\delta}$. Where $C>1$ is a universal constant.
\end{theorem}

\begin{corollary} \label{elementary}
 Let $dim(S)> C_k$ then $S$ is not a $\delta-SG_k$
configuration i.e. there exists a set of independent points $\{s_1,\ldots,s_k\}$
 and $\geq (1-\delta)n$ points $t$ such that $fl(\{s_1,\ldots,s_k,t\})$ is an
elementary $k - flat$. That is:
 \begin{itemize}
  \item $t\notin fl(\{s_1,\ldots,s_k\})$
  \item $fl(\{s_1,\ldots,s_k,t\}) \cap S = \{s_1,\ldots,s_k,t\}$.
 \end{itemize}

\end{corollary}
Right now we set $\delta$ to be a constant $ < 0.5, C_k = \frac{C^k}{\delta}$. Note
that $C_i<C_{i+1}$.
Using the above theorem we prove the following lemma which will be useful to us
later
\begin{lemma}[Bi-chromatic semi-ordinary line]\label{bichromatic}
 Let $X$ and $Y$ be disjoint finite sets in $\C^r$ satisfying the following
conditions.
\begin{enumerate}
\item $dim(Y)>C_4$.
\item $|Y|\leq  c|X|$ with $c < \frac{1-\delta}{\delta}$.
\end{enumerate}
Then there exists a line $l$ such that $|l\cap Y|=1$ and $|l\cap X|\geq 1$
\end{lemma}

\emph{Proof.}
We consider two cases:

\underline{Case 1 : $c|X|\geq |Y|\geq |X|$}

Since $dim(Y)>C_1$, using the corollary above for $S=X\cup Y, k=1$ we can get a
point $s_1 \in X\cup Y$ for which there exist $(1-\delta)(|X|+|Y|)$ points $t$
in $X\cup Y$
such that $t\notin fl\{s_1\}$ and $fl\{s_1,t\}$ is elementary.
If $s_1\in X$ then $(1-\delta)(|X|+|Y|)-|X| \geq (1-2\delta)|X|>0$ of these
flats intersect $Y$ and thus we get such a line $l$. If $s_1\in Y$ then
$(1-\delta)(|X|+|Y|)-|Y| \geq ((1-\delta)(\frac{1}{c}+1) -1)|Y| >0$ of these
flats intersect $X$ giving us the required line $l$ with $|l\cap X|=1$ and
$|l\cap Y|=1$.\\

\underline{Case 2: $|Y|\leq |X|$}

 Now choose a subset $X_1\subseteq X$ such that $|X_1|=|Y|$. Now using the same
argument
as above for $S = X_1\cup Y$ there is a point $s_1\in X_1\cup Y$ such that
$(1-\delta) (|X_1|+|Y|)= 2(1-\delta) |Y| = 2(1-\delta) |X_1|$ flats through
it are elementary in $X_1\cup Y$. If $s_1\in Y$ $(1-2\delta)|Y|>0$ of these
flats intersect $X_1$. If $s_1\in X_1$,
$(1-2\delta)|X_1| >0$ of these flats intersect $Y$. In both these above
possibilities the flat intersects $Y$ and $X_1$ in
exactly one point each. But it may contain more points from $X\setminus X_1$ so
we can find a line $l$ such that $|l\cap Y|=1$
and $|l\cap X|\geq 1$.

\section{A Method of Reconstructing Linear Forms}\label{Identifier}

In a lot of circumstances one might reconstruct a linear form (up to scalar multiplication) inside $V=Lin_\F[\B{x}]$ from it's projections
(up to scalar multiplication) onto some subspaces of $V$. For example consider a linear form $L=a_1x_1+a_2x_2+a_3x_3 (\in Lin_
\F[x_1,x_2,x_3])$ with $a_3\neq 0$,
and assume we know scalar multiples of projections of $L$ onto the spaces $\F x_1$ and $\F x_2$ i.e. we know $L_1=\alpha(a_2x_2+a_3x_3)$ and
$L_2=\beta(a_1x_1+a_3x_3)$ for some $\alpha,\beta\in \F$. Scale these projections to $\tilde L_1 = x_3+\frac{a_2}{a_3}x_3$ and
$\tilde L_2 = x_3 + \frac{a_1}{a_3}x_3$. Using these two define a linear form $x_3+\frac{a_1}{a_3}x_1 + \frac{a_2}{a_3}x_2$. This is
a scalar multiple of our original linear form $L$. We generalize this a little more below.\\

Let $\B{x}\equiv (x_1,\ldots,x_r)$, $\MB = \{l_1,\ldots,l_r\}$ be a basis for $V=Lin_{\F}[x_1,\ldots,x_r]$. For $i\in
\{0,1,2\}$, let $S_i$ be pairwise disjoint non empty subsets of $\MB$
such that $S_0\cup S_1\cup S_2 = \MB$. Let $W_i=sp(S_i)$ and $W_i^\prime =
\bigoplus\limits_{j\neq i}W_j$. Clearly $V=W_0\oplus W_1\oplus W_2 = W_i\oplus
W_i^\prime, i\in \{0,1,2\} $.

\begin{lemma}\label{reconlin}
Assume $L\in V$ is a linear form such that
\begin{itemize}
\item $\pi_{W_2}(L) \neq 0$
\item  For $i\in \{0,1\}, L_i=\beta_i \pi_{W_i^\prime}(L)$ are known for some
non-zero scalars $\beta_i$.
\end{itemize}
Then $L$ is unique up to scalar multiplication and we can construct a scalar
multiple $\tilde{L}$ of $L$.
\end{lemma}
\emph{Proof.}
Let $L=a_1l_1 + \ldots + a_rl_r, a_i\in \F$. Since $\pi_{W_2}(L)\neq 0$, there
exists $l_j\in S_2$ such that $a_j\neq 0$. Let $\tilde{L} = \frac{1}{a_{j}}L$.
For $i\in \{0,1\}$, re-scale
$L_i $ to get $\tilde{L_i}$  making sure that coefficient of $l_j$ is $1$ in
them. Thus for $i=0,1$
 \[
 \pi_{W_i^\prime}(\tilde{L}) = \tilde{L_i}
 \]
 Since $W_0^\prime = W_1\oplus W_2$ and $W_1^\prime = W_0\oplus W_2$  by
comparing coefficients we can get $\tilde{L}$.

\underline{\bf (Algorithm)} \label{linformrecon} Assume we know $S_0,S_1,S_2$ and therefore the
basis change matrix to convert vector representations from $\MS$ to $\MB$. It takes $poly(r)$ time to
convert $[v]_{\MS}$ to $[v]_{\MB}$. Given $L_i$ in the basis $\MB$ it takes $poly(r)$
time(by a linear scan) to find $l_j\in S_2$ with $a_j\neq 0$. This $l_j$ has a
non zero coefficient in both $L_0,L_1$. After this we
just rescale $L_i$ to get $\tilde{L_i}$ such that coefficient of $l_j$ is $1$. Then since $\tilde{L_i} =
\pi_{W_i^\prime}(\tilde{L})$ the coefficient of $l_t$ in $\tilde{L}$ is as
follows :

\begin{displaymath}
   = \left\{
     \begin{array}{lr}
       \text{ coefficient of }l_t\text{ in } \tilde{L_1} & : l_t\in S_0\\
       \text{ coefficient of }l_t\text{ in } \tilde{L_0} & : l_t\in S_1\\
       \text{ coefficient of }l_t\text{ in } \tilde{L_0} = \text{ coefficient of
}l_t\text{ in } \tilde{L_1} & : l_t\in S_2
     \end{array}
   \right.
\end{displaymath}
Finding the right coefficients using this also takes $poly(r)$ time.\\

Next we try and use this to reconstruct $\Pi\Sigma$ polynomials. This case is slightly more complicated and so
we demand that the projections have some special form. In particular the projections onto one subspace preserves pairwise linear
independence of linear factors and onto the other makes all linear factors scalar multiples of each other.
\begin{corollary}\label{polyrecon}
Let $S_i,W_i, i\in \{0,1,2\}$ be as above and $P\in \Pi\Sigma_\F[x_1,\ldots,x_r]$
such that
\begin{enumerate}
\item $\pi_{W_2}(P)\neq 0$
\item For $i\in \{0,1\}$ there exists $\beta_i (\neq 0) \in \F$ such that
$P_0=\beta_0 \pi_{W_0^\prime}(P) = p^{t}$ and $P_1=\beta_1
\pi_{W_1^\prime}(P)=d_1\ldots d_t$. are known i.e.
$p,d_j$ $(j\in [t])$ and $t$ are known.
\end{enumerate}
Then $P$ is unique upto scalar multiplication and we can construct a scalar
multiple $\tilde{P}$ of $P$.
\end{corollary}

\emph{Proof.}
Let $P = L_1\ldots L_t$ with $L_i\in Lin_\F[\B{x}]$. There exists $\beta^j_i,
i\in \{0,1\}, j\in [t]$, such that $\beta^j_0 \pi_{W_0^\prime}(L_j)= p$ and
$\beta^j_1 \pi_{W_1^\prime}(L_j) = d_j$.
Since $p,d_j$ are known by above Lemma \ref{reconlin} we find a scalar multiple
$\tilde{L_j}=\beta^j L_j$ of $L_j$
and therefore find a scalar multiple $\tilde{P} = \tilde{L_1}\ldots\tilde{L_t}$
of $P$. Note that this method also tells us that such a $P$ is unique up to scalar
multiplication. Since we've used the above Algorithm \ref{linformrecon} at most $t$
times with $t\leq deg(P)$, it takes $poly(deg(P),r)$ time to find $\tilde P$.

This corollary is the backbone for reconstructing $\Pi\Sigma$ polynomials from their projections. But first we formally
define a \emph{"Reconstructor"}

\begin{definition}[Reconstructor] \label{Reconstructor} Let $S_i,W_i, i\in
\{0,1,2\}$ be as above. Let $Q$ be a standard $\Pi\Sigma$ polynomial and $P$ be
a standard $\Pi\Sigma$ polynomial
dividing $Q$ with $Q=PR$. Then $(Q,P,S_0,S_1,S_2)$ is called a
\emph{Reconstructor} if:

\begin{itemize}
\item $\pi_{W_2}(P)\neq 0$.
\item $\pi_{W_0^\prime}(P) = \alpha p^t$, for some linear form $p$.
\item Let $l\mid R$ be a linear form and $\pi_{W_2}(l)\neq 0$ then $gcd
(\pi_{W_2}(P),\pi_{W_2}(l)) =1$.

\end{itemize}
\end{definition}

{\bf Note :}

 Let $L_1,L_2$  be two LI linear forms dividing $P$ , then one can show
 \[
 L_1,L_2 \text{ are LI } \Leftrightarrow \pi_{W_1^\prime}(L_1),
\pi_{W_1^\prime}(L_2) \text{ are LI }
 \]
 To see this first observe that the second bullet implies for $i\in [2], L_i\in W_0 + p \Rightarrow
sp(\{L_1,L_2\})\subseteq W_0+p$.

 If $\pi_{W_1^\prime}(L_1), \pi_{W_1^\prime}(L_2)$ are LD then
 \[
 sp(\{L_1,L_2\}) \cap W_1 \neq \{0\}
  \]
 $\Rightarrow (W_0+p) \cap W_1 \neq \{0\}$. Since $W_0\cap W_1 = \{0\}$ we get
that $p\in W_0\oplus W_1 = W_2^\prime
 \Rightarrow \pi_{W_2}(p)=0\Rightarrow \pi_{W_2}(P)=0$ contradicting the first
bullet.\\

 Geometrically the conditions just mean that all linear forms dividing $P$ have
LD projection ($=\gamma p$ for some non zero $\gamma \in \F$) w.r.t.
the subspace $W_0^\prime$ and LI linear forms $p_1,p_2$ dividing $P$  have LI
projections (w.r.t. subspace $W_1^\prime)$. Also no linear form $l$ dividing $R$
belongs to $fl(S_0 \cup S_1 \cup \{p\})$.\\

 We are now ready to give an algorithm to reconstruct $P$ using $\pi_{W_0^\prime}(Q)$ and
 $\pi_{W_1^\prime}(Q)$ by gluing appropriate projections corresponding to $P$.
To be precise:\\

\begin{claim} \label{reconalgoclaim}
 Let $Q,P$ be standard $\Pi\Sigma$ polynomials and $P\mid Q$. Assume $(Q, P,S_0,S_1,S_2)$ is
a \emph{Reconstructor}. If we know both $\pi_{W_0^\prime}(Q)$ and
$\pi_{W_1^\prime}(Q)$. Then we can reconstruct $P$.
\end{claim}

\emph{Proof.} Here is the algorithm:

\IncMargin{1em}
\begin{algorithm}[H]
\SetKwData{Left}{left}\SetKwData{This}{this}\SetKwData{Up}{up}
\SetKwFunction{Union}{Union}\SetKwFunction{FindCompress}{FindCompress}
\SetKwInOut{Input}{input}\SetKwInOut{Output}{output}
%\Function{IdentifyFactors}
\Input{$\pi_{W_0^\prime}(Q)\in \Pi\Sigma[\B{x}], \pi_{W_1^\prime}(Q) \in \Pi\Sigma[\B{x}],S_0,S_1,S_2$}
\Output{ a $\Pi\Sigma$ polynomial $P\mid Q$ }
\BlankLine

$bool flag$, $\Pi\Sigma$ polynomial $P_0,P_1;$\;
Factor $\pi_{W_0^\prime}(Q) = \gamma \prod\limits_{i\in [s]} c_i^{m_i}$, $c_i$'s pairwise LI and normal, $\gamma\in \F$\;
Factor $\pi_{W_1^\prime}(Q) = \delta d_1\ldots d_m$, $\delta\in \F$ and $d_j$ normal\;
\For { $i \in [s]$ \&\& $\pi_{W_1^\prime}(c_i) \neq 0$}{
  $flag=true, P_0=c_i^{m_i}$\;
  \tcp{Assuming projection w.r.t. $W_0^\prime$ to be $c_i^{m_i}$.}
  \For {$j\in [s]$ \&\& $j\neq i$ \&\& $\pi_{W_1^\prime}(c_j)\neq 0$}{
      \If { $gcd(\pi_{W_1^\prime}(c_i), \pi_{W_1^\prime}(c_j))\neq 1$} {
	  $flag = false$\;
	  }
      }
      \If { $flag==true$}{
	  $P_1=1$\;
	  }
      \For{$j \in [m]$}{
	  \If{ $\pi_{W_0^\prime}(d_j)\neq 0$ \& \& $\{\pi_{W_0^\prime}(d_j), \pi_{W_1^\prime}(c_i)\}$ are LD}{
	      $P_1 = P_1d_j$ \;
	      \tcp{This steps collects projection w.r.t. $W_1^\prime$ in $P_1$.}
	    }
	  }
      \If {$(deg(P_1)=m_i)$ \&\& $(P_0,P_1)$ give $\tilde P=\beta P$ using Corollary \ref{polyrecon} }{
	  Make $\tilde P$ \emph{standard} w.r.t. the standard basis $\MS$ to get $P$\;
	    Return {$P$\;}
	  }
	
}
Return {$1$\;}

\caption{Reconstructing Linear Factors}\label{reconalgo}
\end{algorithm}\DecMargin{1em}

\subsection{Explanation}
\begin{itemize}
 \item The algorithm takes as input projections $\pi_{W_0^\prime}(Q)$ and
$\pi_{W_1^\prime}(Q)$ along with the sets $S_i,i=0,1,2$ which form a partition
of a basis $\MB$. We know that there exists a polynomial $P\mid Q$ such that
$(Q,P,S_0,S_1,S_2)$ is a reconstructor and so we try to compute the projections
$\pi_{W_0^\prime}(P),\pi_{W_1^\prime}(P)$.
\item If one assumes that $\pi_{W_0^\prime}(Q) = \gamma \prod\limits_{i\in [s]} c_i^{m_i}$ with the $c_i$'s co-prime, then
by the properties of a reconstructor the projection (of a scalar multiple of $P$) onto $W_0^\prime$  say
$P_0 = \beta_0\pi_{W_0^\prime}(P)$ (for some $\beta_0$) has to be equal to $c_i^{m_i}$ for some $i$. We do this
assignment inside the first for loop.

\item The third property of a reconstructor implies that when we project further to $W_2$, it should not get any more
factors and so we check this inside the second for loop
by going over all other factors $c_j$ of $\pi_{W_0^\prime}(Q)$ and checking if $c_i,c_j$ become LD on
projecting to $W_2$ (i.e. by further projecting to $W_1^\prime$).

\item Now to find (scalar multiple of) the other projections i.e. $P_1 = \beta_1\pi_{W_1^\prime}(P)$ (for some $\beta_1$), we go through
$\pi_{W_1^\prime}(Q)$ and find $d_k$ such that
$\{\pi_{W_1^\prime}(c_i) , \pi_{W_0^\prime}(d_k)\}$ are LD (i.e. they are projections of the same linear form). We collect the product of all such
$d_k$'s. If the choice of $c_i$ were correct then all $d_k$'s would be obtained correctly.
\item The last \emph{"if"} statement just checks that the number of $d_k$'s found above is the same as $m_i$ since $P_0=c_i^{m_i}$ tells us that
the degree of $P$ was $m_i$. We recover a scalar multiple of $P$ using the algorithm explained in Corollary \ref{polyrecon}
and then make it standard to get $P$.

\end{itemize}

\subsection{Correctness}
The correctness of our algorithm is shown by the lemma below.

\begin{claim}\label{returnreconalgo}
 If $(Q,P,S_0,S_1,S_2)$ is a reconstructor for non-constant $P$, then Algorithm \ref{reconalgo} returns $P$.
\end{claim}
\emph{Proof.}
$(Q,P,S_0,S_1,S_2)$ is a reconstructor therefore
\begin{itemize}
 \item $\pi_{W_2}(P)\neq 0$
 \item $\pi_{W_0^\prime}(P) = \delta p^t$
 \item $q\mid \frac{Q}{P} \Rightarrow gcd(\pi_{W_2}(q), \pi_{W_2}(P))=1$
\end{itemize}
\begin{enumerate}
\item It is clear that for one and only one value of $i$, $c_i$ divides $p$. Fix this $i$. Let $Q=PR$, if $c_i^{m_i}\nmid \pi_{W_0^\prime}(P)$
 then $c_i \mid l$ for some linear form $l\mid \pi_{W_0^\prime}(R)$. Condition $3$ in definition of Reconstructor implies that
 $gcd(\pi_{W_2}(P),\pi_{W_2}(l))=1$ but $\pi_{W_2}(c_i)$ divides both of them giving us a contradiction. Since $\pi_{W_0^\prime}(P)$
 has just one linear factor $\Rightarrow \pi_{W_0^\prime}(P)$ is
 a scalar multiple of $c_i^{m_i}$ for some $i$.

 \item Assume the correct $c_i^{m_i}$ has been found. Now let $d_j\mid \pi_{W_1^\prime}(Q)$ such that
 $\{\pi_{W_2}(c_i), \pi_{W_2}(d_j)\}$ are LD. then we can show that $d_j\mid \pi_{W_1^\prime}(P)$.
 Assume not, then for some linear form $l\mid R = \frac{Q}{P}$, $d_j\mid \pi_{W_1^\prime}(l)$.
 $\pi_{W_0^\prime}(d_j)\neq 0$ (which we checked) $\Rightarrow \pi_{W_2}(l)\neq 0$. So we get
 $\pi_{W_2}(c_i)\mid \pi_{W_2}(l) (\neq 0)$
 and so $\pi_{W_2}(c_i) \mid gcd(\pi_{W_2}(P),\pi_{W_2}(l))$ which is therefore $\neq 1$ and condition $3$ of Definition \ref{Reconstructor}
 is violated. So whatever $d_j$ we collect will be a factor of $\pi_{W_1^\prime}(P)$ and we will collect all of them since they are all
 present in $\pi_{W_1^\prime}(Q)$.
 \item We know from proof of Corollary \ref{polyrecon} that if we know $c_i,m_i$ and $d_j$'s correctly then we can recover a
 scalar multiple of $P$ correctly. But $Q,P$ are standard so we return $P$ correctly.
\end{enumerate}

In fact we can show that if we return something it has to be a factor of $Q$.

\begin{claim}\label{reconalgoconverse}
 If Algorithm \ref{reconalgo} returns a $\Pi\Sigma$ polynomial $P$, then $P\mid Q$
\end{claim}

\begin{itemize}
\item If the algorithm returns $1$ from the last return statement, we are done. So let's assume it returns something from
the previous return statement.
 \item So $flag$ has to be true at end $\Rightarrow$ there is an $i\in [s]$
such that $P_0 = c_i^{m_i}$ with the conditions that $\pi_{W_1^\prime}(c_i)\neq 0$ and $gcd(c_i,c_j)=1$ for $j\neq i$. It also means that for exactly
$m_i$ of the $d_j$'s (say $d_1,\ldots,d_{m_i}$) $\{\pi_{W_1^\prime}(c_i), \pi_{W_0^\prime}(d_j)\}$ are LD and $P_1 = d_1\ldots d_{m_i}$.
\item Since $c_i^{m_i} \mid \pi_{W_0^\prime}(Q)$, there exists a factor $\tilde P\mid Q$ of degree $m_i$
such that $\pi_{W_0^\prime}(\tilde P) = c_i^{m_i}$ and $\pi_{W_1^\prime}(c_i)\neq 0$. This $\Rightarrow \pi_{W_2}(\tilde P)\neq 0$.
Clearly $\pi_{W_1^\prime}(\tilde P) \mid \pi_{W_1^\prime}(Q) = d_1\ldots d_m$, hence for all linear factors $\tilde p$
of $\tilde P$, $\pi_{W_1^\prime}(\tilde p)$ should be some $d_j$ with the condition that
$\{\pi_{W_0^\prime}((\pi_{W_1}^\prime)(\tilde p)), \pi_{W_1^\prime}(c_i)\}$ should be LD. The only choice we have are $d_1,\ldots,d_{m_i}$.
So $\pi_{W_0^\prime}(\tilde P) = d_1\ldots d_{m_i}$. All conditions of Corollary \ref{polyrecon} are true and so $\tilde P$
is uniquely defined (up to scalar multiplication) by the reconstruction method given in Corollary \ref{polyrecon}.
So what we returned was actually a factor of $Q$.
\end{itemize}

\subsection{Time Complexity}

Factoring $\pi_{W_0^\prime}(Q),\pi_{W_1^\prime}(Q)$ takes $poly(d)$ time (using Kaltofen's Factoring from \cite{KalTr90}).
The nested for loops run $\leq d^3$ times.
Computing projections with respect to the known decomposition $W_0\oplus W_1\oplus W_2 =\F^r$ of linear forms
over $r$ variables takes $poly(r)$ time. Computing $gcd$ and linear independence of linear forms takes $poly(r)$ time.
The final reconstruction of $P$ using $(P_0,P_1)$ takes $poly(d,r)$ time as has been explained in Corollary \ref{polyrecon}.
Multiplying linear forms to $\Pi\Sigma$ polynomial takes $poly(d^r)$ time. Therefore overall the algorithm takes
$poly(d^r)$ time. In our application $r=O(1)$ and therefore the algorithm takes $poly(d)$ time.

\section{Random Linear Transformations}\label{randomtransform}
This section will prove some results about linear independence and non-degeneracy under random transformations on $\F^r$. This will be required to make our input 
non-degenerate. From here onwards we fix a natural number $N\in \N$
and assume $0<k<r$. Let $T\subset \F^r$ be a finite set with $dim(T)= r$. Next
we consider two $r\times r$ matrices $\Omega,\Lambda$. Entries $\Omega_{i,j},\Lambda_{i,j}$
are picked independently from the uniform distribution on $[N]$. For any basis $\MB$ of $\F^r$ and vector $v\in \F^r$, let $[v]_\MB$
denote the co-ordinate vector of $v$ in the basis $\MB$. If $\MB = \{b_1,\ldots,b_r\}$ then $[v]^i_\MB$ denotes the $i$-th co-ordinate in
$[v]_\MB$. Let $\MS = \{e_1,\ldots,e_r\}$ be the standard
basis of $\F^r$. Let $E_j = sp(\{e_1,\ldots,e_j\})$ and $E_j^\prime = sp(\{e_{j+1},\ldots,e_r\})$, then
$\F^r = E_j \oplus E_j^\prime$. Let $\pi_{W_{E_j}}$ be the orthogonal projection onto $E_j$.
For any matrix $M$, we denote the matrix of it's co-factors by $co(M)$.
We consider the following events :

\begin{itemize}
\item $\ME_0 = \{\Omega \text{ is not invertible }\}$
\item $\ME_1 = \{\exists t(\neq 0)\in T$  : $\pi_{W_{E_1}}(\Omega(t))= 0 \}$
\item $\ME_2 = \{ \exists \{t_1,\ldots,t_r\} \text{ LI vectors in }  T : \{\Omega(t_1),\ldots,\Omega(t_r)\} \text{ is LD }\}$
\item $\ME_3 = \{\exists \{t_1,\ldots,t_r\} \text{ LI vectors in }  T : \{\Omega(t_1),\ldots,\Omega(t_k),\Lambda\Omega(t_{k+1}), \ldots , \Lambda\Omega(t_r) \} 
\text{ is LD } \}$
\item \label{newmatrix} When $t_i,\Lambda,\Omega$ are clear we define the matrix $M =[M_1 \ldots M_r]$
 with columns $M_i$ given as :
\[
   M_i= \left\{
     \begin{array}{lr}
        [\Omega(t_i)]_{\MS} : i\leq k\\

        [\Lambda\Omega(t_i)]_{\MS} : i>k
     \end{array}
   \right.
\]
$M$ corresponds to the linear map
\[
e_i\mapsto \Omega(t_i) \text{ for } i\leq k \text{ and } e_i\mapsto \Lambda\Omega(t_i) \text{ for } i>k
\]

$\ME_4 = \{ \{\exists \{t_1,\ldots,t_r\} \text{ LI vectors in }  T $ and $ t\in T\setminus sp(\{t_1,\ldots,t_k\}) :
[co(M)[\Omega(t)]_{\MS}]^{k+1}_\MS=0 \}$
\item $\ME_5 = \ME_4 \mid \ME_3^c$
\end{itemize}

Next we show that the probability of all of the above events is small. Before doing that let's explain the events. This will give an intuition to why the events 
have low probabilities.
\begin{itemize}
\item $\ME_0$ is the event where $\Omega$ is not-invertible. Random Transformations should be invertible.
\item $\ME_1$ is the event where there is a non-zero $t\in T$
such that the projection to the first  co-ordinate (w.r.t. $\MS$) of $\Omega$ applied on $t$ is $0$. We don't expect this for
a random linear transformation. Random Transformation on a non-zero vector should give a non-zero coefficient of $e_1$.
\item $\ME_2$ is the event such that $\Omega$ takes a basis to a LD set i.e. $\Omega$ is not invertible (random linear operators are invertible).
\item $\ME_3$ is the event such that for some basis applying $\Omega$ to the first $k$ vectors and $\Lambda\Omega$ to the last $n-k$ vectors gives a LD set. So 
this operation is not-invertible. For
ranrom maps this should not be the case.
\item  $\ME_4$ is the event that there is some basis $\{t_1,\ldots,t_r\}$ and
$t$ outside $sp(t_1,\ldots,t_k)$ such that the $(k+1)^{th}$ co-ordinate
of $co(M)[\Omega(t)]_\MS$ w.r.t the standard basis is $0$. If $M$ were invertible, clearly the set
$ \MB = \{\Omega(t_1),\ldots,\Omega(t_k),\Lambda\Omega(t_{k+1}), \ldots , \Lambda\Omega(t_r) \}$
would be a basis and $co(M)$ will be a scalar multiple
of $M^{-1}$. So we are asking if the $(k+1)^{th}$ co-ordinate of $\Omega(t)$
in the basis $\MB$ is $0$. For random $\Omega,\Lambda$ we would
expect $M$ to be invertible and this co-ordinate to be non-zero.
\end{itemize}

Now let's formally prove everything. We will repeatedly use the popular Schawrtz-Zippel Lemma which the reader can find in \cite{Sax09}.

\begin{claim}
 $Pr[\ME_1] \leq \frac{|T|}{N^r}$
\end{claim}

\emph{Proof.}
 Fix a non-zero $t = \left(\begin{array}{c} a_1\\.\\.\\a_r\end{array}\right)$ with $a_i\in \F$ and let $\Omega = (\Omega_{i,j}), 1\leq i,j\leq
r$. Then the first co-ordinate of $\Omega(t)$ is $\Omega_{1,1}a_1 + \Omega_{1,2}a_2+\ldots + \Omega_{1,r}a_r$. Since $t\neq 0$, not all $a_i$ are
$0$ and this is therefore not an identically zero
polynomial in $(\Omega_{1,1},\ldots,\Omega_{1,r})$. Therefore by Schwartz-Zippel lemma $ Pr[[\Omega(t)]^1_{\MS}= 0] \leq \frac{1}{N^r} $. Using a union bound inside $T$ we get $ Pr[ \exists t (\neq
0)\in T :
 [\Omega(t)]^1_{\MS} =0] \leq
  \frac{|T|}{N^r}$.

\begin{claim}
 $Pr[\ME_2]\leq \frac{r}{N^{r^2}}$
\end{claim}
\emph{Proof.}
  Clearly $\ME_2 \subseteq \ME_0$ and so
 $Pr[\ME_2]\leq Pr[\ME_0]$. $\ME_0$ corresponds to the polynomial equation $det(\Omega)=0$. $det(\Omega)$ is
 a degree $r$ polynomial in $r^2$ variables and is also not identically zero, so using Schwartz-Zippel lemma we get
 $Pr[\ME_2]\leq Pr[ \ME_0] \leq \frac{r}{N^{r^2}}$.

\begin{claim}
 $Pr[\ME_3]\leq {|T|\choose r}\frac{2r}{N^{2r^2}}$
\end{claim}
\emph{Proof.}
 Fix an LI set $t_1,\ldots,t_r$. The set $\{\Omega(t_1),\ldots,\Omega(t_k), \Lambda\Omega(t_{k+1}),\ldots \Lambda\Omega(t_r)\}$
 is LD iff the $r\times r$ matrix $M$ formed by writing these vectors (in basis $\MS$) as columns (described in part \ref{newmatrix} above)
 has determinant $0$.
 $M$ has entries polynomial (of degree $\leq 2$) in $\Omega_{i,j}$ and $\Lambda_{i,j}$ and so $det(M)$ is a polynomial in
 $\Omega_{i,j},\Lambda_{i,j}$ of degree $\leq 2r$.
 For $\Omega=\Lambda = I$ (identity matrix) this matrix just becomes the matrix formed by the basis $\{t_1,\ldots,t_r\}$ which has
 non-zero determinant and so $det(M)$ is not the identically zero polynomial.  By Schwartz-Zippel lemma $Pr[det(M)=0]\leq \frac{2r}{N^{r^2}N^{r^2}} =
 \frac{2r}{N^{2r^2}}$. Now we vary the LI set $\{t_1,\ldots,t_r\}$, there are $\leq {|T|\choose r}$ such sets and so by a union bound
 $Pr[\ME_3]\leq {|T|\choose r} \frac{2r}{N^{2r^2}}$.

\begin{claim}
 $Pr[\ME_4]\leq {|T|\choose r+1}\frac{2r-1}{N^{2r^2}}$
\end{claim}
\emph{Proof.}
 Fix an LI set $t_1,\ldots,t_r$ and a vector $t\notin sp(\{t_1,\ldots,t_k\})$. Let $t = \sum\limits_{i=1}^r a_it_i$. Since
 $t\notin sp(\{t_1\ldots,t_k\})$, $a_s\neq 0$ for some $s\in \{k+1,\ldots,r\}$.
 Let $\MB = \{\Omega(t_1),\ldots,\Omega(t_k), \Lambda\Omega(t_{k+1}),\ldots \Lambda\Omega(t_r)\}$.
 Let $M$ be the matrix whose columns are from $\MB$
 (Construction has been explained in part \ref{newmatrix} above). We know that the co-factors of a matrix are
 polynomials of degree $\leq r-1$ in the matrix elements.
 In our matrix $M$ all entries are polynomials of degree $\leq 2$ in $\Omega_{i,j},\Lambda_{i,j}$, so all entries of $co(M)$ are polynomials of degree
 $\leq 2r-2$ in $\Omega_{i,j}, \Lambda_{i,j}$. Thus $[co(M)[\Omega(t)]_\MS]^{k+1}_{\MS} =
 \sum\limits_{i=1}^r co(M)_{k+1,i}[\Omega(t)]^i_\MS$ is a polynomial of degree $\leq 2r-1$. This polynomial is not identically zero. Define
 $\Omega$ to be the matrix (w.r.t. basis $\MS$) of the linear map $\Omega(t_i) = e_i$ and $\Lambda$ to be the matrix (w.r.t. basis $\MS$) of the map
\begin{displaymath}
  \Lambda = \left\{
     \begin{array}{lr}
       \Lambda(e_i)=e_i : i\notin\{s,k+1\}\\
 \Lambda(e_s) = e_{k+1}\\
 \Lambda(e_{k+1})=e_s\\
     \end{array}
   \right.
\end{displaymath}
With these values the set $\MB$ becomes $\{e_1,\ldots,e_k,e_s,e_{k+2},\ldots,e_{s-1},e_{k+1},e_{s+1},\ldots,e_r\}$. If one now
looks at $M$ i.e. the matrix formed using entries of $\MB$ as columns it's just the permutation matrix that flips $e_s$ and $e_{k+1}$.
This matrix is the inverse of itself and so has determinant $=\pm1$, thus $co(M) = \pm M^{-1} = \pm M$. Therefore $co(M)[\Omega(t)]_\MS = \pm M \left
(\begin{array}{c} a_1\\.\\.
\\a_r\end{array}\right) = \pm \left(\begin{array}{c} a_1\\.\\a_k\\a_s\\a_{k+2}\\.\\a_{s-1}\\a_{k+1}\\.a_{s+1}\\.\\a_r\end{array}\right)$. Since $a_s \neq 0$, 
we get $[co(M)[\Omega(t)]_\MS]^{k+1}_\MS \neq 0$. So the polynomial is not identically
zero and we can use Schwartz-Zippel Lemma to say that $Pr[[co(M)[\Omega(t)]_\MS]^{k+1}_\MS = 0] \leq \frac{2r-1}{N^{r^2}N^{r^2}} = \frac{2r-1}
{N^{2r^2}}$. Now we vary $\{t_1,\ldots,t_r,t\}$ inside $T$ and use union bound to show $Pr[\ME_4]\leq {|T|\choose r+1}\frac{2r-1}{N^{2r^2}}$.

Even though this is just basic probability we include the following:
\begin{claim}
 $Pr[\ME_5] \leq {|T|\choose r}\frac{2r-1}{N^{2r^2}-{|T|\choose r}2r}$
\end{claim}
\emph{Proof.}
 $Pr[\ME_5] = Pr[\ME_4\mid \ME_3^c] = \frac{Pr[\ME_4\cap \ME_3^c]}{Pr[\ME_3^c]} \leq \frac{Pr[\ME_4]}{Pr[\ME_3^c]}
 \leq {|T|\choose r+1}\frac{\frac{2r-1}{N^{2r^2}}}{1-{|T|\choose r}\frac{2r}{N^{2r^2}}}  = {|T|\choose r+1}\frac{2r-1}
 {N^{2r^2}-{|T|\choose r}2r}$

In our application of the above $r = O(1), |T| = poly(d), N = 2^{d}$ and so all probabilities are very small as $d$ grows. So we will
assume that none of the above events occur. By union bound that too will have small probability and so with very
high probability $\ME_0,\ME_1,\ME_2,\ME_3,\ME_4,\ME_5$ do not occur.

\section{Set $\MC$ of Candidate Linear Forms}\label{findcandidate}
This section deals with constructing a $poly(d)$ size set $\MC$ which
contains each $l_{ij}, (i,j)\in \{0,1\}\times [M]$. First we define the
set and prove a bound on it's size.
\subsection{Structure and Size of $\MC$}
Let's recall $f=G(\alpha_0T_0 + \alpha_1 T_1)$ and define two other
polynomials:

\[
g=\frac{f}{G} = \alpha_0T_0+\alpha_1T_1
\]
\[
h =\frac{f}{Lin(f)} = \frac{g}{Lin(g)}
\]
Assume $deg(h) = d_h$
\begin{definition}\label{candidatedef}
Our candidate set is defined as:
\[
\MC \eqdef \{l = x_1-a_2x_2-\ldots- a_rx_r \in Lin_{\F}[\B{x}] :
h(a_2x_2+\ldots + a_rx_r,x_2,\ldots,x_r) \in \Pi\Sigma^{d_h}_\F[x_2,\ldots,x_r]
\}
\]
(for definition of $\Pi\Sigma^{d_h}_\F[x_2,\ldots,x_r]$ See Section \ref{notation} )

\end{definition}
In the claim below we show that linear forms dividing polynomials $T_i, i=0,1$
are actually inside $\MC$ (first part of claim). The remaining linear forms in
$\MC$ (which we call
\emph{``spurious''}) have a nice structure (second part of claim). In the third
part of our claim we arrive at a bound on the size of $\MC$. Recall the definition of $c_\F(k)$
from Theorem \ref{rankbound}.

\begin{claim}\label{candidate}
The following are true about our candidate set $\MC$.
\begin{enumerate}
\item $\ML(T_i)\subseteq \MC, i=0,1$.
\item \label{candidateprop} Let $k=c_{\F}(3)+2$ and suppose $\{ l_{j} ; j\in
[k]\} \subset \ML(T_i)$ are LI . Then for any
$l\in \MC\setminus (\ML(T_0)\cup \ML(T_1))$, there exists $j\in [k]$
such that $fl(\{l,l_{j}\})\cap \ML(T_{1-i}) \neq \phi$ i.e. the line joining $l$
and $l_{j}$ does not intersect the set $\ML(T_{1-i})$.
\item $|\MC|\leq M^4+2M\leq d^4 + 2d.$
\end{enumerate}
\end{claim}

\emph{Proof.}
Let's first recall the definition of our candidate set
\[
\MC \eqdef \{l = x_1-a_2x_2-\ldots- a_rx_r \in Lin_{\F}[\B{x}] :
h(a_2x_2+\ldots + a_rx_r,x_2,\ldots,x_r) \in \Pi\Sigma^{d_h}_\F[x_2,\ldots,x_r]
\}
\]
Also recall that
\[
 h = \frac{g}{Lin(g)} = \frac{f}{Lin(f)}
\]

\begin{enumerate}
\item Let $l = x_1-a_2x_2-\ldots- a_rx_r \in \ML(T_{1-i})$. Let's denote the tuple
$v\equiv (a_2x_2+\ldots + a_rx_r,x_2,\ldots,x_r)$.
Since $gcd(T_0,T_1)=1$ and $l\mid T_{1-i}$ we know that $l\nmid T_i$ and therefore $Lin(g)(v)\neq 0$. We can then compute
\[
h(v) = \frac{\alpha_{i}T_{i}(v)}
{Lin(g)(v)}
= \alpha_{i}H_1(v)\ldots H_{d_h}(v)\in \Pi\Sigma^{d_h}_{\F}[x_2,\ldots,x_r]
\]
where $H_j \in Lin_{\F}[x_2,\ldots,x_r]$. So $\ML(T_i)\subseteq \MC$ for $i=0,1$.

\item Consider $l=x_1-a_2x_2-\ldots- a_rx_r \in \MC\setminus (\ML(T_0)\cup \ML(T_1))$ and assume that
$sp(\{l,l_j\})\cap \ML(T_{1-i})=\phi$ for all $j\in[k]$. We know that
 \[
 g(v) = Lin(g)(v) H_1(v)\ldots H_{d_h}(v) = \alpha_0T_0(v) + \alpha_1T_1(v)
\]

Let $g^\prime $ be the following identically zero $\Sigma\Pi\Sigma(3)[x_2,\ldots,x_r]$
polynomial (with circuit $\MC^\prime$)

 \[
 g^\prime = Lin(g)(v) H_1(v)\ldots H_{d_h}(v) -\alpha_0T_0(v) -
\alpha_1T_1(v)
\]

We know
\[
 \MC^\prime = gcd(\MC^\prime) Sim(\MC^\prime) \Rightarrow Sim(\MC^\prime)\equiv
0
\]
Recall that $l_j(v)\mid T_{i}(v)$, therefore
the $l_j(v)$ cannot be factors of $gcd(\MC^\prime)$
because if they did then there exist pair $l_{j}, l_{(1-i)t}$ such that
$\{l_{j}(v), l_{(1-i)t}(v)\}$ is LD or in other words $sp(\{l,l_j\})\cap \ML(T_{1-i})\neq \phi$ and we have a contradiction.
Also the set $\{l_j(v) : j\in [k]\}$ has dimension $\geq k-1$ since
the dimension could fall only
by $1$ when we go modulo a linear form (project to hyperplane). This means that $rank(Sim(\MC^\prime))\geq k-1\geq
c_{\F}(3)+1$.

{\bf If $Sim(\MC^\prime)$ were not minimal} $\Rightarrow \MC^\prime$ is not minimal $\Rightarrow$ one of it's gates would be $0$.
Since $l\notin \ML(T_0)\cup\ML(T_1) \Rightarrow \alpha_0T_0(v)+\alpha_1T_1(v) \equiv 0
\Rightarrow$ for every $j\in [k]$ there exist $l_{(1-i)j}\mid T_{1-i}$ such that $l_{(1-i)j}(v),l_j(v)$ are LD.
$\Rightarrow sp(\{l,l_j\})\cap \ML(T_{1-i}) \neq \phi$ for $j\in [k]$,
a contradiction to our assumption.

{\bf If $Sim(\MC^\prime)$ were minimal}, we have an identically zero simple
minimal circuit $Sim(\MC^\prime)$ with $rank(Sim(\MC^\prime))\geq c_{\F}(3)+1 $
contradicting Theorem \ref{rankbound}.

So our assumption is wrong and $sp(\{l,l_j\})\cap \ML(T_{1-i}) \neq \phi$ for
some $j\in [k]$.
\item  Let $l\in \MC\setminus (\ML(T_0)\cup \ML(T_1))$. Consider a set $\{l_1,\ldots,l_{k+2}\}\subset \ML(T_i)$
of $k+2$ LI linear forms. By the above argument there exist three distinct elements in this set say $l_1,l_2,l_3$ such that
$sp(\{l_j,l\})\cap \ML(T_{1-i})\neq \phi$ for $j\in [3]$. Let $\{l_1^\prime,l_2^\prime,l_3^\prime\} \subset \ML(T_{1-i})$ such that
$l_j^\prime \in sp(\{l_j,l\})$ for $j\in [3]$. Then $gcd(l_j,l_j^\prime)=1$ implies that $l\in sp(\{l_j,l_j^\prime\})$ for $j\in [3]$.
 Since $l,l_j,l_j^\prime$ are all standard (coefficient of $x_1$ is $1$), Lemma \ref{spantoflat} tells us
 \[
  l\in fl(\{l_j,l_j^\prime\})
 \]
 for $j\in [3]$. So $l$ lies on the lines $\vec{L_j} = fl(\{l_j,l_j^\prime\})$ for $j\in [3]$. At least two of
 these lines should be distinct otherwise $dim(\{l_1,l_2,l_3\})\leq 2$ which is a contradiction. So $l$ is the
 intersection of these two lines.
There are $M^2$ such lines and so $M^4$ such intersections. If $l\in \ML(T_0)\cup \ML(T_1)$ we have $\leq 2M$ other possibilities. So
$|\MC|\leq M^4+2M = O(d^4)$.

\end{enumerate}

Let's now give an algorithm to construct this set.
\subsection{Constructing the set $\MC$}

Here is an algorithm to construct the set $\MC$. An explanation is given in the lemma below.

\begin{algorithm}[H]\label{candidatealgo}

\SetKwData{Left}{left}\SetKwData{This}{this}\SetKwData{Up}{up}
\SetKwInOut{Name}{FunctionName}\SetKwFunction{FindCompress}{FindCompress}
\SetKwInOut{Input}{input}\SetKwInOut{Output}{output}
\Name{Candidates}
\Input {$f\in \Sigma\Pi\Sigma_\F(2)[\B{x}]$}
\Output {Set $\MC$ of Linear Forms}
Define $\MC=\phi;$\;
Use polynomial factorization from \cite{KalTr90} to find $Lin(f)$\;
 Consider polynomial $h=\frac{f}{Lin(f)}$\;
 Let $a_2,\ldots,a_r$ be variables.\;
 Compute coefficient vector {\bf b} of $h(a_2x_2+\ldots+a_rx_r, x_2,\ldots,x_r)$.\;
 Consider the polynomials $\{F_i, i\in [m]\}$ constructed in Corollary \ref{variety}.\;
 Using your favorite algorithm (e.g. Buchberger's \cite{Buchberger76}) to solve polynomial equations, find all complex solutions to the system
$\{F_i({\bf b})=0, i\in [m]\}$.\;
 For each solution $(a_2,\ldots,a_r) \in \F^r$ do :  $\MC = \MC \cup \{(1,a_2,\ldots,a_r)\}$\;

\Return{$\MC$\;}

\caption{Set $\MC$ of candidate linear forms}
\end{algorithm}

\begin{lemma}
Given a polynomial $f \in \F[x_1,\ldots,x_r]$ of degree $d$ in
$r$ independent variables
which admits a $\Sigma\Pi\Sigma_\F(2)[x_1,\ldots,x_r]$-representation : $f =
\prod\limits_{i\in [d-M]}G_i(\alpha_0\prod\limits_{j\in[M]}l_{0j} +
\alpha_1\prod\limits_{k\in[M]}l_{1k} )$ such that $G_t,l_{ij} (t\in [d-M], i\in \{0,1\}, j\in [M])$ are standard w.r.t. the
standard basis $\{x_1,\ldots,x_n\}$ then we can find in deterministic time $poly(d)$,
the corresponding candidate set $\MC$ (see Definition \ref{candidatedef}) described above.
\end{lemma}

\emph{Proof.}
The proof also contains an explanation of the algorithm above
\begin{itemize}
\item Let $l = x_1-a_2x_2-\ldots -a_rx_r \in \MC$ be a candidate linear form. We know
that $h(a_2x_2+\ldots + a_rx_r,x_2,\ldots,x_r)\in
\Pi\Sigma^{d_h}_\F[x_2,\ldots,x_r]\subset \Pi\Sigma^{d_h}_\C[x_1,\ldots,x_r]$.
\item Using Theorem \ref{variety} we know that
$h(a_2x_2+\ldots + a_rx_r,x_2,\ldots,x_r) \in
\Pi\Sigma^{d_h}_\C[x_2,\ldots,x_r]\Leftrightarrow$ for the coefficient vector ${\bf b}$ of $h(a_2x_2+\ldots + a_rx_r,x_2,\ldots,x_r)$ inside
$\C[x_2,\ldots,x_r]$ satisifes $F_1({\bf b})=\ldots=F_m({\bf b})=0$ for the polynomials $\{F_i : i\in [m]\}$ obtained in Corollary \ref{variety}.
.
\item For any $t\leq d_h$, computing $(a_2x_2+\ldots + a_rx_r)^t$ requires $poly(t^r)$ time and it also has
$poly(t^r)$  terms and degree $t$. Multiplying such powers to other variables  and
adding $poly(d_h^r)$ many such expressions also requires $poly(d_h^r)$ time.
Hence computing the coefficient vector {\bf b} takes polynomial time since $r$ is a constant. Each co-ordinate of
this coefficient vector is a polynomial in $r-1$ variables $(a_2,\ldots,a_r)$ of degree $poly(d_h^r)$.

\item Now we think of the $a_i$'s as our unknowns and obtain them by solving the
polynomial system $\{F_i({\bf b}) =0, i\in [m]\}$. The number of polynomials
is $m=poly(d^r)$ and degrees are $poly(d)$. $F_i$'s are polynomials in $poly(d^r)$ variables. Expanding $F_i({\bf b})$
will clearly take $poly(d^r)$ time and now we will have $poly(d^r)$ polynomials in $r$ variables of degrees $poly(d^r)$. Note that
$r=O(1)$ and so we need to solve $poly(d)$ polynomials of degree $poly(d)$ in constant many variables. Also Claim \ref{candidate} implies that the number
 of solutions $\leq M^4+2M = O(poly(d))$. So using Buchberger's algorithm \cite{Buchberger76} we can solve the system
 for $(a_2,\ldots,a_r)$ in $poly(d)$ time. Once we have the solutions we consider only those linear
forms which are in $\F[x_1,\ldots,x_r]$ and add them to $\MC$.
\end{itemize}

\section{Proofs from Subsection \ref{reducefactors}}

\begin{claim} \label{spuriousliproof}
Let $(S = \{l_{1}\ldots, l_{k}\},D)$ be a Detector pair in $\ML(T_i)$. Let
$l_{k+1}\in D$. For a $standard$ linear form $l\in V$, if $l\mid g$ then
$l\notin sp(\{l_{1},\ldots,l_{k}\})$ .
\end{claim}
\emph{Proof.}
Assume $l\mid g$ and $l\in sp(\{l_{1},\ldots,l_{k}\})$. Let $W = sp(\{l\})$,
extend it to a basis and in the process obtain $W^\prime$ such that $W\oplus
W^\prime = V$. We get
\[
\pi_{W^\prime}(\alpha_0T_0 + \alpha_1T_1) =0
\]
$\pi_{W^\prime}(\alpha_iT_i)\neq 0$ (i.e. $l\nmid T_0T_1$), otherwise $l$ divides both $T_0,T_1$ and
$gcd(T_0,T_1)$ won't be $1$. So we have an equality of non zero $\Pi\Sigma$
polynomials
\[
 \alpha_0\prod\limits_{j=1}^M\pi_{W^\prime}(l_{0j}) =
-\alpha_1\prod\limits_{j=1}^M\pi_{W^\prime}(l_{1j})
\]

Therefore there
exists a permutation $\theta : [M] \rightarrow [M]$  such that
$\{\pi_{W^\prime}(l_{(1-i)j}), \pi_{W^\prime}(l_{i\theta(j)})\}$ are LD
$\Rightarrow l\in sp(\{l_{(1-i)j}, l_{i\theta(j)} \})$. Since $l\nmid T_0T_1$
this also means that $l_{(1-i)j}\in sp(\{l,l_{i\theta(j)}\})$ and $l_{i\theta(j)}\in sp(\{l,l_{(1-i)j}\})$.\\

In particular there is
an $l_{k+1}^\prime\in \ML(T_{1-i})$ such that $l_{k+1}^\prime\in
sp(\{l,l_{k+1}\})$ and $l_{k+1}\in sp(\{l,l_{k+1}^\prime\})$.\\

  Since $l\in sp(\{l_{1},\ldots,l_{k}\})\Rightarrow l_{k+1}^\prime\in
sp(\{l_{1},\ldots,l_{k},l_{k+1}\})$. All linear forms here are standard(i.e. coefficient of $x_1$ is $1$)
  and so by Lemma \ref{spantoflat}, $l_{k+1}^\prime \in fl(\{l_1,\ldots,l_k,l_{k+1}\})$. Below we use
  the definition of detector pair and get
  \[
  l_{k+1}^\prime\in fl(\{l_{1},\ldots,l_{k},l_{k+1}\})\cap \ML(T_{1-i})\subseteq
fl(\{l_{1},\ldots,l_{k}\})
  \]
  And $l_{k+1}\in sp(\{l,l_{k+1}^\prime\})\Rightarrow l_{k+1}\in
sp(\{l_1,\ldots,l_k\})$ which is a contradiction to $(S,D)$ being a detector pair..

\begin{claim}\label{spuriousextraproof}
Let $l \in Lin_\F[\B{x}]$ be $standard$ such that $l \mid g$ and
$\MC$ be the candidate set. Assume $(S = \{l_{1},\ldots,l_{k}\}, D(\neq \phi))$
is a Detector pair in $\ML(T_i)$. Then $|\ML(T_{1-i}) \cap (fl(S\cup\{l\}) \setminus
fl(S))|\geq 2$. That is the flat $fl(\{l_{1},\ldots,l_{k},l\})$ contains
at least two distinct points from $\ML(T_{1-i})(\subseteq \MC)$ outside $fl(\{l_1,\ldots,l_k\})$.
\end{claim}

\emph{Proof.}
From the previous claim we know that $\{l_{1},\ldots,l_{k},l\}$ is an LI set.
Also like above we know there exists $l_{j}^\prime \in \ML(T_{1-i}), j\in [3]$
such that $l_{j} \in sp(\{l, l_{j}^\prime\}), l_{j}^\prime \in
sp(\{l, l_{j}\})$. Since $\{l_{1},l_{2},l_{3}\}$ are LI,
at least two of the $l_{j}^\prime$'s, $j\in [3]$ must be distinct, otherwise
$sp(\{l_{1},l_{2},l_{3}\})\subset sp(\{l,l_{1}^\prime\})$ which is not possible
as LHS has dimension $3$ and RHS has dimension $2$. Thus there exist two
distinct $l_{1}^\prime, l_{2}^\prime \in sp(\{l_{1},l_{2},l_{3},l\})\subset
sp(\{l_{1},\ldots,l_{k},l\})$. Note that $l_1,\ldots,l_k,l,l_1^\prime,l_2^\prime$ are all standard (i.e. coefficient of $x_1$ is $1$)
and so by Lemma \ref{spantoflat}
\[
 l_j^\prime \in fl(\{l_1,\ldots,l_k,l\})
\]
for $j\in [2]$.\\

If for any $j\in [2]$, $l_j^\prime \in
sp(\{l_1,\ldots,l_k\})$ then $l\in sp(\{l_j,l_j^\prime\}) \Rightarrow l\in sp(\{l_1,\ldots,l_k\})$
which is a contradiction. This also shows that $l_j^\prime \notin fl(\{l_1,\ldots,l_k\})$ for $j\in [2]$.\\

 From what we showed above we may conclude:
\[
 l_j^\prime \in fl(\{l_1,\ldots,l_k,l\})\setminus fl(\{l_1,\ldots,l_k\})
\]
for $j\in [2]$. Hence Proved.

\begin{lemma}\label{prooffilteredfactor}
The following are true:
\begin{enumerate}
\item If $l\mid I$ (i.e. $l$ was identified) then $l\in \ML(G)\setminus
\ML(g) $.
\item \label{retainedfactor}If $l\mid G$ (i.e. $l$ was retained) then
$(fl(\{l_{1},\ldots,l_{k},l\})\setminus fl(\{l_{1},\ldots,l_{k}\})) \cap
(\ML(T_{1-i})\cup (\ML(T_i)\setminus D)) \neq \phi $ that is

$(fl(\{l_{1},\ldots,l_{k},l\})\setminus fl(\{l_{1},\ldots,l_{k}\}))$ contains
a point from $\ML(T_i)\setminus D$ or $\ML(T_{1-i})$.

\item \label{retaineddetector} If $l\mid G$ and $l_{k+1}\in D$ then
$l \notin sp(\{l_{1},\ldots,l_{k},l_{k+1}\})$.
    \end{enumerate}
\end{lemma}

\emph{Proof.}
\begin{enumerate}

\item Assume $l\mid I$ (i.e. $l$ was identified) and $l\mid g$. Then by Claim
\ref{spuriousli} we know that $\{l_{1},\ldots,l_{k},l\}$ are LI and so the
first $"if"$ condition is true. By Claim \ref{spuriousextra} we know that there are two
other points $\{l_1^\prime,l_2^\prime\} \subset \MC \cap (fl(\{l_{1},\ldots,l_{k},l\})\setminus fl(\{l_{1},\ldots,l_{k}\}))$, so the second
$"if"$ condition will also be true and thus $l$ will not be identified which is a
contradiction. Therefore $l\in \ML(G)\setminus\ML(g)$.
\item Assume $l\mid G$ (i.e. $l$ was not identified). This means both $"if"$ statements were
true for $l$. Thus $\{l_{1},\ldots,l_{k},l\}$
is LI. Also there exist distinct $\{l_1^\prime, l_2^\prime\} \in \MC \cap (fl(\{l_{1},\ldots,l_{k},l\})\setminus
fl(\{l_{1},\ldots,l_{k}\}))$. If
\[
l_1^\prime \in (\ML(T_{1-i})\cup (\ML(T_i)\setminus D)) \text{ or }
l_2^\prime  \in (\ML(T_{1-i})\cup (\ML(T_i)\setminus D))
 \]
 we are done so assume both are in
 \[
 \MC\setminus((\ML(T_{1-i})\cup (\ML(T_i)\setminus D)))) = (\MC\setminus
(\ML(T_i)\cup \ML(T_{1-i})))\cup D
 \]
 If one of them say $l_1^\prime \in \MC\setminus(\ML(T_i)\cup \ML(T_{1-i}))$, then by
Part \ref{candidateprop} of Claim \ref{candidate}, for some $j\in [k]$,  $
sp(\{l_1^\prime,l_{j}\})\cap \ML(T_{1-i}) \neq \phi$. Let $\tilde l_{j}\in
sp(l_1^\prime,l_{j})\cap \ML(T_{1-i}) \Rightarrow$
 \[
\tilde l_{j}\in sp(\{l_1^\prime ,l_{j}\})\subseteq sp(\{l_{1},\ldots,l_{k},l\})
 \]
 Since all linear forms $\tilde l_j,l_1,\ldots,l_k,l$ are standard (coefficient of $x_1$ is $1$) by Lemma \ref{spantoflat}
 \[
  \tilde l_j \in fl(\{l_{1},\ldots,l_{k},l\})
 \]

 Also $\tilde l_{j},l_{j}$ are LI and $\tilde l_{j}\in sp(\{l_1^\prime ,l_{j}\})$ together imply $l_1^\prime\in sp(\{l_{j},\tilde l_{j}\})$.
 Note that $l_1^\prime \notin fl(\{l_{1},\ldots,l_{k}\})\Rightarrow l_1^\prime\notin sp(\{l_{1},\ldots,l_{k}\})$ which along with
 $l_1^\prime\in sp(\{l_{j},\tilde l_{j}\})$ will then give
  \[
  \tilde l_{j}\notin sp(\{l_{1},\ldots,l_{k}\})
  \]

  So we found $\tilde l_{j}\in \ML(T_{1-i})\cap
(fl(\{l_{1},\ldots,l_{k},l\})\setminus fl(\{l_{1},\ldots,l_{k}\})) $ and we
are done.\\

So the only case that remains
now is that $l_1^\prime,l_2^\prime \in D$.  Let's complete the proof in the following steps
\begin{itemize}
 \item $l_1^\prime\in fl(\{l_{1},\ldots,l_{k},l\})\setminus fl(\{l_{1},\ldots,l_{k}\})
\Rightarrow l\in sp(\{l_{1},\ldots,l_{k},l_1^\prime\})$
\item Using the above bullet, $l_2^\prime \in fl(\{l_1,\ldots,l_k,l\})\Rightarrow l_2^\prime \in sp(\{l_{1},\ldots,l_{k},l_1^\prime\})$. Linear forms
$l_2^\prime, l_1,\ldots,l_k,l$ are standard (coefficient of $x_1$ is $1$) so using Lemma \ref{spantoflat}, $l_2^\prime \in fl(\{l_{1},\ldots,l_{k},l_1^\prime\})$
\item $l_2^\prime\in D \Rightarrow l_2^\prime\notin fl(\{l_{1},\ldots,l_{k}\})$
\item The above two bullets and $\{l_1^\prime,l_2^\prime\}\subset \ML(T_i)$ tell us that $fl(\{l_1,\ldots,l_k,l_1^\prime\})$
is not elementary which is a contradiction.
\end{itemize}
So atleast one of $l_1^\prime,l_2^\prime$ is inside $\ML(T_{1-i})\cup (\ML(T_i)\setminus D)$

\item Let $l_{k+1}\in D$ and $l\in sp(\{l_{1},\ldots,l_{k},l_{k+1}\})$. Since $l,l_1,\ldots,l_k,l_{k+1}$ are standard, by Lemma
\ref{spantoflat},
$l\in fl(\{l_{1},\ldots,l_{k},l_{k+1}\})$. Clearly $l\notin fl(\{l_{1},\ldots,l_{k}\})$ otherwise it would get identified at
the first $"if"$. Therefore $l\in fl(\{l_1,\ldots,l_k,l_{k+1}\})\setminus fl(\{l_1,\ldots,l_k\})$
By Part $2$ above let $l_1^\prime
\in (fl(\{l_{1},\ldots,l_{k},l\})\setminus fl(\{l_{1}\ldots,l_{k}\})) \cap (\ML(T_{1-i})\cup
(\ML(T_i)\setminus D))$. So $l_1^\prime \in \ML(T_{1-i})$ or $l_1^\prime \in \ML(T_i)\setminus D$.\\

This tells us that $l_1^\prime \in sp(\{l_{1},\ldots,l_{k},l_{k+1}\})\setminus fl(\{l_{1},\ldots,l_{k}\})$.
All linear forms $l_1^\prime,l_1,\ldots,l_k,l_{k+1}$ are standard (i.e. coefficients of $x_1$ is $1$) so by Lemma \ref{spantoflat}
we get that $l_1^\prime \in fl(\{l_{1},\ldots,l_{k},l_{k+1}\})\setminus fl(\{l_{1},\ldots,l_{k}\})$. Now using
the definition of detector pair $l_1^\prime \notin \ML(T_{1-i})$ since $fl(\{l_{1},\ldots,l_{k},l_{k+1}\})\cap \ML(T_{1-i})
\subseteq fl(\{l_{1},\ldots,l_{k}\})$ . The flat $fl(\{l_1,\ldots,l_k,l_{k+1}\})$ is elementary
in $\ML(T_i)$, so $l_1^\prime$ can belong here only if $l_1^\prime=l_{k+1}$ which is not possible since $l_1^\prime \notin D$.
So we have a contradiction. Hence Proved.

\end{enumerate}

\begin{lemma} \label{detectorexpansionproof}
Let $(S=\{l_{1},\ldots,l_{k}\},D)$ be a detector in $\ML(T_i)$. For each
$(l,l_j) \in \MC \times S$ define the space $U_{\{l,l_j\}} = sp(\{l,l_j\})$.
Extend $\{l,l_j\}$ to a basis and in the process obtain $U_{\{l,l_j\}}^\prime$
such that $V = U_{\{l,l_j\}}\oplus U_{\{l,l_j\}}^\prime$. Define the set:
\[
 X = \{l\in \MC : \pi_{U^\prime_{\{l,l_j\}}}(f) \neq 0, \text{  for all }
l_j\in S\}
\]
Then $D\subset X\subset \ML(T_i)$.
\end{lemma}

\emph{Proof.}
\underline{($D\subset X$) : } Consider $l_{k+1} \in D$. Since $D\subset
\ML(T_i)
\Rightarrow l_{k+1}\in \MC$. Assume $l_{k+1}\notin X$, so there exists a $j\in
[k]$ such that
$\pi_{U^\prime_{\{l_{k+1},l_j\}}}(f)=0$. That is:
\[
\pi_{U_{\{l_{k+1},l_j\}}^\prime}(G(\alpha_0T_0+\alpha_1T_1))=0.
\]So
\[
\prod\limits_{t\in
[N_1]}\pi_{U_{\{l_{k+1},l_j\}}^\prime}(G_t)(\alpha_0\prod\limits_{s\in
[M]}\pi_{U_{\{l_{k+1},l_j\}}^\prime}(l_{0s})+ \alpha_1\prod\limits_{s\in
[M]}\pi_{U_{\{l_{k+1},l_j\}}^\prime}(l_{1s})) = 0
\]
Now
\[ l_{j}\in \ML(T_i)\Rightarrow
\pi_{U_{\{l_{k+1},l_j\}}^\prime}(T_i)=0\Rightarrow
\prod\limits_{t\in [N_1]}\pi_{U_{\{l_{k+1},l_j\}}^\prime}(G_t)\prod\limits_{s\in
[M]}\pi_{U_{\{l_{k+1},l_j\}}^\prime}(l_{(1-i)s})= 0.
\]
Since $G_t \mid G$, by Part \ref{retaineddetector} of
Lemma \ref{filteredfactor} $\pi_{U_{\{l_{k+1},l_j\}}^\prime}(G_t)\neq 0$ for all
$t\in [N_1]$. If
for some $s\in [M]$,
$\pi_{U_{\{l_{k+1},l_j\}}^\prime}(l_{(1-i)s})=0$ then
$l_{(1-i)s}\in sp(\{l_{j},l_{k+1}\})\Rightarrow l_{(1-i)s}\in
sp(\{l_{1},\ldots,l_{k},l_{k+1}\})
\Rightarrow l_{(1-i)s}\in sp(\{l_{1},\ldots,l_{k}\}) $ (by definition of
Detector
Pair in \ref{detectorset}).

\[
 l_{(1-i)s} \in sp(\{l_j,l_{k+1}\}) \text{ and } \{l_{(1-i)s},l_j\} \text{ LI }
\Rightarrow l_{k+1}\in sp(\{l_{(1-i)s},l_j\})
\]

This means $l_{k+1}\in sp(\{l_{1},\ldots,l_{k},l_{(1-i)s}\})\subset
sp(\{l_1,\ldots,l_k\})$ which is a contradiction to $l_{k+1}\in D$.
So $\pi_{U_{\{l_{k+1},l_j\}}^\prime}(f)\neq 0$ for all $j\in [k]
 \Rightarrow l_{k+1}\in X$. Therefore $D\subset X$.\\

\underline{($X\subset \ML(T_i)$) : } Consider $l\in X$. We need to show $l\in
\ML(T_{i})$. We already know $l\in \MC$.
\begin{itemize}
\item  If $l\in \ML(T_{1-i})$, then $\pi_{U_{\{l,l_j\}}^\prime}
(f) = 0$ for all
$j\in [k]$ since $l\mid T_{1-i}$ and $l_j\mid T_i$.
Contradiction to $l\in X$.

\item If $l\in \MC\setminus (\ML(T_i)\cup \ML(T_{1-i}))$  by Part
\ref{candidateprop} of Claim \ref{candidate} we know that there exists
$j\in [k]$ such that $sp(\{l_{j},l\})\cap \ML(T_{1-i}) \neq \phi$. Let
$l_{j}^\prime \in sp(\{l_{j},l\})\cap \ML(T_{1-i})$. We show that
$sp(\{l_j^\prime,l_j\}) = sp(\{l_j,l\}) = U_{\{l_j,l\}}$.
\begin{itemize}
\item $l_j^\prime \in sp(\{l_j,l\})\Rightarrow sp(\{l_j^\prime,l_j\})\subset
sp(\{l_j,l\})$.	
\item Let $l_j^\prime = \alpha l_j + \beta l$. We know that $\{l_j,l_j^\prime\}$
are LI since $l_j\in \ML(T_i)$ and $l_j^\prime \in \ML(T_{1-i})$. So $\beta\neq
0 \Rightarrow l\in sp(\{l_j^\prime,l_j\})\Rightarrow sp(\{l,l_j\}) \subset
sp(\{l_j^\prime,l_j\}) \Rightarrow sp(\{l,l_j\})=sp(\{l_j^\prime,l_j\})$.
\end{itemize}
Use the same extension for $sp(\{l,l_j\})=sp(\{l_j^\prime,l_j\})= U_{\{l_j,l\}}$ to get
$\pi_{U_{\{l,l_j\}}^\prime}(f)=\pi_{U_{\{l_j^\prime,l_j\}}^\prime}
(f) =0$ (since $l_j^\prime \mid T_{1-i}$ and $l_j\mid T_i$).
Contradiction to $l\in X$.
\end{itemize}

Therefore $l\in \ML(T_i) \Rightarrow X\subset \ML(T_i)$.

\section{Proofs from Subsection \ref{hardcase}}

\begin{claim}\label{calculationproof}
The following is true

\[
  \frac{(2-v(\delta,\theta))}{v(\delta,\theta)}\leq \frac{1-\delta}{\delta}
\]

\end{claim}

\emph{Proof.}
Note that
\[
 \frac{(2-v(\delta,\theta))}{v(\delta,\theta)} =
  \begin{cases}
      \hfill \frac{1+\delta + \theta}{1-\delta-\theta}    \hfill & \text{ if
$|\ML(T_{0})|\leq \theta |\ML(T_1)|$} \\
      \hfill \frac{3-(1-\delta)(1+\theta)}{(1-\delta)(1+\theta)-1} \hfill &
\text{ if $\theta |\ML(T_1)| < |\ML(T_0)| \leq |\ML(T_1)|$} \\
  \end{cases}
\]
By simple computation $\delta \in (0, \frac{7-\sqrt{37}}{6})$ gives
\[
3\delta^2-7\delta +1 >0 \Rightarrow 0 <
\frac{3\delta}{1-\delta} < 1-3\delta < 1 \Rightarrow \frac{1+\delta +
\theta}{1-\delta-\theta}<
\frac{1-\delta}{\delta}
\]

Also
\[
\theta > \frac{3\delta}{1-\delta} \Rightarrow
\frac{3-(1-\delta)(1+\theta)}{(1-\delta)(1+\theta)-1} <
\frac{1-\delta}{\delta}
\]

\begin{lemma}\label{largedetectorproof}
Let $k=c_{\F}(3)+2$ (see definition of $c_\F(k)$ in Theorem \ref{rankbound}). Fix $\delta, \theta$ in range given in Claim \ref{calculation} above .
Then for some $i\in \{0,1\}$ there exists a Detector Pair
$(S=\{l_{1},\ldots,l_{k}\},D)$ in $\ML(T_i)$  with
$|D|\geq v(\delta,\theta) \max(|\ML(T_{0})|,|\ML(T_{1})|)$.

\end{lemma}

\emph{Proof.}
We assume $|\ML(T_0)|\leq \ML(T_1)$. The other case gives the same result for(maybe) a different value of $i$ .
 We will consider linear forms as points in the space $\F^r$. Let's consider the
two cases used in the definition of $v(\delta,\theta)$.
 \begin{itemize}
  \item \underline{{\bf Case 1 : $|\ML(T_{0})|\leq \theta |\ML(T_1)|$ ( i.e. $\ML(T_0)$ is
much smaller ) $\Rightarrow v(\delta,\theta) = 1-\delta-\theta$ : }}

   Since $dim(\ML(T_1)) \geq r-1 \geq C_{2k-1} > C_k$ (See Section \ref{incidence}
for definition of $C_{k}$) by Corollary \ref{elementary} there exists a set $S$ of $k$ LI points say $S =\{l_{1},\ldots,l_{k}\}
\subseteq \ML(T_1)$ and a set
  $Z \subseteq \ML(T_1)$ of size $\geq (1-\delta)|\ML(T_1)|$ such that for any $l_{k+1}\in Z$
  \begin{itemize}
  \item $l_{k+1}\notin fl(\{l_{1},\ldots,l_{k}\})$.
  \item $fl(\{l_{1},\ldots,l_{k},l_{k+1}\})$ is elementary in $\ML(T_1)$.
  \end{itemize}
  Next we define our set $D$ according to the condition we needed in the definition of detector (See Subsection \ref{detectorset}).
\[
 D\eqdef \{ l_{k+1} \in Z : fl(\{l_{1},\ldots,l_{k},l_{k+1}\}) \cap \ML(T_{0})
\subset fl(\{l_{1},\ldots,l_{k}\})\}
\]
In the following lines we will show that this set $D$ has large size, to be precise:
\[
 |D|\geq (1-\delta-\theta)|\ML(T_1)|
\]
We do this in steps:
\begin{enumerate}
\item First we define a special subset of $Z$
\[
\tilde Z = \{l_{k+1}\in Z : (fl(\{l_1,\ldots,l_{k+1}\})\setminus fl(\{l_1,\ldots,l_k\})) \cap \ML(T_0) \neq \phi\}
\]
We claim that $Z\setminus \tilde Z \subset D$. Let $l_{k+1}\in Z\setminus \tilde{Z}\Rightarrow (fl(\{l_1,\ldots,l_{k+1}\})
\setminus fl(\{l_1,\ldots,l_k\})) \cap \ML(T_0) = \phi \Rightarrow fl(\{l_1,\ldots,l_{k+1}\})\cap \ML(T_0)\subset fl(\{l_1,\ldots,l_k\})$ and so $l_{k+1}\in D$.
 \item Next we show that for distinct $l_{k+1},\tilde l_{k+1}\in Z (\subseteq \ML(T_1))$
 \[
 (fl(\{l_{1},\ldots,l_{k},l_{k+1}\})\setminus fl(\{l_{1},\ldots,l_{k}\}))\cap (fl(\{l_{1},\ldots,l_{k},\tilde l_{k+1}\})\setminus fl(\{l_{1},\ldots,l_{k}\})) =\phi
 \]
If not then there exist scalars $\mu_j,\nu_j,j\in [k+1]$ such that
\[
\nu_1 l_1 + \ldots \nu_k l_k + \nu_{k+1}l_{k+1} = \mu_1 l_1 + \ldots \mu_k l_k + \mu_{k+1}\tilde l_{k+1}
\]
with $\nu_{k+1}\neq 0$ implying that $l_{k+1}\in sp(\{l_{1},\ldots,l_{k},\tilde l_{k+1}\})$.
Since all linear forms are $standard$ this implies
$l_{k+1}\in fl(\{l_{1},\ldots.l_{k},\tilde l_{k+1}\})$ (See Lemma \ref{spantoflat}). Also $l_{k+1}\in Z\Rightarrow l_{k+1}\notin fl(\{l_1,\ldots,l_k\})$. 
Together this
means that $l_{k+1}\in fl(\{l_1,\ldots,l_k,\tilde l_{k+1}\})\setminus fl(l_1,\ldots,l_k)$ and we arrive at a contradiction to $fl(\{l_1,\ldots,l_k,\tilde 
l_{k+1}\})$
being elementary.
\item From what we showed above every $l\in \ML(T_0)$ can belong to at most one of the sets $fl(\{l_1,\ldots,l_{k+1}\})\setminus fl(\{l_1,\ldots,l_k\})$
with $l_{k+1}\in Z$ (since intersection
between two such sets is $\phi$) and therefore there can be at most $|\ML(T_0)|$ such $l_{k+1}$'s in $\tilde Z$ $\Rightarrow$ $|\tilde Z| \leq |\ML(T_0)|$.
\end{enumerate}

So we get :
\[
 |D|\geq |Z|-|\ML(T_{0})|\geq (1-\delta-\theta)|\ML(T_1)|
\]
$(S,D)$ is a detector pair in $\ML(T_1)$ by the choice of $Z$ and $D$.
  \item \underline{{\bf Case 2 : $\theta |\ML(T_1)| < |\ML(T_0)| \leq |\ML(T_1)|$ ( i.e.
sizes are comparable ) $\Rightarrow v(\delta,\theta) = (1-\delta)(1+\theta)-1$ : }}

Since $dim(\ML(T_0)\cup \ML(T_{1})) = r > C_{2k-1}$, by Corollary \ref{elementary} we know that
there exist $2k-1$ independent points $l_1,\ldots,l_{2k-1} \in \ML(T_0)\cup
\ML(T_{1})$ and
a set $Z\subseteq \ML(T_0)\cup \ML(T_{1})$ of size $\geq
(1-\delta)(|\ML(T_0)|+|\ML(T_{1})|) $ such that for
all $l\in Z$
\begin{itemize}
 \item $l\notin fl(\{l_1,\ldots,l_{2k-1}\})$.
 \item $fl(\{l_1,\ldots,l_{2k-1}, l\})$ is elementary in $\ML(T_0)\cup
\ML(T_{1})$.
\end{itemize}

By pigeonhole principle, $k$ of the $\{l_j\}_{j=1}^{2k-1}$ points must belong to
either $\ML(T_0)$ or
$\ML(T_{1})$. Let's assume they belong to $\ML(T_i)$ (for some $i\in \{0,1\}$)
(say the points are $l_{1},\ldots,l_{k}$), then consider $D = Z\cap \ML(T_i)$.
Clearly for every
$l\in D$, $l\notin fl(\{l_{1},\ldots,l_{k}\})$ and
$fl(\{l_{1},\ldots,l_{k},l\})$ is elementary
in $\ML(T_0)
\cup \ML(T_{1})$. This immediately tells us that
$(S = \{l_{1},\ldots,l_{k}\},D)$ satisfies all properties of being a
detector pair in $\ML(T_i)$. We defined $D = Z\cap \ML(T_i)$. Since $Z\subseteq \ML(T_i)\cup \ML(T_{1-i})$ we have
$Z = (Z\cap \ML(T_i))\cup (Z\cap \ML(T_{1-i})) \subset D \cup \ML(T_{1-i})$ giving
\[
|D| + |\ML(T_{1-i})| \geq |Z| \Rightarrow |D| \geq |Z| - |\ML(T_{1-i})|\geq
(1-\delta)(|\ML(T_0)|+|\ML(T_{1})|)-|\ML(T_{1-i})|
\]
\[
\geq
((1-\delta)(1+\theta)-1)\max(|\ML(T_0)|,|\ML(T_1)|)
 \]

 \end{itemize}

 Combining the two cases we see that for some $i\in \{0,1\}$ there exists a Detector set
$(S=\{l_{1},\ldots,l_{k}\},D)$ in $\ML(T_i)$  with
$|D|\geq v(\delta,\theta) \max(|\ML(T_{0})|,|\ML(T_{1})|)$.

\begin{lemma}\label{findreconstructorproof}
The following are true:
\begin{enumerate}
 \item $dim(\pi_{W_0^\perp}({\ML(U_{1-i})}))> C_4$
 \item $\pi_{W_0^\perp}({\ML(U_{1-i})})\cap \pi_{W_0^\perp}({D}) = \phi$
 \item $|\pi_{W_0^\perp}({\ML(U_{1-i})})| \leq
\frac{1-\delta}{\delta}|\pi_{W_0^\perp}({D})|$
\end{enumerate}
\end{lemma}

\emph{Proof.}
\begin{enumerate}
 \item Since $dim({\ML(U_{1-i})})\geq r-1$ we get
  $dim(\pi_{W_0^\perp}({\ML(U_{1-i})})) \geq r-1-k >C_4$.
  \item Assume $\exists$ $d_1 \in D, u  \in \ML(U_{1-i})$ such that
$\pi_{W_0^\perp}({d})=\pi_{W_0^\perp}({u})
 \Rightarrow \exists \lambda,\nu\in \F$ such that $\nu d_1+\lambda u \in
W_0^\perp$. Since $\pi_{\tilde W_0}(d_1)\neq 0$ both $\nu,\lambda\neq 0$. Thus
   $u\in sp(\{l_1,\ldots,l_k,d_1\})
 \Rightarrow u \in fl(\{l_{1},\ldots,l_{k},d_1\})$ (using Lemma \ref{spantoflat} since all linear forms involved are
 \emph{standard} i.e. have coefficient of $x_1$ equal to $1$).
Also $u\in \ML(G T_{1-i})\Rightarrow u\in
fl(\{l_{1},\ldots,l_{k},d_1\})\cap (\ML(G)\cup \ML(T_{1-i}))$.
We know from Part \ref{retainedfactor} of Lemma \ref{filteredfactor} that
$fl(\{l_{1},\ldots,l_{k},d_1\})\cap \ML(G) = \phi \Rightarrow u \in
fl(\{l_{1},\ldots,l_{k},d_1\})\cap
\ML(T_{1-i})\subseteq fl\{l_{1},\ldots,l_{k}\}$ because $(S,D)$ was a detector pair.
But $u \in fl(\{l_{1},\ldots,l_{k}\})\Rightarrow d_1 \in sp(\{l_{1},\ldots,l_{k}\})$ which is
a contradiction because $d_1 \in D$ and $(S,D)$ is a detector pair.
  \item We first plan to show $\pi_{W_0^\perp}({\ML(U_{1-i})})\subset
\pi_{W_0^\perp}({\ML(T_{1-i})}) \cup
\pi_{W_0^\perp}({\ML(T_i)\setminus D})$.
 Clearly $U_{1-i}\mid G T_{1-i} \Rightarrow \ML(U_{1-i})\subset
\ML(G T_{1-i})\Rightarrow
 \pi_{W_0^\perp}({\ML(U_{1-i})})\subset
\pi_{W_0^\perp}({\ML(G T_{1-i})}) \subset \pi_{W_0^\perp}({\ML(G)})
\cup \pi_{W_0^\perp}({\ML(T_{1-i})})$. Now consider any $l\in \ML(G)$.
We know that $(S_0 = \{l_{1},\ldots,l_{k}\},D)$ is a detector pair, so by Part \ref{retainedfactor}
of Lemma \ref{filteredfactor} we get
 \[
  (fl(\{l_{1},\ldots,l_{k},l\})\setminus fl(\{l_{1},\ldots,l_{k}\})) \cap
(\ML(T_{1-i})\cup
  (\ML(T_i)\setminus D)) \neq \phi
 \]

So there exists $l^\prime \in \ML(T_{1-i})\cup (\ML(T_i)\setminus D)$ such that
$\pi_{W_0^\perp}(l), \pi_{W_0^\perp}(l^\prime)$ are both non-zero and
are LD $\Rightarrow \pi_{W_0^\perp}( l)=
\pi_{W_0^\perp}( l^\prime)$
implying that $\pi_{W_0^\perp}({\ML(G)})\subset
\pi_{W_0^\perp}({\ML(T_{1-i})\cup (\ML(T_i)\setminus D)})$ giving us
$\pi_{W_0^\perp}({\ML(U_{1-i})})\subset
\pi_{W_0^\perp}({\ML(T_{1-i})}) \cup
\pi_{W_0^\perp}({\ML(T_i)\setminus D})$ and therefore
\[
|\pi_{W_0^\perp}({\ML(U_{1-i})})|\leq|\pi_{W_0^\perp}({\ML(T_{1-i})})|
+ |\pi_{W_0^\perp}({\ML(T_i)\setminus D})|
\]
Now we try to show $|\pi_{W_0^\perp}({\ML(T_i)\setminus D})| =
|\pi_{W_0^\perp}({\ML(T_i)})|-|D|$
 \begin{enumerate}
 \item It's straightforward to see $\pi_{W_0^\perp}({\ML(T_i)}) =
\pi_{W_0^\perp}( D)\cup \pi_{W_0^\perp}({\ML(T_i)\setminus D})$. Also
$\pi_{W_0^\perp}({\ML(T_i)\setminus D}) \cap
\pi_{W_0^\perp}( D)=\phi$. If not then there exists $l^\prime \in
\ML(T_i)\setminus D, l^{\prime\prime} \in D$ such that
 $0\neq \pi_{W_0^\perp}( {l^{\prime\prime}}) = \pi_{W_0^\perp}(
l^\prime)\Rightarrow  \pi_{W_0^\perp}(l^{\prime\prime}),\pi_{W_0^\perp}(l^\prime)$ are
 LD $\Rightarrow l^\prime \in sp\{l_{1},\ldots,l_{k},l^{\prime\prime}\}\setminus
sp\{l_{1},\ldots,l_{k}\} \Rightarrow$ (by Lemma \ref{spantoflat}),  $ l^\prime \in
fl\{l_{1},\ldots,l_{k},l^{\prime\prime}\}\setminus
fl\{l_{1},\ldots,l_{k}\}$ which is a contradiction to
 the flat being elementary inside $\ML(T_i)$. So $|\pi_{W_0^\perp}({\ML(T_i)})| =
|\pi_{W_0^\perp}( D)| +  |\pi_{W_0^\perp}({\ML(T_i)\setminus D})|$.
 \item $\pi_{W_0^\perp}$ is injective on $ D$. Let
$\pi_{W_0^\perp}({l^\prime})=\pi_{W_0^\perp}({l^{\prime\prime}})$ for LI forms
$\{l^\prime, l^{\prime\prime}\}\subset D$, then
 $l^\prime \in sp(\{l_1,\ldots,l_k,l^{\prime\prime}\})\Rightarrow$ (by Lemma \ref{spantoflat}), $ l^\prime \in
fl(\{l_{1},\ldots,l_{k},l^{\prime\prime}\})$ and clearly $l^\prime \notin
fl\{l_{1},\ldots,l_{k}\}$ (since it's in $D$),
 which is again a contradiction to the flat being elementary , thus
$|\pi_{W_0^\perp}( D)| = | D| = |D|$ (since $D$ is a set of $normal$ linear forms ).
 \end{enumerate}
 Combining these with Claim \ref{calculation} and Lemma \ref{largedetector} we get
 \[
 |\pi_{W_0^\perp}({\ML(U_{1-i})})|\leq 2 \max(|\ML(T_0)|, |\ML(T_1)|)-|D|\leq
(2-v(\delta,\theta))\max(|\ML(T_0)|, |\ML(T_1)|)
\]
$\Rightarrow$
\[
 \frac{|\pi_{W_0^\perp}({\ML(U_{1-i})})|}{|\pi_{W_0^\perp}( D)|}
\leq
\frac{(2-v(\delta,\theta))}{v(\delta,\theta)}\leq \frac{1-\delta}{\delta}
\]

\end{enumerate}

\section{Proofs from Section \ref{highdimrecon}}
Our field $\F$ has characteristic zero. For simplicity let's assume it is an extension of $\Q$ and therefore contains $\Z$.
All random selections are done from the set $[N]=\{1,\ldots, N\}$.
\begin{lemma}\label{linindrandom}
Let $\F^n$ be the $n$ dimensional vector space over $\F$. Suppose $v_i :
i\in [n]$ are vectors in $\F^n$ with each co-ordinate chosen independently from the
uniform distribution on $[N]$. Consider the event
\[
 \ME = \{\{v_1,\ldots,v_n\} \text{ are LI }\}
\]
Then $Pr[\ME]\geq 1-\frac{n}{N}$.
\end{lemma}
\emph{Proof.}
Each $v_i\in \F^n$ is chosen such that each co-ordinate is chosen uniformly
randomly from the set $[N]$.
Let $v_i$ be the vector $(V_{i,1},\ldots,V_{i,n})$. Consider the matrix
$\tilde V = (V_{i,j})$. The $v_i$'s will be linearly
independent if and only if $\tilde V$ is invertible i.e. $det(V_{i,j})\neq 0$. Note that
$det(V_{i,j})$ is not the zero polynomial since the monomial $V_{1,1}V_{2,2}\ldots V_{n,n}$
has coefficient $1$. Now we can use Schwartz-Zippel Lemma \cite{Sax09} on this
polynomial to yield:
  \[
   Pr[det(\tilde V)=0]
\leq \frac{n}{N}
  \]
  Therefore $ Pr[\ME] = Pr[det(\tilde V)\neq 0] \geq
1-\frac{n}{N}$.

\begin{lemma}\label{linindproj}
Assume conditions in the previous lemma. For a fixed $r$, consider the subspaces $V = sp\{v_1,\ldots,v_r\}$
and $V^\prime = sp\{v_{r+1},\ldots,v_n\}$. Let's assume that that $\ME$ occurs i.e. $\{v_1,\ldots,v_n\}$ are LI. So $dim(V)=r$.
We know $\F^n = V\oplus V^\prime$.  Let $\pi_{V} : \F^n \rightarrow V$ be the
orthogonal projection onto $V$ under this decomposition . Let $T \subset \F^n$ be finite. Consider the event
\[
 \mathcal{F} = \{\exists \text{ an LI set } \{l_1,\ldots,l_r\}\subset T \text{ such that } \{\pi_V(l_1),\ldots,\pi_V(l_r)\} \text{ is LD }\}
\]
Then $Pr[\mathcal{F}]\leq {|T|\choose r}\{\frac{n}{N} + \frac{r(n-1)}{N}\}$

\end{lemma}

\emph{Proof.}
Fix $\{l_1,\ldots,l_r\}\subset T$ an LI set. Extend it to get a basis $\{l_1,\ldots,l_n\}$ of $\F^n$.
Let $l_i = \sum\limits_{j\in [n]}L_{i,j}e_j$ and $L$ be the matrix $(L_{i,j})$. From the discussion
above we have $\tilde V=(V_{i,j})$. Now let $P_r$ be the $n\times n$ matrix
\[
P_r = \begin{bmatrix}
    I_r       & 0_{r,n-r} \\
    0_{n-r,r}      & 0_{n-r,n-r}
\end{bmatrix}
\]
where $I_r$ is the $r\times r$ identity matrix and $0_{p,q}$ is the $p\times q$ matrix with all $0$ entries. Also for
any $n\times n$ matrix $A$, define $M_r(A)$ to be the principal $r\times r$ minor of $A$. Consider the equation given by
\[
 det(M_r(P_r L co(\tilde V))) =0
\]
where $co(\tilde V)$ is the co-factor matrix of $\tilde V$. Since entries of $co(\tilde V)$ are polynomials in the $V_{i,j}$'s and $L$ is a fixed matrix,
the entries of $P_rLco(\tilde V)$ are polynomials in $V_{i,j}$'s. So $det(M_r(P_r L co(\tilde V)))$ is a polynomial in $V_{i,j}$'s.
This polynomial can't be identically $0$. Choose $V_{i,j}=L_{i,j}$, then since $\tilde V$ is invertible, $Lco(\tilde V) = det(L) I$ giving
$P_rLco(\tilde V) = det(L) P_r \Rightarrow det(M_r(P_r L co(\tilde V))) = det(L)\neq 0$. Degree of the polynomial $det(M_r(P_rLco(\tilde V)))$ is clearly
$\leq r(n-1)$. Therefore by Schwartz Zippel Lemma
\[
 Pr[det(M_r(P_r L co(\tilde V))) =0] \leq \frac{r(n-1)}{N}
\]
Consider the set
\[
 S(\{l_1,\ldots,l_r\}) = \{(V_{i,j}) : det(\tilde V)\neq 0 , det(M_r(P_r L co(\tilde V))\neq 0\}
\]
On this set $S(\{l_1,\ldots,l_r\})$, $\{v_1,\ldots,v_n\}$ is a basis and we have the following matrix equations :
\[
 \begin{bmatrix}
    v_1 \\
    . \\
    . \\
    v_n
\end{bmatrix} =  \tilde V  \begin{bmatrix}
    e_1 \\
    . \\
    . \\
    e_n
\end{bmatrix} \text{ and }
\begin{bmatrix}
    l_1 \\
    . \\
    . \\
    l_n
\end{bmatrix} =  L  \begin{bmatrix}
    e_1 \\
    . \\
    . \\
    e_n
\end{bmatrix}
\Rightarrow
\begin{bmatrix}
    l_1 \\
    . \\
    . \\
    l_n
\end{bmatrix} =  L\tilde V^{-1}  \begin{bmatrix}
    v_1 \\
    . \\
    . \\
    v_n
\end{bmatrix}
\]
and so
\[
 \begin{bmatrix}
    \pi_V(l_1) \\
    . \\
    \pi_V(l_r)
\end{bmatrix} =  \frac{1}{det(\tilde V)}M_r(P_rLco(\tilde V))  \begin{bmatrix}
    v_1 \\
    . \\
    v_r
\end{bmatrix}
\]

Therefore $\{\pi_V(l_1),\ldots ,\pi_V(l_r)\}$ is an LI set. Now $S(\{l_1,\ldots,l_r\})^c = \{ (V_{i,j}) : det(\tilde V)=0$ \text{ or } $det(M_rLco(M))=0 \} 
\Rightarrow
Pr[S(\{l_1,\ldots,l_r\})^c] \leq \frac{n}{N} + \frac{r(n-1)}{N}$. Next we vary $\{l_1,\ldots,l_r\}$ and apply union bound to get
\[
 Pr[\mathcal{F}] \leq \sum\limits_{\{l_1,\ldots,l_r\}\subset T}S(\{l_1,\ldots,l_r\})^c \leq {|T|\choose r}\{\frac{n}{N} + \frac{r(n-1)}{N}\}
\]

In our application $|T|=poly(d)$ and $r$ is a constant, so we choose $N=2^{d+n}$ and make this probability very small.

\begin{lemma}\label{lagrangeinterp}
Let $f|_{V}(\B{X}) = \sum\limits_{\{\B{\alpha} :
|\B{\alpha}|=d\}}a_{\B{\alpha}}\B{X}^{\B{\alpha}}$ be a homogeneous multivariate
polynomial of degree $d$ in $r$ variables $X_1,\ldots,X_r$. Let $p_i : 1\leq i
\leq {d+r-1 \choose r-1}$
be randomly chosen points in $V$ ( dimension $r$ random subspace of $\F^n$ chosen in the above lemmas).
Then with high probability one can
find all the $a_{\B{\alpha}}$.
\end{lemma}
\emph{Proof.}
We evaluate the polynomial at each of the $p_i$'s. So we have ${d+r-1 \choose
r-1}$ evaluations.
The number of coefficients is also ${d+r-1 \choose r-1}$ so we get a linear
system in the coefficients where the
matrix ($X$) entries are just monomials evaluated at the $p_i$'s. Since $f$ is
not identically zero clearly there exist values for the points $p_i$'s such that
the determinant of this matrix is non zero polynomial so it cannot be
identically zero. Now the degree of the determinant polynomial is bounded by
$d{d+r-1 \choose r-1} \leq poly((d+r)^{r})$.
So by Schwarz Zippel lemma
\[
Pr[a_{\B{\alpha}} \text{ is recovered correctly }] = Pr[det(X)\neq 0] \geq
1-\frac{poly(d^r)}{N}
\]

\newpage

\bibliography{cccpaper}
\bibliographystyle{plain}

%\addcontentsline{toc}{section}{Appendices}{Bibligoraphy}

%%%%%%%%%%%%%%%%%%%%%%%%%%%%%
\end{document}